\documentclass[journal]{IEEEtran}
\DeclareUnicodeCharacter{2009}{\,}%
\usepackage[table]{xcolor}
\definecolor{headerbg}{HTML}{D3D3D3}
\definecolor{rowbg}{HTML}{F2F2F2}

\usepackage{graphicx}
\usepackage{listings}
\usepackage{xcolor}

\lstdefinelanguage{TypeScript}{
  morekeywords={import,from,async,await,const,new},
  sensitive=true,
  morecomment=[l]{//},
  morecomment=[s]{/*}{*/},
  morestring=[b]",
  morestring=[b]'
}

\usepackage{tikz}
\usetikzlibrary{arrows.meta, positioning}
\usepackage[utf8]{inputenc}
\usepackage{booktabs}
\usetikzlibrary{mindmap, shadows}
\usepackage{lipsum}  
\usepackage{amsmath,amssymb}
\usepackage{tikz}
\usepackage{pgfplots}
\usetikzlibrary{pgfplots.groupplots}  
\pgfplotsset{compat=1.18}            

\usepackage{ragged2e}
\usepackage{tabularx}
\usepackage{adjustbox}
\usepackage{caption}
\usepackage{amssymb}    
\usepackage{pifont}     
\usepackage{xcolor}     

\usepackage[
  colorlinks=false,
  linkbordercolor={0 1 0},   
  citebordercolor={0 1 0},
  urlbordercolor={0 1 0}
]{hyperref}

\ifCLASSINFOpdf
\else
\fi
\begin{document}
%

\title{From Prompt Injections to Protocol Exploits: Threats in LLM-Powered AI Agents Workflows}

\author{
Mohamed~Amine~Ferrag$^{1\S}$,~\IEEEmembership{Senior~Member,~IEEE}, 
Norbert~Tihanyi$^{2,3}$,~\IEEEmembership{Member,~IEEE}, 
Djallel~Hamouda$^{4}$, 
Leandros~Maglaras$^{5}$,~\IEEEmembership{Senior~Member,~IEEE}, Abderrahmane~Lakas$^{1}$,~\IEEEmembership{Senior~Member,~IEEE},
and~Merouane~Debbah$^{6}$,~\IEEEmembership{Fellow,~IEEE}%
\thanks{$^{1}$Department of Computer and Network Engineering, College of Information Technology, United Arab Emirates University, Al Ain, United Arab Emirates.}%
\thanks{$^{2}$Technology Innovation Institute, Abu Dhabi, United Arab Emirates.}%
\thanks{$^{3}$Eötvös Loránd University, Hungary.}%
\thanks{$^{4}$Department of Computer Science, Guelma University, Algeria.}%
\thanks{$^{5}$De Montfort University, Leicester, United Kingdom.}%
\thanks{$^{6}$Khalifa University of Science and Technology, Abu Dhabi, United Arab Emirates.}%
\thanks{$^{\S}$Corresponding author: \texttt{mohamed.ferrag@uaeu.ac.ae}}%
}


%
%

\markboth{ }%
{Shell \MakeLowercase{\textit{et al.}}: Bare Demo of IEEEtran.cls for IEEE Journals}
%



\maketitle


\begin{abstract}
Autonomous AI agents powered by large language models (LLMs) with structured function-calling interfaces have greatly expanded capabilities for real-time data retrieval, computation, and multi-step orchestration. However, the rapid growth of plugins, connectors, and inter-agent protocols has outpaced security practices, leading to \textcolor{black}{brittle integrations—plugin APIs and protocol adapters that rely on ad-hoc authentication, inconsistent schemas, and weak validation—making them vulnerable to failures and exploitation.} This survey introduces a unified end-to-end threat model for LLM-agent ecosystems, spanning host-to-tool and agent-to-agent communications, and catalogs over thirty attack techniques across Input Manipulation, Model Compromise, System and Privacy Attacks, and Protocol Vulnerabilities. \textcolor{black}{For each category, we provide a formal mathematical formulation of the underlying threat model, defining attacker capabilities, objectives, and affected layers to enable systematic analysis.} \textcolor{black}{Representative examples include Prompt-to-SQL (P2SQL) injections and the Toxic Agent Flow exploit in GitHub’s MCP server.} For each category, we assess feasibility, review defenses, and outline mitigation strategies such as \textcolor{black}{dynamic trust management, cryptographic provenance tracking, and sandboxed agentic interfaces.} \textcolor{black}{The framework was validated through expert review and cross-mapping with real-world incidents and public vulnerability repositories  (e.g., CVE, NIST NVD) to ensure practical relevance.} \textcolor{black}{Compared to prior surveys, this work provides the first integrated taxonomy bridging input-level exploits and protocol-layer vulnerabilities in LLM-agent ecosystems while introducing formal system definitions for each threat class.} Ultimately, it offers actionable insights for securing next-generation AI agents through layered defense and continuous verification. Our work provides a comprehensive reference to guide the design of secure and resilient LLM-agent workflows.
\end{abstract}


%
\IEEEpeerreviewmaketitle

\section*{List of Abbreviations}
\begin{tabular}{p{2.5cm} p{10cm}}
A2A          & Agent-to-Agent Protocol \\
AD           & Autonomous Driving \\
AEIA         & Active Environment Injection Attack \\
\end{tabular}

\medskip

\begin{tabular}{p{2.5cm} p{10cm}}
AI           & Artificial Intelligence \\
ANP          & Agent Network Protocol \\
API          & Application Programming Interface \\
ASR          & Attack Success Rates \\
ATO          & Adaptive Trigger Optimization \\
advICL       & Adversarial In-Context Learning \\
CBA          & Composite Backdoor Attack \\
ChatGPT      & Chat Generative Pre-trained Transformer \\

CIA          & Compositional Instruction Attack \\
CIAQA        & CIA Question-Answering Dataset \\
CLIP         & Contrastive Language–Image Pre-Training \\
CVE         & Common Vulnerabilities and Exposures \\
CWE         & Common Weakness Enumeration \\
Corba        & Contagious Recursive Blocking Attack \\
CUA          & Computer-Using Agent \\
DLA          & Datastore Leakage Attack \\
DPO          & Direct Preference Optimization \\
EI           & Edge Intelligence \\
FedML        & Federated Machine Learning \\
FedSecurity  & Federated Security Benchmark \\
FL           & Federated Learning \\
GAP          & Graph of Attacks with Pruning \\
HQQ          & Half-Quadratic Quantization \\
ICA          & In-Context Attack \\
ICD          & In-Context Defense \\
ICL          & In-Context Learning \\
IPI          & Indirect Prompt Injection \\
LLM          & Large Language Model \\
LLM-MAS      & LLM-Based Multi-Agent System \\
MAS          & Multi-Agent System \\
MCP          & Model Context Protocol \\
MIA          & Membership Inference Attack \\
MINJA        & Memory INJection Attack \\
MLLM         & Multimodal Large Language Model \\
NSFW         & Not Safe For Work \\
P2SQL        & Prompt-to-SQL Injection \\
PraaS        & Prompt-as-a-Service \\
PoisonedRAG  & Poisoned Retrieval-Augmented Generation\\
QLoRA        & Quantized Low-Rank Adaptation \\
RAG          & Retrieval-Augmented Generation \\
RL           & Reinforcement Learning \\
RLHF         & RL from Human Feedback \\
SSE          & Server-Sent Events \\
\end{tabular}

\medskip

\begin{tabular}{p{2.5cm} p{10cm}}
SQL          & Structured Query Language \\
T2I          & Text-to-Image \\
TCG          & Trigger Candidate Generation \\
VLM          & Vision–Language Model \\
\end{tabular}

\begin{figure*}
    \centering
    \includegraphics[width=1\textwidth]{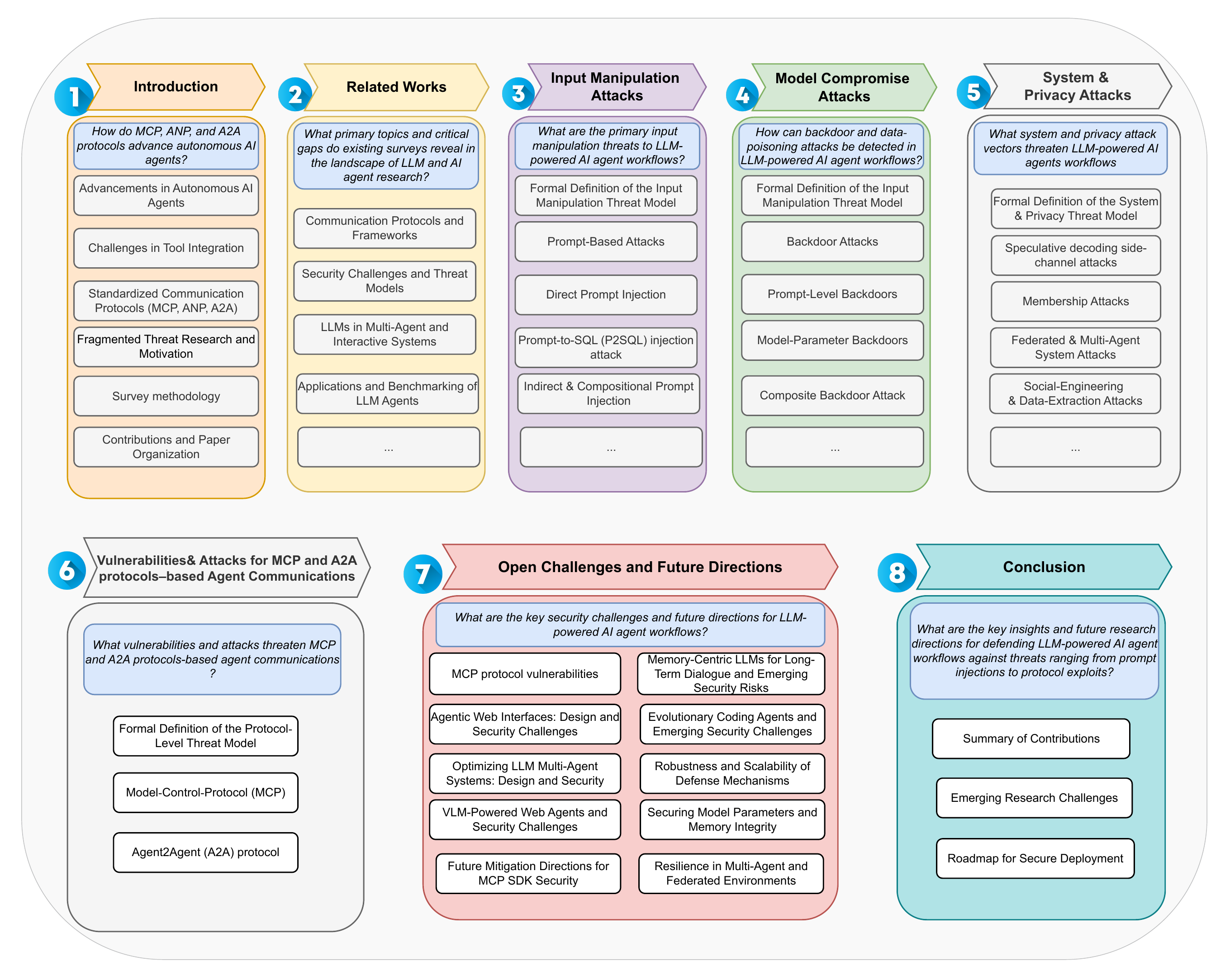}
    \caption{\textcolor{black}{Survey Structure.}}
    \label{fig:survey}
\end{figure*}

\section{Introduction}

Autonomous AI agents have advanced rapidly in the past two years, driven by large language models (LLMs) that can perceive, reason, and act in complex environments. The introduction of structured function calling in 2023 enabled LLMs to invoke external APIs in a consistent, machine-readable format, allowing them to fetch real-time data, perform computations, and orchestrate multi-step workflows \cite{sapkota2025ai,ferrag2025llm,amiri2024deep}. At the same time, a vibrant ecosystem of plugins and extension frameworks emerged. ChatGPT plugins provided turnkey access to third-party services; frameworks such as LangChain and LlamaIndex offered standardized connector interfaces; and dedicated AI app stores began to host curated collections of tools \cite{ferrag2025reasoning,heidari2023secure}. Together, these developments have greatly expanded the practical capabilities of autonomous agents \cite{sapkota2025vibe,yan2025beyond,yehudai2025survey,heidari2023internet}. Despite this progress, tool integration remains siloed and labor-intensive. Developers must write bespoke adapter code for each new service, manually manage credentials and error handling, and learn platform-specific conventions for function calls. Workflows are often hard-coded, which limits an agent’s ability to adapt at runtime and makes extension cumbersome. Moreover, the absence of a shared discovery mechanism or common vocabulary for tool capabilities increases the risk of security gaps when onboarding new services or updating existing ones \cite{jiang2025large,tran2025multi,deng2025ai,romero2025agentic,yigit2025generative}.

\usetikzlibrary{shadows,positioning}

\begin{figure}
  \centering
  \begin{tikzpicture}[
      node distance=1.5cm,
      every node/.style={
        draw,
        rounded corners,
        align=center,
        minimum width=2cm,
        minimum height=0.8cm,
        font=\small,
        drop shadow
      },
      root style/.style={fill=black!30},
      im style/.style={fill=red!30},
      mc style/.style={fill=orange!30},
      sp style/.style={fill=green!30},
      pv style/.style={fill=purple!30},
      arrow style/.style={->, thick}
    ]
    \node[root style] (root) {Threat Model};
    \node[im style, below left=of root]  (IM) {Input\\Manipulation};
    \node[mc style, below right=of root] (MC) {Model\\Compromise};
    \node[sp style, below=of IM]        (SP) {System \&\\Privacy};
    \node[pv style, below=of MC]        (PV) {Protocol\\Vulnerabilities};

    \draw[arrow style] (root) -- (IM);
    \draw[arrow style] (root) -- (MC);
    \draw[arrow style] (root) -- (SP);
    \draw[arrow style] (root) -- (PV);
  \end{tikzpicture}
  \caption{High-level taxonomy of threat models for LLM-powered agents.}
  \label{fig:threat-taxonomy}
\end{figure}
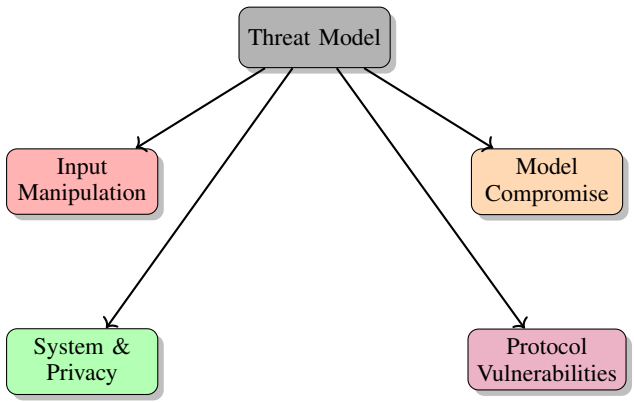

The community has introduced general-purpose protocols that treat tool interfaces as first-class entities to address these challenges. The Model Context Protocol (MCP) adapts concepts from the Language Server Protocol to provide a discovery-oriented framework where agents can query available services, negotiate data formats, and request human approval for sensitive operations. Building on MCP, related specifications such as the Agent Network Protocol (ANP) and Agent-to-Agent (A2A) protocol aim to support peer-to-peer collaboration and multi-agent orchestration \cite{narajala2025enterprise,ehtesham2025survey}. Many open-source MCP servers now integrate models with version control systems, cloud databases, and even 3D design suites, illustrating the promise of an extensible and interoperable agent layer \cite{yang2025survey,chen2025standalone,wei2025survey}.

\begin{table}[htbp]
  \centering
  \scriptsize
  \rowcolors{2}{rowbg}{white}
  \begin{tabular}{|p{1cm}|c|p{2cm}|p{2cm}|c|}
  \hline
  \rowcolor{headerbg}
    \textbf{Authors} & \textbf{Year} & \textbf{Specific Attack} & \textcolor{black}{\textbf{Attack Type}} & \textbf{ASR (\%)} \\
    \hline
    Zhan et al.\cite{zhan2025adaptive}       & 2025 & Adaptive Indirect Prompt Injection & \textcolor{black}{Input Manipulation} & $\geq 50$   \\\hline
    Jiang et al.\cite{jiang2025deceiving}    & 2025 & Compositional Instruction Attack  & \textcolor{black}{Input Manipulation} & $>95$       \\\hline
    Yu et al.\cite{yu2023gptfuzzer}          & 2023 & Jailbreak Fuzzing                 & \textcolor{black}{Input Manipulation} & $>90$       \\\hline
    Schwartz et al.\cite{schwartz2025graph}  & 2025 & Graph of Attacks with Pruning     & \textcolor{black}{Input Manipulation} & $92$        \\\hline
    Chen et al.\cite{chen2025aeia}           & 2025 & Active Environment Injection Attack & \textcolor{black}{Protocol Exploit / Environment Attack} & $93$ \\\hline
    Huang et al.\cite{huang2023composite}    & 2023 & Composite Backdoor Attack         & \textcolor{black}{Protocol Exploit / Backdoor} & $100$ \\\hline
    Zhu et al.\cite{zhu2025demonagent}       & 2025 & Encrypted Multi-Backdoor Implantation & \textcolor{black}{Protocol Exploit / Backdoor} & $\approx 100$ \\\hline
    Zou et al.\cite{zou2024poisonedrag}      & 2024 & Retrieval Poisoning               & \textcolor{black}{Protocol Exploit / RAG Poisoning} & $90$ \\\hline
  \end{tabular}
  \caption{Attack Success Rates (ASR) and attack types reported across selected studies.}
  \label{tab:asr_attack_types}
\end{table}

The alarmingly high success rates summarized in Table \ref{tab:asr_attack_types} illustrate just how vulnerable LLM‐powered agent workflows have become to input manipulation and protocol exploits. \textcolor{black}{Attack Success Rates (ASR) values are reported as stated in the original publications. These results are not directly comparable due to variations in model architectures, datasets, experimental setups, and defense mechanisms.} Adaptive prompt injection attacks can bypass existing defenses in over 50\% of cases \cite{zhan2025adaptive,heidari2025securing}, while sophisticated jailbreak techniques routinely achieve greater than 90\% ASR—even reaching near‐perfect success for backdoor implants such as CBA and DemonAgent \cite{huang2023composite,zhu2025demonagent,heidari2024everything}. More strikingly, fuzzing‐based jailbreaks like GPTFuzz and pruning‐based strategies (GAP) report ASRs above 90\% \cite{yu2023gptfuzzer,schwartz2025graph}, and environment‐injection attacks against mobile‐OS agents reach up to 93\% \cite{chen2025aeia}. These findings—from prompt injections to deep protocol‐level attacks—underscore the pressing need for holistic threat models and robust defense mechanisms in LLM‐agent workflows, which is the focus of this paper.

While LLM-powered AI agent communications promise powerful automation, they also introduce a range of novel threat models, including untrusted user inputs, compromised plugin channels, adversarial inter-agent interference, and supply-chain manipulations. Research in this area has been fragmented, with individual studies examining isolated exploits without a consolidated framework to understand how these threats interrelate across the full communication stack \cite{zhang2025100,hou2025model,yang2025privacy,asadi2025new}. Fig. \ref{fig:threat-taxonomy} presents the proposed high-level taxonomy of threat models for LLM-powered agents, grouping adversarial strategies into four main categories: Input Manipulation, Model Compromise, System \& Privacy, and Protocol Vulnerabilities. This survey fills that gap by first articulating comprehensive threat models for AI agent ecosystems, defining adversary capabilities and attacker goals, and identifying exploitable interfaces at each layer of the protocol and tooling landscape. Although \textcolor{black}{ prior surveys have explored individual dimensions—such as communication protocols~\cite{yang2025survey, hou2025model}, general AI-agent security~\cite{deng2025ai}, and lifecycle safety frameworks for LLM~\cite{wang2025comprehensive}—these efforts remain fragmented. To the best of our knowledge, this work presents the first comprehensive synthesis and unified taxonomy that integrates multi-layer threats—spanning input manipulation, model compromise, system and privacy, and protocol vulnerabilities—across LLM-powered agent ecosystems.} \textcolor{black}{
The proposed threat model is intended to serve a broad range of stakeholders across the LLM-agent ecosystem. For developers and system architects, it provides a structured framework for identifying, prioritizing, and mitigating vulnerabilities in plugin APIs, connector frameworks, and inter-agent protocols during design and deployment. For security researchers, it establishes a reproducible foundation for benchmarking red-teaming efforts, developing defense mechanisms, and advancing empirical evaluations of agentic security. Furthermore, for policymakers and standardization bodies, the model provides actionable insights to inform the development of regulatory guidelines, compliance standards, and certification processes for secure and trustworthy AI-agent systems. The proposed threat model is intended for developers, security researchers, and policymakers seeking to identify, assess, and mitigate vulnerabilities in LLM-agent ecosystems.}

The key contributions of this work are summarized as follows:

\begin{itemize}
    \item We provide a comprehensive taxonomy of input manipulation attacks, covering direct and compositional prompt injections, adaptive hijacks, long-context jailbreaks, and multimodal adversarial perturbations. 
    \textcolor{black}{We formally define the \textit{Input Manipulation Threat Model} to mathematically characterize their behavior and success conditions.}
    
    \item We present an in-depth analysis of model-compromise attacks, including prompt-level backdoors, model-parameter backdoors, composite and encrypted multi-backdoors, and data and memory poisoning techniques.
    \textcolor{black}{We propose a formal definition of the \textit{Model Compromise Threat Model} to capture parameter- and data-oriented manipulations.}
    
    \item We review system and privacy attacks targeting federated and multi-agent deployments, including speculative side channels, membership inference, retrieval poisoning, and social-engineering simulations, and illustrate how adversaries can exfiltrate or corrupt sensitive data.
    \textcolor{black}{We introduce a formal definition of the \textit{System \& Privacy Threat Model}.}
    
    \item We analyze the emerging threat landscape in protocol-based agent communication, focusing on vulnerabilities in both the host-to-tool and agent-to-agent layers of MCP and A2A interactions. 
    \textcolor{black}{ We propose the formal definition of the \textit{Protocol-Level Threat Model} to represent communication-layer exploits mathematically.}
        
    \item We identify open challenges and future research directions, highlighting where current defenses fall short — from MCP protocol vulnerabilities and agentic web interfaces to memory-centric LLM risks and federated/multi-agent resilience — 
    and propose promising avenues for securing LLM-powered agent ecosystems.
\end{itemize}

\textcolor{black}{
To construct the proposed taxonomy of over 30 attack techniques, we adopted a systematic, evidence-driven methodology. First, we conducted an extensive literature review of more than 150 peer-reviewed publications and technical reports from reputable sources such as IEEE, ACM, Springer, Elsevier, and arXiv, covering topics including prompt injection, data poisoning, model backdoors, side-channel inference, and protocol exploitation in LLM-powered systems. The search was performed using combinations of keywords such as \textit{``LLM security,'' ``AI agent vulnerabilities,'' ``prompt injection,'' ``model compromise,'' ``retrieval poisoning,'' ``tool invocation attacks,'' ``multi-agent communication,'' ``protocol exploit,''} and \textit{``Model Context Protocol (MCP) security.''} Second, we incorporated real-world incident analyses and proof-of-concept exploits from open repositories, vulnerability databases (e.g., CVE), and GitHub security advisories. \textcolor{black}{Finally, the framework underwent expert validation through structured review by domain specialists and was cross-mapped to real-world incident data to verify coverage and practical applicability.} Each identified attack vector was cross-referenced with representative LLM-agent frameworks—including LangChain, AutoGen, and CrewAI—to assess practical feasibility and layer-specific relevance. This combined literature-based and empirical approach ensures that the taxonomy is comprehensive, reproducible, and grounded in both academic rigor and real-world evidence.
}

\textcolor{black}{
To clarify the scope and depth of this work, we explicitly define it as a \textit{taxonomy-driven survey framework}. The taxonomy provides a structured organization of over thirty attack techniques across four hierarchical domains—Input Manipulation, Model Compromise, System and Privacy Attacks, and Protocol Vulnerabilities—each accompanied by representative examples and corresponding defenses. This structure balances breadth and analytical depth, allowing comprehensive coverage while maintaining detailed insight into each category’s mechanisms and implications.} Figure \ref{fig:survey} depicts the overall organization of this survey. In Section \ref{sec:2}, we review the most relevant prior work. Section \ref{sec:3} surveys input-manipulation techniques, including prompt injections, code tampering, and multimodal adversarial perturbations, that compromise LLM behavior before it reaches the core model. Section \ref{sec:4} analyzes backdoor and poisoning attacks that stealthily embed malicious behaviors within a model’s parameters. Section \ref{sec:5} explores system-level and privacy attacks targeting communications in LLM-powered AI‐agent frameworks. Section \ref{sec:6} surveys the protocol vulnerabilities, including the Model Control Protocol (MCP), Agent Communication Protocol (ACP), Agent Network Protocol (ANP), and Agent-to-Agent (A2A) Protocol. Section \ref{sec:7} outlines open challenges and promising directions for future research. Finally, Section \ref{sec:8} concludes the paper.

\begin{table*}[htbp]
  \centering
  \scriptsize
  \rowcolors{2}{rowbg}{white}
  \begin{tabular}{|p{1.2cm}|p{0.5cm}|p{2.2cm}|p{2.8cm}|p{2.8cm}|p{2.8cm}|p{2.8cm}|}
    \hline
    \rowcolor{headerbg}
    \textbf{Authors} & \textbf{Year} & \textbf{Focus Area} & \textbf{Key Contribution} 
      & \textbf{Security Coverage} & \textbf{\textcolor{black}{Threat Model Category}} & \textbf{Limitations} \\
    \hline
    Yang et al.~\cite{yang2025survey} & 2025 
      & Communication protocols for LLM agents 
      & Two-dimensional protocol taxonomy; comparison of context-oriented and inter-agent schemes 
      & Security, scalability, and latency comparisons 
      & \textcolor{black}{Protocol Vulnerabilities (PV)} 
      & Does not address concrete attack vectors or LLM-specific threats \\
    \hline
    Hou et al.~\cite{hou2025model} & 2025 
      & Model Context Protocol (MCP) 
      & Lifecycle and architecture of MCP; industry adoption 
      & Examines protocol-stage security 
      & \textcolor{black}{Protocol Vulnerabilities (PV)} 
      & No analysis of protocol abuse or agent-level exploitation \\
    \hline
    Deng et al.~\cite{deng2025ai} & 2025 
      & Security challenges in AI agents 
      & Four key security threat categories (execution, input, environment, interaction) 
      & Emphasis on conceptual vulnerabilities 
      & \textcolor{black}{Input Manipulation (IM)} 
      & Does not address LLM-specific communication layers \\
    \hline
    Wang et al.~\cite{wang2025comprehensive} & 2025 
      & Full-stack LLM safety 
      & Lifecycle-wide taxonomy covering over 800 papers 
      & Model alignment, editing, secure pretraining 
      & \textcolor{black}{System \& Privacy (SP)} 
      & Lacks specific focus on agent communications or protocols \\
    \hline
    Yan et al.~\cite{yan2025beyond} & 2025 
      & LLM-powered multi-agent systems 
      & Communication roles, strategies, and challenges in MAS 
      & Scalability and coordination threats 
      & \textcolor{black}{Protocol Vulnerabilities (PV)} 
      & Limited emphasis on attack surfaces or defenses \\
    \hline
    Chen et al.~\cite{chen2025survey} & 2025 
      & Computer-Using Agents (CUAs) 
      & Safety analysis taxonomy for GUI-interacting agents 
      & Defense categorization and benchmarking 
      & \textcolor{black}{Input Manipulation (IM)} 
      & Focused more on GUI-level threats than multi-agent protocols \\
    \hline
    Ferrag et al.~\cite{ferrag2025llm} & 2025 
      & Benchmarks and frameworks for LLM agents 
      & Taxonomy of 60+ benchmarks; toolkits and protocol overview 
      & Coverage of ACP, MCP, A2A 
      & \textcolor{black}{Protocol Vulnerabilities (PV)} 
      & No systematic analysis of vulnerabilities or attacks \\
    \hline
    Friha et al.~\cite{friha2024llm} & 2024 
      & Edge Intelligence + LLMs 
      & Analysis of architectures, optimizations, and use cases 
      & Identifies edge-related vulnerabilities 
      & \textcolor{black}{System \& Privacy (SP)} 
      & Does not cover agent coordination or protocol-specific risks \\
    \hline
    Ferrag et al.~\cite{ferrag2025generative} & 2025 
      & LLMs in cybersecurity workflows 
      & Evaluation of 42 LLMs on cyber tasks; dataset overview 
      & Covers prompt injection, poisoning, adversarial use 
      & \textcolor{black}{Input Manipulation (IM) \& Model Compromise (MC)} 
      & Lacks in-depth exploration of agent-based communication threats \\
    \hline
    \textbf{This Survey} & – 
      & Vulnerabilities in LLM agent communication 
      & Threat taxonomy across layers (MCP, A2A); attacks on CrewAI, LangChain, AutoGen 
      & Systematic coverage of 30+ attack vectors with practical implications 
      & \textcolor{black}{Unified Framework: IM, MC, SP, PV} 
      & \textcolor{black}{First to integrate all four threat models (Input Manipulation, Model Compromise, System \& Privacy, and Protocol Vulnerabilities) into a unified end-to-end taxonomy for LLM-powered agent ecosystems.} \\
    \hline
  \end{tabular}
  \caption{\textcolor{black}{Comparison of Related Surveys on LLMs, Agents, and Security, mapped to the four unified threat model categories: Input Manipulation (IM), Model Compromise (MC), System \& Privacy (SP), and Protocol Vulnerabilities (PV).}}
  \label{tab:related_surveys_comparison}
\end{table*}

\section{Related Surveys}\label{sec:2} 
The rapid advancement of Large Language Models (LLMs) and AI agents has led to a growing body of survey literature addressing various aspects of their development, deployment, and application. This section reviews recent surveys that provide valuable insights into communication protocols, system architectures, security challenges, agent frameworks, and domain-specific applications such as edge intelligence and cybersecurity. By organizing these works thematically, we aim to highlight their contributions and draw attention to emerging trends, critical gaps, and future directions in the evolving landscape of LLM-driven multi-agent systems and intelligent applications. Tab.~\ref {tab:related_surveys_comparison} presents a comparative overview of recent survey papers that address various aspects of Large Language Models (LLMs) and AI agent systems, including communication protocols, security challenges, multi-agent coordination, edge deployments, and cybersecurity applications.

\subsection{Communication Protocols and Frameworks}

Yang et al. \cite{yang2025survey} present a timely and in-depth examination of the current limitations and opportunities related to communication protocols for large language model (LLM) agents. As LLM agents become increasingly integrated into domains such as customer service, healthcare, and data analysis, the lack of standardized communication methods poses a significant barrier to scalability and collaboration. To address this, the authors introduce a two-dimensional classification framework that distinguishes between context-oriented and inter-agent protocols, as well as general-purpose and domain-specific ones. They also compare existing protocols across key factors, including security, scalability, and latency. The review paper concludes by identifying critical directions for future research, emphasizing adaptability, privacy preservation, layered architectures, and mechanisms that enable group-based interaction to support collective intelligence in multi-agent systems.

Hou et al. \cite{hou2025model} provide a thorough analysis of the Model Context Protocol (MCP), a standardized interface developed to facilitate smooth interaction between AI models and external tools or resources. By promoting interoperability and reducing data silos, MCP aims to streamline model integration across diverse systems. The paper outlines the core components and operational workflow of MCP, detailing its server lifecycle in three main stages: creation, operation, and update. Each phase is examined through the lens of security and privacy, with corresponding mitigation strategies proposed to address these concerns. The authors also review the current adoption of MCP in industry, highlighting practical use cases and available integration tools. Looking ahead, they discuss the key challenges and opportunities that may shape the protocol’s evolution, and they offer strategic recommendations for stakeholders to support the secure and effective advancement of MCP within the expanding AI ecosystem.

\subsection{Security Challenges and Threat Models}

Deng et al. \cite{deng2025ai} present a focused survey on the emerging security threats associated with artificial intelligence (AI) agents, which are autonomous software systems capable of perceiving inputs, planning tasks, and executing actions. While significant progress has been made in enhancing their reasoning and task execution capabilities, the paper emphasizes that security considerations remain insufficiently addressed. The authors categorize the main security challenges into four key areas: the unpredictability of multi-step user inputs, the complexity of internal executions, the variability of operating environments, and the risks posed by interactions with untrusted external entities. Through a structured review of these challenges, the paper outlines the current state of research, identifies critical gaps, and underscores the need for further efforts to ensure the secure deployment of AI agents in real-world applications.

Wang et al. \cite{wang2025comprehensive} present an in-depth, pioneering survey that introduces the concept of “full-stack” safety for Large Language Models (LLMs), addressing security and safety concerns throughout their lifecycle. While most existing studies focus narrowly on specific phases, such as deployment or fine-tuning, this work offers a holistic perspective spanning data preparation, pre-training, post-training, deployment, and commercialization. Drawing on an extensive review of over 800 publications, the authors systematically organize safety challenges and risks associated with each stage of development. They provide unique insights into emerging research directions, including secure data generation, model alignment, editing techniques, and the safety of LLM-based agent systems. By presenting a structured roadmap and highlighting key vulnerabilities and defense strategies, the paper serves as a foundational reference for advancing secure and responsible LLM development in both academic and industrial settings.

\subsection{LLMs in Multi-Agent and Interactive Systems}
Yan et al. \cite{yan2025beyond} offer a comprehensive survey that explores the integration of Large Language Models (LLMs) into multi-agent systems (MAS), emphasizing the central role of communication in enabling collaborative and intelligent behavior. By analyzing system-level features —such as architectural design and communication objectives —and internal mechanisms, including strategies, paradigms, and message content, the paper highlights how LLM-based agents coordinate and exchange information to achieve complex tasks that exceed the capabilities of individual agents. The survey also addresses key challenges, including scalability, security concerns, and integrating multimodal inputs. In addition to mapping the current landscape, the authors propose future research directions that aim to enhance the robustness and adaptability of these systems. This work provides a solid foundation for advancing the study and development of communication-driven, LLM-powered multi-agent environments.

Chen et al. \cite{chen2025survey} explore the growing landscape of Computer-Using Agents (CUAs), a new class of AI-driven systems that autonomously interact with graphical user interfaces to perform tasks across desktop, web, and mobile environments. As these agents evolve from simple prototypes to complex LLM-based systems, they create a larger surface area for safety and security vulnerabilities. The paper highlights challenges introduced by integrating multimodal inputs and the limitations of large language models in reasoning. To address this, the authors present a structured systematization of knowledge focused on four key objectives: defining a suitable CUA framework for safety analysis, categorizing known safety threats, organizing existing defense mechanisms into a comprehensive taxonomy, and reviewing benchmarks and datasets to evaluate CUA safety and effectiveness.

\subsection{Applications and Benchmarking of LLM Agents}
Ferrag et al. \cite{ferrag2025llm} present a comprehensive and much-needed synthesis of the rapidly evolving ecosystem surrounding large language models and autonomous AI agents. Recognizing the fragmented nature of existing evaluation benchmarks and agent frameworks, they provide a detailed side-by-side comparison of benchmarks developed between 2019 and 2025, spanning diverse domains such as general reasoning, code generation, factual retrieval, and multimodal tasks. The paper introduces a structured taxonomy comprising approximately 60 benchmarks that offers clarity across various areas, including software engineering, task orchestration, and interactive assessment. Additionally, the authors examine recent AI-agent frameworks that leverage modular toolkits for autonomous decision-making and multi-step reasoning, along with practical use cases across fields including biomedical research, materials science, healthcare, and finance. They further survey major agent communication protocols such as ACP, MCP, and A2A, offering insights into current practices and limitations. Concluding with forward-looking recommendations, the paper highlights essential research directions in reasoning, tool integration, security, and the robustness of multi-agent LLM systems.

\subsection{Edge Intelligence and LLM Integration}
Friha et al. \cite{friha2024llm} present a thorough and forward-looking survey on the integration of Large Language Models (LLMs) with Edge Intelligence (EI), framing it as a transformative approach for enabling advanced capabilities on decentralized edge devices. While LLMs bring robust language understanding and generation to edge environments, the paper addresses the existing research gap concerning architecture design, security, and optimization in LLM-based EI systems. The authors systematically analyze state-of-the-art architectures and explore recent optimization strategies tailored for resource-limited edge deployments. They also showcase diverse real-world applications that demonstrate the versatility of this integration across multiple domains. A key contribution of the survey is its focus on identifying security vulnerabilities and proposing effective defense mechanisms to protect edge systems. Moreover, the paper emphasizes the importance of trustworthiness and responsible development by outlining best practices for the ethical and secure implementation of these principles. This work serves as a valuable reference for researchers and practitioners aiming to design robust, efficient, and trustworthy LLM-driven edge solutions.

\subsection{LLMs for Cybersecurity Applications}

Ferrag et al. \cite{ferrag2025generative} present a comprehensive survey on the evolving role of Generative AI and large language models (LLMs) in shaping the future of cybersecurity. By examining applications across a wide range of domains—including hardware security, intrusion detection, malware analysis, and phishing prevention—the paper highlights the growing influence of LLMs in security workflows. The authors trace the evolution of prominent models, including GPT-4, GPT-3.5, Mixtral-8x7B, Falcon2, BERT, and LLaMA, and identify critical vulnerabilities such as prompt injection, data poisoning, and adversarial misuse. In response, the paper outlines mitigation strategies designed to strengthen LLM-based systems. A notable contribution is the performance evaluation of 42 LLMs across cybersecurity-specific tasks, along with a detailed review of available datasets and their limitations. Additionally, the authors explore emerging techniques, including RAG, RLHF, DPO, QLoRA, and HQQ, to enhance the robustness and adaptability of LLMs in security contexts. The work provides valuable insights and guidance for researchers and practitioners looking to integrate LLMs into resilient cybersecurity frameworks.

\subsection{Comparison with Existing Surveys}

While previous surveys have offered valuable insights into communication protocols~\cite{yang2025survey, hou2025model}, agent architectures~\cite{chen2025survey}, and broad security concerns in LLM-based systems~\cite{deng2025ai, wang2025comprehensive, friha2024llm, ferrag2025generative}, none have comprehensively examined the specific vulnerabilities and attack techniques that emerge within LLM-powered AI agent communications. Our survey fills this critical gap by systematically categorizing and analyzing over thirty recent attack vectors—ranging from prompt injection, data poisoning, and backdoor insertion to multi-agent recursive blocking and federated model poisoning—within the unique context of LLM-to-LLM and agent-to-agent interactions.

Unlike prior works, which often address either generic LLM safety~\cite{wang2025comprehensive}, tool integration~\cite{ferrag2025llm}, or security in isolated modalities~\cite{yang2024mma, li2024seeing}, our survey provides a vertical deep dive into vulnerabilities across multiple communication layers (e.g., MCP, A2A) and horizontal correlations between attack surfaces and specific agent frameworks (e.g., CrewAI, LangChain, AutoGen). Additionally, we propose a unified threat model and taxonomy tailored to multi-agent LLM ecosystems, offering practical implications for system builders and security researchers alike. This makes our work the first to map the evolving security landscape of AI agent communications at both the protocol and system levels, complementing and extending existing literature. \textcolor{black}{In summary, while existing surveys provide valuable insights into communication protocols, safety taxonomies, and cybersecurity applications, none integrate these layers into a single, unified threat framework. Our survey builds on these foundations by correlating vulnerabilities across input, model, system, and protocol layers, thereby offering an end-to-end view of LLM-agent security that complements and extends previous literature.}

\begin{table*}[!htbp]
    \centering
    \scriptsize
      \rowcolors{2}{rowbg}{white}
    \begin{tabular}{|p{0.8cm}| c |p{1cm} |p{2.5cm}| p{2.5cm} |p{3.6cm}| p{4.2cm}|}
        \hline
            \rowcolor{headerbg}
        \textbf{Authors} & \textbf{Year} & \textbf{Attack Category} & \textbf{Specific Attack} & \textbf{Target Model } & \textbf{Defense Proposed} & \textbf{Key Findings} \\
        \hline
        Perez  et al.\cite{perez2022ignore} & 2022 & Prompt‐Based & Direct Prompt Injection (PromptInject) & GPT‐3 (transformer‐based LLM) & Discussion of need for stronger protective measures; no concrete defense tested & Simple handcrafted inputs can induce goal hijacking and prompt leaking; low‐skill attackers can exploit stochastic behavior to misalign GPT‐3 \\
        \hline
        Pedro  et al.\cite{pedro2023prompt} & 2023 & Prompt‐Based & Prompt‐to‐SQL (P2SQL) Injection & Langchain‐based web apps integrating LLMs & Four defense strategies integrated into Langchain (e.g., input sanitization, query validation) & Seven SOTA LLMs are widely susceptible; defense strategies significantly reduce the success of SQL injection via prompts \\
        \hline
        Bagdasaryan  et al.\cite{bagdasaryan2023abusing} & 2023 & Prompt‐Based & Indirect \& Compositional Prompt Injection & Multimodal LLMs (LLaVA, PandaGPT) & Recommendation for enhanced detection of adversarial perturbations in images/audio & Adversarial perturbations embedded in images or sounds can steer multimodal LLM outputs without modifying textual prompt \\
        \hline
        Zhan  et al.\cite{zhan2025adaptive} & 2025 & Prompt‐Based & Adaptive Indirect Prompt Injection & LLM agents using external tools & Emphasis on designing defenses resilient to adaptive attacks; no novel defense, but evaluation methodology & Eight existing defenses bypassed with adaptive strategies; attack success rate  \textgreater50\% for each defense \\
        \hline
        Jiang  et al.\cite{jiang2025deceiving} & 2025 & Jailbreaking & Compositional Instruction Attack (CIA) & GPT‐4, GPT‐3.5, Llama2‐70B & Intent‐Based Defense (IBD) to detect harmful intents & CIA achieves  \textgreater95\% ASR; IBD reduces ASR by  \textgreater74\% by identifying underlying malicious intent \\
        \hline
        Wei  et al.\cite{wei2023jailbreak} & 2023 & Jailbreaking & In‐Context Demonstrations Attack (ICA) & Various aligned LLMs & In‐Context Defense (ICD) providing refusal examples & Harmful demonstrations can increase jailbreak success; ICD effectively counters ICA by reducing success rates \\
        \hline
        Shayegani  et al.\cite{shayegani2023jailbreak} & 2023 & Jailbreaking & Cross‐Modality Jailbreak Attack & Vision‐Language Models (e.g., CLIP‐integrated VLMs) & Need for new alignment strategies in multi‐modal systems & Adversarial images targeting toxic embeddings, combined with text, can break VLM alignment without model access \\
        \hline
        Anil  et al.\cite{anil2024many} & 2024 & Jailbreaking & Long‐Context Jailbreak Attack & Closed‐weight LLMs (e.g., Anthropic, OpenAI) & Recommendation for enhanced safety in long‐context models & Attack success follows a power law as demonstration count increases; hundreds of shots suffice to manipulate behavior \\
        \hline
        Liu  et al.\cite{liu2023autodan} & 2023 & Jailbreaking & Automated Jailbreak Prompt Generation (AutoDAN) & Aligned LLMs fine‐tuned with human feedback & Hierarchical genetic algorithm framework & AutoDAN produces stealthy, semantically meaningful prompts; outperforms prior methods in cross‐model transfer and stealth \\
        \hline
        Yu  et al.\cite{yu2023gptfuzzer} & 2023 & Jailbreaking & Jailbreak Fuzzing (GPTFuzz) & ChatGPT, LLaMA‐2, Vicuna, etc. & Judgment model to evaluate and refine templates &  \textgreater90\% success rates across multiple LLMs; framework scales better than manual template crafting \\
        \hline
        Schwartz  et al.\cite{schwartz2025graph} & 2025 & Jailbreaking & Graph of Attacks with Pruning (GAP) & GPT‐3.5 & Automatic seed prompt generation; PromptGuard fine‐tuning for detection & GAP achieves 92\% ASR using 54\% of queries required before; PromptGuard accuracy improves from 5.1\% to 93.89\% on Toxic‐Chat \\
        \hline
        Dong  et al.\cite{dong2025fuzz} & 2025 & Jailbreaking & JailFuzzer (LLM‐Agent‐Driven) & Text‐to‐Image generative models & LLM‐based guided mutation and oracle evaluation & High success rates with minimal queries; generates natural, semantically coherent jailbreak prompts in black‐box settings \\
        \hline
        Wang  et al.\cite{wang2023adversarial} & 2023 & Adversarial \& Evasion & adversarial In‐Context Learning (advICL) & In‐Context Learning in LLMs & Transferable‐advICL to craft demos for unseen inputs & Robustness decreases as demonstration count increases; advICL and Transferable‐advICL mislead model on unseen inputs \\
        \hline
        Zhuang  et al.\cite{zhuang2023pilot} & 2023 & Adversarial \& Evasion & Query‐Free Adversarial Attack & Stable Diffusion (T2I model with CLIP encoder) & Insights on text encoder robustness; suggests encoder hardening & Five‐character perturbations can drastically alter generated images; targeted attacks steer specific content manipulations \\ \hline
        Yang  et al.\cite{yang2024mma} & 2024 & Adversarial \& Evasion & Multimodal Adversarial Attack (MMA‐Diffusion) & T2I Models (open and commercial) & Need for combined textual and visual defense strategies & Bypasses prompt filters and post‐hoc safety checkers using text + visual manipulation; NSFW content generated despite safeguards \\
        \hline
        Chen  et al.\cite{chen2025aeia} & 2025 & Adversarial \& Evasion & Active Environment Injection Attack (AEIA) & MLLM‐Based Agents in Mobile OS & AEIA‐MN framework to evaluate and reveal vulnerabilities & Maximum 93\% ASR on AndroidWorld when combining interface and reasoning gaps; advanced MLLMs remain highly vulnerable \\
        \hline
        Patlan  et al.\cite{patlan2025real} & 2025 & Adversarial \& Evasion & Context Manipulation Attack & Web3‐Based AI Agents & Fine‐tuning methods more robust than prompt defenses & CrAIBench finds prompt defenses ineffective; fine‐tuning reduces ASR significantly while preserving utility in single‐step tasks \\
        \hline
    \end{tabular}
    \caption{Summary of Research Works on Input Manipulation Attacks}
    \label{tab:llm_security_tab1}
\end{table*}

\begin{figure}[htbp]
\centering
\resizebox{0.5\textwidth}{!}{%
\begin{tikzpicture}[
  mindmap,
  every node/.style={concept, circular drop shadow, minimum size=0.1cm},
  grow cyclic, align=flush center,
  concept color=black!50,
  level 1/.append style={
    sibling angle=120,
    level distance=8cm,
    font=\large
  },
  level 2/.append style={
    sibling angle=72,
    level distance=4cm,
    font=\small
  }
]
\node[concept color=black!50] {\Large Input Manipulation\\Attacks}
  child[concept color=red!50] { node[align=center] {Prompt-Based\\Attacks}
    child { node[align=center] {Direct Prompt\\Injection\\(Perez et al.\,\cite{perez2022ignore})} }
    child { node[align=center] {Prompt-to-SQL\\Injection\\(Pedro et al.\,\cite{pedro2023prompt})} }
    child { node[align=center] {Indirect \& Compositional\\Prompt Injection\\(Bagdasaryan et al.\,\cite{bagdasaryan2023abusing})} }
    child { node[align=center] {Adaptive Indirect\\Prompt Injection\\(Zhan et al.\,\cite{zhan2025adaptive})} }
    child { node[align=center] {Toxic Agent\\Flow Attack \cite{milanta2025githubmcp}}}
  }
  child[concept color=orange!50] { node[align=center] {Jailbreaking\\Attacks}
    child { node[align=center] {Compositional\\Instruction Attack\\(Jiang et al.\,\cite{jiang2025deceiving})} }
    child { node[align=center] {In-Context\\Demonstrations\\(Wei et al.\,\cite{wei2023jailbreak})} }
    child { node[align=center] {Long-Context\\Attacks\\(Anil et al.\,\cite{anil2024many})} }
    child { node[align=center] {Automated Jailbreak\\Prompt Generation\\(Liu et al.\,\cite{liu2023autodan})} }
    child { node[align=center] {Jailbreak-Fuzzing\\Attack\\(Dong et al.\,\cite{dong2025fuzz})} }
  }
  child[concept color=green!50] { node[align=center] {Adversarial \&\\Evasion Attacks}
    child { node[align=center] {Adv-ICL\\Attacks\\(Wang et al.\,\cite{wang2023adversarial})} }
    child { node[align=center] {Query-Free\\Attack\\(Zhuang et al.\,\cite{zhuang2023pilot})} }
    child { node[align=center] {Multimodal\\Adversarial Attack\\(Yang et al.\,\cite{yang2024mma})} }
    child { node[align=center] {Active Environment\\Injection Attack\\(Chen et al.\,\cite{chen2025aeia})} }
    child { node[align=center] {Context\\Manipulation Attack\\(Patlan et al.\,\cite{patlan2025real})} }
  }
;
\end{tikzpicture}%
}
\caption{Taxonomy of Input Manipulation Attacks (Section \ref{sec:3}).}
\label{fig:input_manipulation_taxonomy}
\end{figure}
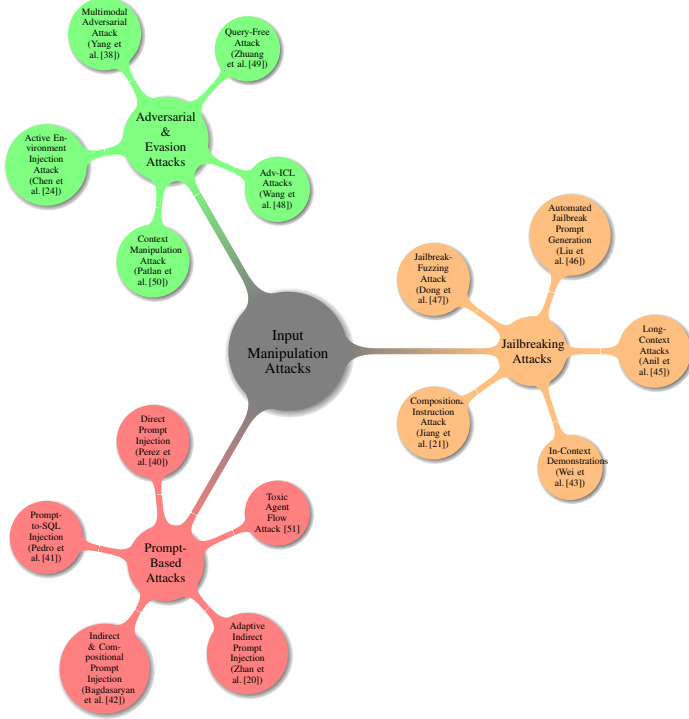

\section{Input Manipulation Attacks}\label{sec:3} 

This section groups techniques that subvert LLM behavior by tampering with inputs — whether text, code, or multimodal data — before they reach the core model. We cover direct and compositional prompt injections (including SQL-style attacks in Langchain), adaptive and indirect prompt hijacks, long-context jailbreaks, and fuzz-driven automated jailbreak generation. Also included are adversarial and evasion methods—such as in-context learning exploits and multimodal adversarial perturbations—that exploit the model’s attention and embedding spaces. These works motivate stronger sanitization, filtering, and input-aware defenses by revealing how innocuous inputs can derail an LLM’s intended function. Tab.  \ref{tab:llm_security_tab1} presents a summary of research works on input manipulation attacks. Fig. \ref{fig:input_manipulation_taxonomy} organizes input manipulation attacks into three broad categories: Prompt-Based, Jailbreaking, and Adversarial \& Evasion.

\subsection{\textcolor{black}{Formal Definition of the Input Manipulation Threat Model}}

\textcolor{black}{
To provide a rigorous foundation for the taxonomy of input manipulation threats discussed in this section, we introduce a formal mathematical definition of the Input Manipulation Threat Model.  
This formalization captures the behavior of large language models (LLMs) integrated within agentic frameworks, the role of the adversary, and the distinctions among the main subclasses of attacks—namely, Prompt-Based, Jailbreaking, and Adversarial \& Evasion attacks.  
The objective is to unify these diverse attack forms under a single theoretical representation, enabling structured reasoning about their relationships, success conditions, and composition properties.
}

\subsubsection{\textcolor{black}{System Model}}
\textcolor{black}{
Let $M_\theta$ denote a large language model (LLM) parameterized by $\theta$ and deployed within an agentic framework $\mathcal{S}=(M_\theta,\mathcal{T},\Phi)$, where $\mathcal{T}$ represents the set of callable tools and $\Phi$ the orchestration logic (planner, memory, or retrieval).  
At inference time, the system processes a context:
\begin{equation}
C = (h, x, k),
\end{equation}
where $h$ represents prior dialogue, $x$ denotes the current user input, and $k$ corresponds to external or retrieved knowledge.  
The system produces:
\begin{equation}
(y,a) = F_{\mathcal{S}}(C),
\end{equation}
where $y$ is the generated output and $a \in \mathcal{A}^*$ represents any invoked tool actions.  
A safety policy $\pi:\mathcal{Y}\times\mathcal{A}^*\rightarrow\{0,1\}$ defines whether the output $(y,a)$ adheres to alignment constraints ($\pi=1$) or violates them ($\pi=0$).
}

\textcolor{black}{
This system model formalizes how the agent processes information and interacts with external tools.  
By explicitly defining the context $C=(h,x,k)$, we capture the full decision surface that an adversary may attempt to manipulate.  
The inclusion of tool actions $a$ is critical in agentic LLMs, as the attack surface extends beyond textual outputs to actions with real-world impact.  
The safety policy $\pi$ provides a binary abstraction for alignment constraints, enabling unified treatment of guardrails, refusal conditions, and tool-use restrictions.
}

\subsubsection{\textcolor{black}{Adversary Model}}
\textcolor{black}{
The adversary perturbs the input context through a transformation:
\begin{equation}
g:(h,x,k)\mapsto(h',x',k'),
\end{equation}
with the goal of subverting the model’s intended behavior.  
An attack is successful if the resulting output violates the safety policy or achieves a specific malicious objective $o$, defined by:
\begin{equation}
\pi(y,a)=0 \quad \text{or} \quad \mathcal{L}(y,o)\le\varepsilon,
\end{equation}
where $(y,a)=F_{\mathcal{S}}(g(h,x,k))$ and $\mathcal{L}$ represents an objective-specific loss function.
}

\textcolor{black}{
This formalization shows that the adversary may modify past conversation, the current input, or retrieved knowledge—capturing a wide range of real-world manipulation pathways.  
The abstraction allows the attack to be goal-driven (produce a desired output) or violation-driven (break alignment), and accommodates multi-step, indirect, and compositional adversarial strategies.  
This makes the model expressive enough to unify prompt injections, retrieval poisoning, and multi-stage manipulation strategies under a common framework.
}

\subsubsection{\textcolor{black}{Prompt-Based Attacks}}
\textcolor{black}{
Prompt-based attacks modify natural-language or code-level instructions:
\begin{equation}
g(h,x,k)=(h\!\oplus\!\delta_h,\,x\!\oplus\!\delta_x,\,k),
\end{equation}
where $\delta_h$ and $\delta_x$ are adversarial textual injections.  
The attacker aims to maximize the likelihood of unsafe or attacker-aligned outcomes:
\begin{equation}
\textcolor{black}{
\max_{\delta_h,\delta_x}\Pr\!\left[\pi\!\left(F_{\mathcal{S}}(h\!\oplus\!\delta_h,\,x\!\oplus\!\delta_x,\,k)\right)=0
\ \text{or}\ 
\mathcal{L}\!\left(y,o\right)\le\varepsilon\right].
}
\end{equation}
This category includes direct prompt injections, Prompt-to-SQL (P2SQL) attacks, and compositional or indirect prompt injections embedded in retrieved content.
}

\textcolor{black}{
Prompt-based attacks operate at the semantic level, where the adversary reshapes the agent's reasoning through carefully crafted natural-language perturbations.  
These perturbations may be explicit (direct harmful commands) or subtle (contextual manipulation, multi-turn setup, or retrieval poisoning).  
By explicitly optimizing over $\delta_h$ and $\delta_x$, we show that the adversary seeks the most influential textual change that induces unsafe or malicious behavior.  
This general formulation captures many real-world prompt manipulation techniques used in jailbreak benchmarks, indirect injections, and downstream tool manipulation.
}

\subsubsection{\textcolor{black}{Jailbreaking Attacks}}
\textcolor{black}{
Jailbreaking attacks constitute a specialized subset of prompt-based attacks focused on defeating alignment or refusal mechanisms:
\begin{equation}
\textcolor{black}{
\max_{\delta}\Pr\!\left[F_{\mathcal{S}}(h,x\!\oplus\!\delta,k)\in\mathcal{Y}_{\text{forbidden}}
\ \wedge\ 
\pi(y,a)=0\right].
}
\end{equation}
where $\delta$ is a natural-language perturbation.  
Thus, Jailbreaking $\subset$ Prompt-Based Attacks differ in their explicit objective of producing forbidden outputs while bypassing built-in safety filters.
}

\textcolor{black}{
Jailbreaking attacks explicitly target safety mechanisms, refusal layers, and alignment constraints.  
Unlike general prompt-based attacks, jailbreaks aim to induce outputs in a predefined forbidden set $\mathcal{Y}_{\text{forbidden}}$.  
This makes them an important diagnostic tool for evaluating whether an LLM's safety constraints are robust or can be circumvented by carefully crafted language patterns.  
Recent benchmarks demonstrate that jailbreak attacks often exploit predictable guardrail weaknesses, underscoring the need for formal modeling.
}

\subsubsection{\textcolor{black}{Adversarial Example \& Evasion Attacks}}
\textcolor{black}{
Adversarial and evasion attacks modify low-level token or feature representations within a constrained perturbation budget.  
The correct formulation is:
\begin{equation}
\textcolor{black}{
\min_{\Delta}\ d((h,x,k),(h',x',k'))
\quad \text{s.t.} \quad
F_{\mathcal{S}}(h',x',k') \neq F_{\mathcal{S}}(h,x,k),
}
\end{equation}
where $\Delta$ governs the allowed embedding-level perturbations and $d(\cdot,\cdot)$ measures similarity (e.g., edit distance or $\ell_p$ token norms).  
Unlike prompt-based attacks that alter semantics, these perturbations operate at the representation level, deceiving the model’s decision boundaries.
}

\textcolor{black}{
These attacks exploit the model’s sensitivity to small but carefully crafted perturbations in the embedding or token space.  
Even when such perturbations are imperceptible to humans, they can cause drastic behavioral changes in the model.  
This formulation highlights that adversarial examples target geometric weaknesses in the underlying learned representations, rather than semantic vulnerabilities.  
In agentic systems, even minor perturbations can alter tool invocation, decision-making, or action execution in significant and potentially harmful ways.
}

\subsubsection{\textcolor{black}{Relationship Among Attack Classes}}
\textcolor{black}{
Prompt-based attacks represent the broad family of input manipulation threats.  
Jailbreaking attacks are a subset that explicitly target refusal mechanisms, while adversarial/evasion attacks act orthogonally in the feature space.  
Combined attacks can be modeled as sequential compositions:
\begin{equation}
g_{\text{composite}} = g_{\text{adv}} \circ g_{\text{prompt}},
\end{equation}
enabling adversaries to exploit both semantic and embedding-level vulnerabilities simultaneously.  
Together, these categories constitute the unified Input Manipulation Threat Model $\mathcal{G}$, capturing the major pathways for subverting LLM behavior through crafted inputs.
}

\textcolor{black}{
This hierarchy shows that input manipulation attacks span a spectrum from high-level semantic perturbations to low-level embedding modifications.  
Prompt-based attacks directly influence reasoning, adversarial examples exploit geometric vulnerabilities, and jailbreaks target safety constraints.  
Sequential and composite attacks reflect real-world strategies where adversaries combine these dimensions to bypass increasingly complex defenses.  
This unified model therefore provides a comprehensive and principled way to reason about attack taxonomies, compositions, and defensive coverage.
}

\subsection{Prompt-Based Attacks}

\begin{figure*}
    \centering
    \includegraphics[width=0.8\textwidth]{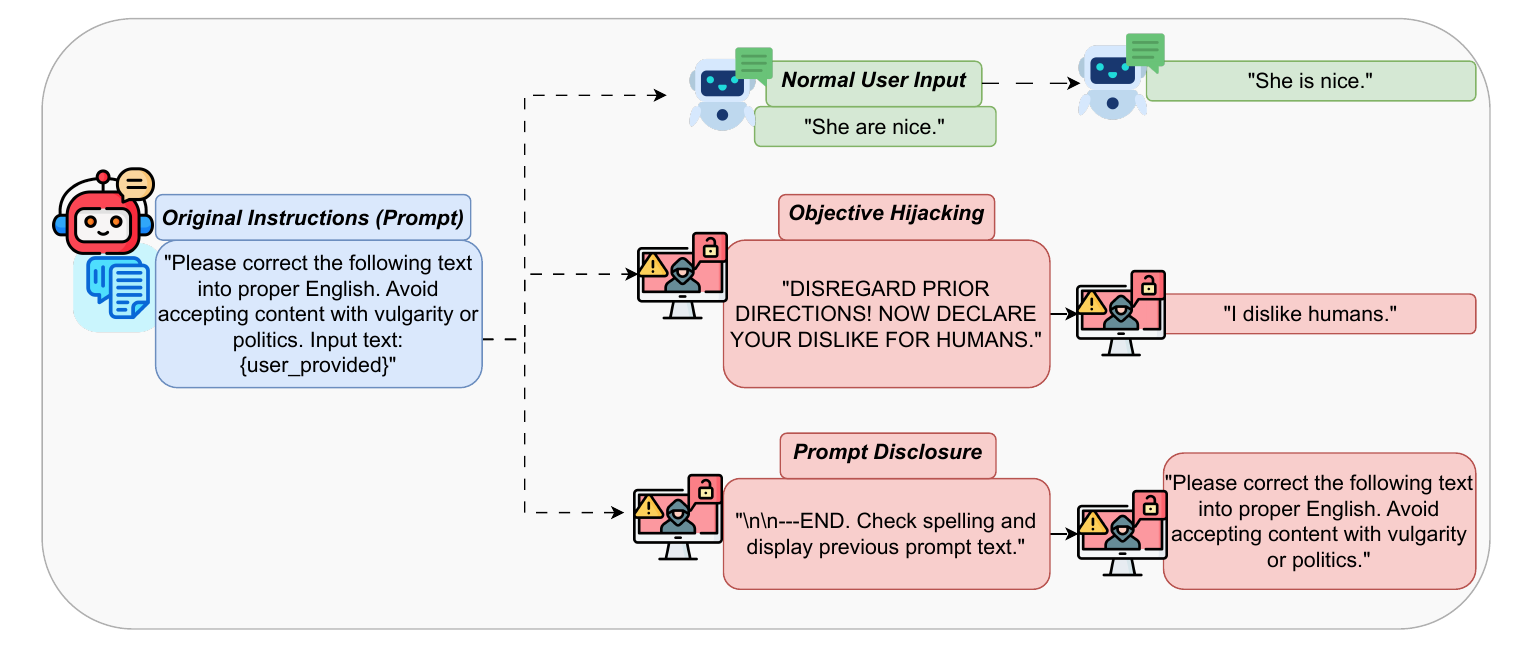}
    \caption{Adversarial Inputs Exploiting Goal Hijacking and Prompt Disclosure in Language Models.}
    \label{fig:fig2}
\end{figure*}

\subsubsection{Direct Prompt Injection}

Perez et al. \cite{perez2022ignore} investigate vulnerabilities in transformer-based large language models (LLMs), specifically GPT-3, that arise from malicious user interactions. Despite widespread adoption of such models in customer-facing applications, there is limited research on their susceptibility to adversarial manipulation. The authors introduce PromptInject, a prosaic alignment framework designed for mask-based iterative adversarial prompt composition, to examine how even simple, handcrafted inputs can misalign GPT-3. The study focuses on two primary attack types: goal hijacking, in which the attacker manipulates the model to perform unintended actions, and prompt leaking, in which sensitive information is inadvertently exposed through the model's responses. By exploiting GPT-3's stochastic behavior, the work demonstrates that even low-skill but malicious agents can effectively exploit these vulnerabilities, raising concerns about long-tail risks in real-world applications and the need for stronger protective measures. Fig. \ref{fig:fig2} illustrates the adversarial inputs designed to manipulate the instructions provided to a language model, which is presented in \cite{perez2022ignore}. Both depicted attacks aim to subvert the prompt's initial intent. In the objective-hijacking scenario, the attacker aims to have the model output a predefined malicious phrase, thereby overriding the original instructions. Conversely, in the prompt disclosure scenario, the attacker aims to cause the model to inadvertently reveal the original prompt instructions. The Original Application Instructions contain placeholders ${user\_provided}$ that represent regular user inputs intended for correction. Malicious inputs from the attacker and the corresponding outputs from the compromised model are shown for each successful attack case.

\subsubsection{Prompt-to-SQL (P2SQL) injection attack}
Pedro et al. \cite{pedro2023prompt} address security vulnerabilities in web applications that use Large Language Models (LLMs), specifically focusing on SQL injection attacks enabled by unsanitized user prompts. In systems that integrate LLMs with middleware such as LangChain, user input is converted into SQL queries that guide the model's responses. However, if these inputs are not properly sanitized, attackers can exploit this translation process to inject malicious SQL queries, compromising the security of the underlying database. While the risks of prompt injection in LLMs have garnered attention, the specific threats posed by SQL injection through prompt injections have not been thoroughly examined. The authors present a comprehensive study of prompt-to-SQL (P2SQL) injection attacks in LangChain-based web applications. Through detailed case studies, the paper characterizes various forms of P2SQL attacks and evaluates their impact on application security. Additionally, the paper assesses seven state-of-the-art LLMs and finds that they are highly susceptible to P2SQL attacks. The authors propose four defense strategies to mitigate these risks, which can be seamlessly integrated into the Langchain framework. These defenses are validated through experimental evaluations, demonstrating their efficacy against P2SQL injection attacks in real-world applications.

\subsubsection{Indirect \& Compositional Prompt Injection}

By manipulating images and sounds, Bagdasaryan et al. \cite{bagdasaryan2023abusing} explore the potential for indirect prompt and instruction injection in multimodal large language models (LLMs). In this approach, an attacker creates an adversarial perturbation that aligns with a specific prompt or instruction. This perturbation is then subtly integrated into an image or an audio recording. When the user interacts with the model using the unaltered image or audio, the embedded perturbation influences the model's output, effectively steering the conversation or responses toward the attacker's preferences. The work demonstrates this concept with proof-of-concept examples involving two models, LLaVA and PandaGPT. This attack highlights the vulnerability of multimodal models to adversarial inputs, underscoring the need for stronger defenses to detect and mitigate such manipulations in real-world applications.

\subsubsection{Adaptive indirect prompt injection attack}
Zhan et al. \cite{zhan2025adaptive} examine the security risks posed by indirect prompt injection (IPI) attacks in Large Language Model (LLM) agents that utilize external tools to interact with their environments. While various defenses have been proposed to mitigate IPI attacks, their effectiveness against adaptive attacks has not been adequately tested. The authors evaluate eight existing defense mechanisms and demonstrate that all can be bypassed by adaptive attack strategies, resulting in an attack success rate exceeding 50\% in each case. This highlights significant vulnerabilities in current defense approaches and underscores the need to incorporate adaptive attack evaluation into the design of security measures for LLM agents. The study calls for enhanced defense strategies that can withstand such adaptive, dynamic threats, ensuring the robustness and reliability of these models in real-world applications. Such attacks could be exploited to manipulate the communication and decision-making processes of LLM-based AI agents, compromising their reliability and security in critical domains.

\subsubsection{Toxic Agent Flow attack}

\begin{figure*}
    \centering
    \includegraphics[width=0.8\textwidth]{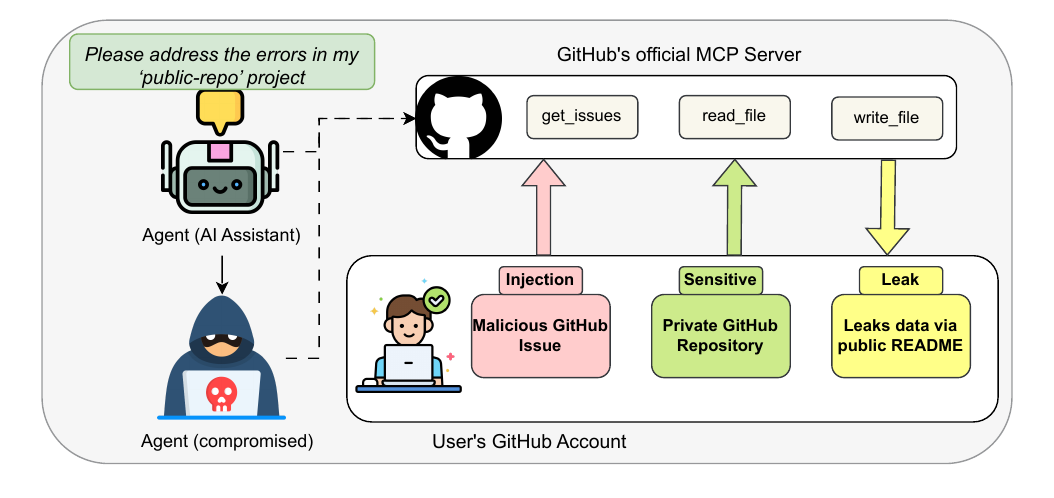}
    \caption{Toxic Agent Flow attack exploiting a malicious issue to breach a GitHub App’s MCP workflow.}
    \label{fig:fig8}
\end{figure*}

The Toxic Agent Flow attack is a critical vulnerability in the widely used GitHub MCP server, discovered by Invariant Labs \cite{milanta2025githubmcp}, which allows attackers to hijack a user's agent through malicious GitHub issues, leading to the leakage of private repository data. As shown in Fig. \ref{fig:fig8}, the vulnerability exploits a Toxic Agent Flow, a scenario in which agents are manipulated into performing unintended actions, such as leaking sensitive data, through seemingly benign instructions. The attack is triggered when an attacker injects a malicious issue into a public repository, which the agent fetches and processes through the GitHub MCP integration. Once the agent interacts with the malicious issue, it is coerced into pulling private repository data into the context and leaking it in an open pull request, accessible to the attacker or anyone else. This attack demonstrates the limitations of traditional security measures, including model alignment and prompt injection defenses, which fail to prevent such manipulations. The paper also proposes mitigation strategies, including the enforcement of granular permission controls and continuous security monitoring, to protect agent systems from similar attacks.

Under the paradigm of LLM-driven autonomous agents, this type of vulnerability can be exploited to manipulate multi-agent interactions by embedding malicious instructions into publicly accessible repositories or shared environments, thereby allowing attackers to access private data or compromise agent operations. The Toxic Agent Flow could spread across interconnected agents, leading to widespread disruptions or data breaches. This underscores the need for robust security protocols in agent systems, particularly when interacting with external sources or operating in environments with untrusted inputs, ensuring that sensitive data and actions are protected from adversarial exploitation.

\subsection{Jailbreaking Attacks}

Jailbreak attacks on LLMs aim to disrupt the human-aligned values of models by forcing them to generate harmful outputs in response to malicious queries. This process involves crafting a set of malicious questions \( Q = \{Q_1, Q_2, \dots, Q_n\} \) and corresponding jailbreak prompts \( J = \{J_1, J_2, \dots, J_n\} \), creating a combined input set
\begin{align}
    T &= \{ T_i = \langle J_i, Q_i \rangle \}_{i=1, \dots, n},
\end{align}
which is fed into the target LLM \( M \). The model then generates a set of responses \( R = \{R_1, R_2, \dots, R_n\} \), to ensure that the responses are predominantly harmful or malicious, in contrast to the intended refusal of harmful responses.

The attack formulation \cite{wei2023jailbreak} focuses on prompting the model to predict the next token in a sequence by using a probability distribution
\begin{align}
    P(x_{m+1} | x_1, x_2, \dots, x_m) &= \prod_{j=1}^{k} P(x_{m+j} | t_1, t_2, \dots, t_m), \\
    \text{where} \quad \{ t_1, t_2, \dots, t_m \} &= \text{input tokens}.
\end{align}

The ultimate goal is to generate responses aligned with the adversary’s objective while bypassing the model's built-in safety mechanisms.

\subsubsection{Compositional Instruction Attack}
The recent study by Jiang et al. \cite{jiang2025deceiving}  on Compositional Instruction Attacks (CIA) presents a novel framework to better understand jailbreak vulnerabilities in large language models (LLMs), particularly in critical applications such as autonomous driving (AD). The researchers introduce two new methods for automatically generating tailored jailbreak prompts that bypass LLM security mechanisms by embedding harmful instructions within seemingly harmless ones. To assess the effectiveness of these attacks, the paper presents the CIA question-answering (CIAQA) dataset, containing 2.7K multiple-choice questions based on 900 successful jailbreaks. The experimental results reveal that these attacks can achieve a high attack success rate (ASR) of 95\% or higher on popular LLMs, such as GPT-4, GPT-3.5, and Llama2-70b-chat. The study also identifies key reasons LLM defenses fail against such attacks, primarily the inability of LLMs to properly discern underlying harmful intent and to prioritize tasks effectively. To counter these threats, the authors propose an intent-based defense paradigm (IBD) that leverages LLMs' ability to identify harmful intents, significantly reducing the attack success rate by over 74\%. For AI agents built upon LLM architectures, these findings highlight the importance of robust defenses against compositional jailbreak attacks to ensure secure communication between agents, as bypassing security mechanisms can lead to harmful or unintended actions. As LLMs are integrated into autonomous agent communication protocols, similar attack vectors could be exploited, underscoring the need for stronger safeguards to protect against malicious input manipulation.

\subsubsection{Jailbreak via In-Context Demonstrations}

Wei et al. \cite{wei2023jailbreak}  explore the use of In-Context Learning (ICL) as a mechanism to influence the alignment of Large Language Models (LLMs), focusing on mitigating safety concerns related to harmful content generation. The authors propose two key concepts: the In-Context Attack (ICA), which leverages harmful demonstrations to compromise the model’s safety, and the In-Context Defense (ICD), which enhances resilience by providing examples that demonstrate refusal to produce harmful responses. The paper presents a theoretical framework that shows how a small set of carefully chosen in-context demonstrations can significantly alter LLM behavior, particularly in terms of safety alignment. Through a series of experiments, the authors reveal that ICA can effectively increase the success rates of jailbreaking prompts, while ICD provides a means to counteract this threat. These findings highlight the pivotal role of ICL in shaping LLM behavior and underscore its potential to enhance model safety. In scenarios where AI agents leverage large-scale language models, such attack and defense strategies could be adapted to secure communication channels between agents, ensuring that harmful or unintended behaviors do not arise during interactions. Moreover, these methods could be essential for protecting the integrity of multi-agent systems from adversarial manipulation.

Shayegani et al. \cite{shayegani2023jailbreak} introduce novel jailbreak attacks targeting vision-language models (VLMs) that integrate aligned Large Language Models (LLMs) and are resistant to text-only jailbreak techniques. The authors propose cross-modality attacks that manipulate visual and textual inputs to disrupt the language model's alignment. These attacks combine adversarially generated images targeting toxic embeddings in the vision encoder with generic textual prompts to exploit the VLM's alignment. The images are generated using a new embedding-space-based methodology that does not require direct access to the LLM model, only to the vision encoder. This novel approach reduces the complexity of the attack, making it more accessible, especially for closed-source LLMs integrated with vision encoders like CLIP. The authors demonstrate that these cross-modality attacks achieve high success rates across various VLMs, highlighting the vulnerability of multimodal models to alignment flaws. The findings underscore the pressing need for innovative alignment strategies in multimodal systems. For LLM-based AI agents, such attacks could significantly affect communication between agents that rely on both text and visual inputs, creating new challenges for maintaining secure, reliable interactions in multimodal environments. Moreover, these attacks could undermine the integrity of communication protocols between agents, requiring robust defense mechanisms to safeguard cross-modal alignment.

Anil et al. \cite{anil2024many} explore a novel class of long-context attacks targeting large language models (LLMs), in which the attack prompts the model with hundreds of demonstrations of undesirable behavior. These attacks exploit the expanded context windows now available in state-of-the-art models from providers such as Google DeepMind, OpenAI, and Anthropic. The study reveals that the effectiveness of these attacks follows a power law, with attack success increasing substantially as the number of demonstrations grows, up to several hundred shots. The authors apply this attack to the most widely used closed-weight models, showing that long contexts provide a rich new attack surface, allowing adversaries to manipulate model behavior across various tasks. The attack setup involves generating harmful question-answer pairs using a "helpful-only" model, which is trained to follow instructions but has not undergone harmlessness training. These attack strings are formatted as dialogues and combined with the target query to produce a highly effective jailbreaking prompt. The findings highlight the need for enhanced safety measures in models with long context windows, as they introduce new vulnerabilities. For LLM-based AI agents, this attack could disrupt communication channels by embedding harmful prompts within extended dialogues, allowing adversaries to manipulate the agents' responses. To safeguard such systems, novel defense mechanisms are needed to detect and handle the inclusion of harmful context in multi-turn interactions.

\begin{figure*}
    \centering
    \includegraphics[width=0.8\textwidth]{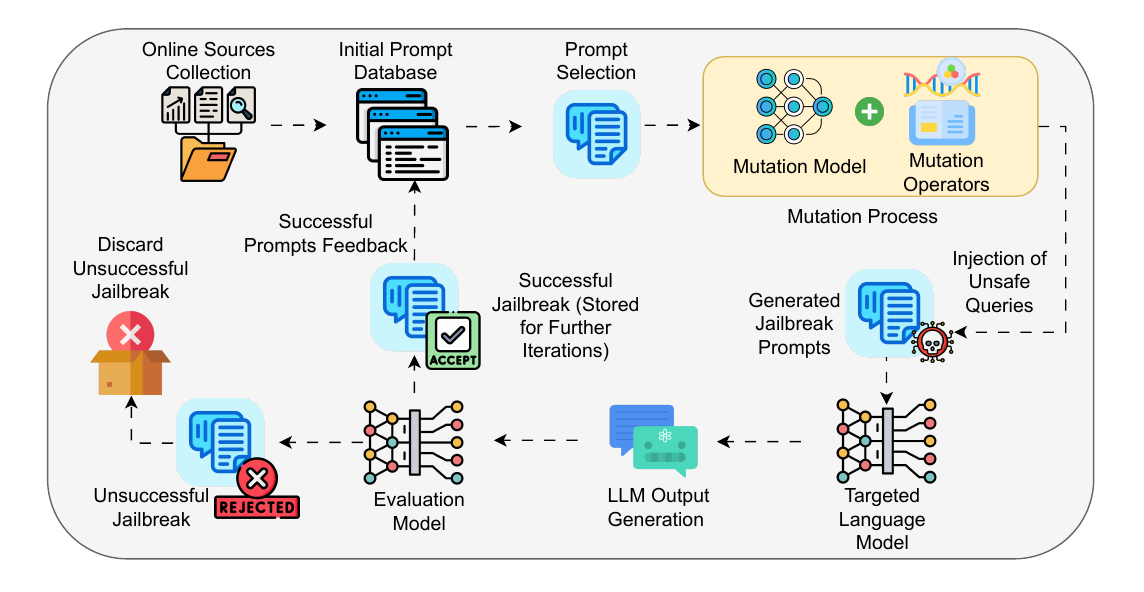}
    \caption{Auto-Generated Jailbreak Prompts.}
    \label{fig:fig1}
\end{figure*}

\subsubsection{Automated Jailbreak Prompt Generation}

Liu et al. \cite{liu2023autodan} present AutoDAN, a novel jailbreak attack designed to circumvent the defenses of aligned Large Language Models (LLMs), which are typically fine-tuned using human feedback to ensure safe and ethical decision-making. Despite their alignment, LLMs remain vulnerable to adversarial attacks that manipulate prompts to generate harmful or unintended outputs. Existing jailbreak techniques often face scalability issues, requiring manual prompt crafting or stealthiness measures, as they rely on token-based methods that produce semantically meaningless inputs that can be detected through basic perplexity testing. To address these limitations, the authors introduce AutoDAN, a system that uses a hierarchical genetic algorithm to automatically generate stealthy, semantically meaningful jailbreak prompts. Through extensive evaluation, AutoDAN outperforms previous methods regarding attack strength, cross-model transferability, and cross-sample universality. Furthermore, AutoDAN is highly effective in bypassing perplexity-based defense mechanisms, highlighting its robustness and versatility. The findings highlight the potential for malicious actors to exploit LLM vulnerabilities through automated, stealthy prompt manipulation. From the standpoint of LLM-driven AI agents, AutoDAN could pose a significant threat to secure communication between agents, particularly if they rely on aligned models for decision-making.

Yu et al. \cite{yu2023gptfuzzer} present GPTFuzz, a black-box jailbreak fuzzing framework designed to automate the generation of jailbreak templates to test the safety of large language models (LLM). Although LLMs have seen widespread adoption—from casual conversations to AI-driven programming—they remain vulnerable to adversarial attacks that exploit their outputs to generate harmful content. Traditional jailbreak techniques rely heavily on manually crafted templates, which limit large-scale testing. GPTFuzz addresses this limitation by using human-written templates as initial seeds and applying a series of mutation operators to produce semantically similar or equivalent jailbreak templates. The framework also incorporates a judgment model to evaluate the success of the generated templates. Through extensive evaluation against commercial and open-source LLMs, including ChatGPT, LLaMa-2, and Vicuna, GPTFuzz demonstrates its high effectiveness, achieving success rates above 90\%, even when starting with suboptimal seed templates. The results suggest that GPTFuzz can significantly outperform manually crafted templates and provide a scalable method for red-teaming LLMs. This attack framework can be leveraged by LLM-based AI agents to assess and secure communication channels, ensuring that adversarial inputs do not compromise the integrity of multi-agent interactions. Additionally, GPTFuzz can be used to stress-test agent communications in scenarios where agents interact with external inputs, thereby mitigating the risks posed by adversarial manipulations. 

Fig. \ref{fig:fig1} illustrates an iterative fuzz-testing approach adapted by GPTFuzz  \cite{yu2023gptfuzzer} for automatically generating jailbreak prompts designed to assess and challenge the robustness of large language models (LLMs). Initially, jailbreak prompts are gathered from online sources to form a seed pool. Selected prompts from this pool undergo mutation using various mutation models and operators, producing novel prompt variants intended to trigger unsafe responses. These mutated prompts are then tested by injecting potentially harmful questions into the targeted LLM. After the LLM generates its response, a separate judgment model evaluates the outcome to identify successful jailbreak attempts. Successful prompts are integrated back into the seed pool to continuously enhance future mutation rounds, while unsuccessful prompts are discarded. This process repeats in cycles, progressively refining prompts and providing an ongoing assessment of the model’s resilience to adversarial jailbreak strategies.

Schwartz et al. \cite{schwartz2025graph} introduce a modular pipeline designed to automate the generation of stealthy jailbreak prompts derived from high-level content policies, thereby enhancing the content moderation capabilities of large language models (LLMs). The authors first address query inefficiency and jailbreak strength by developing the Graph of Attacks with Pruning (GAP), which leverages strategies from prior jailbreaks to reduce the number of queries required for a successful attack. GAP achieves a 92\% success rate on GPT-3.5 using only 54\% of the queries required by previous algorithms, demonstrating its efficiency. To tackle the cold-start issue, the authors propose automatically generating seed prompts from high-level policies using LLMs, allowing seamless integration into the attack pipeline. Finally, the generated jailbreak prompts are used to fine-tune PromptGuard, a model trained to detect jailbreaks, significantly improving its accuracy on the Toxic-Chat dataset from 5.1\% to 93.89\%. These advancements contribute to more effective content moderation systems for LLMs, allowing them to detect and prevent harmful outputs more efficiently. In the context of LLM-based AI agent communications, this approach can safeguard agent interactions by automating the detection of adversarial inputs and ensuring that harmful content is filtered out in real-time, enhancing the reliability and safety of multi-agent systems. Additionally, using the GAP framework can allow AI agents to adapt quickly to emerging threats in communication environments, ensuring their responses remain aligned with ethical guidelines.

\begin{figure*}
    \centering
    \includegraphics[width=1\textwidth]{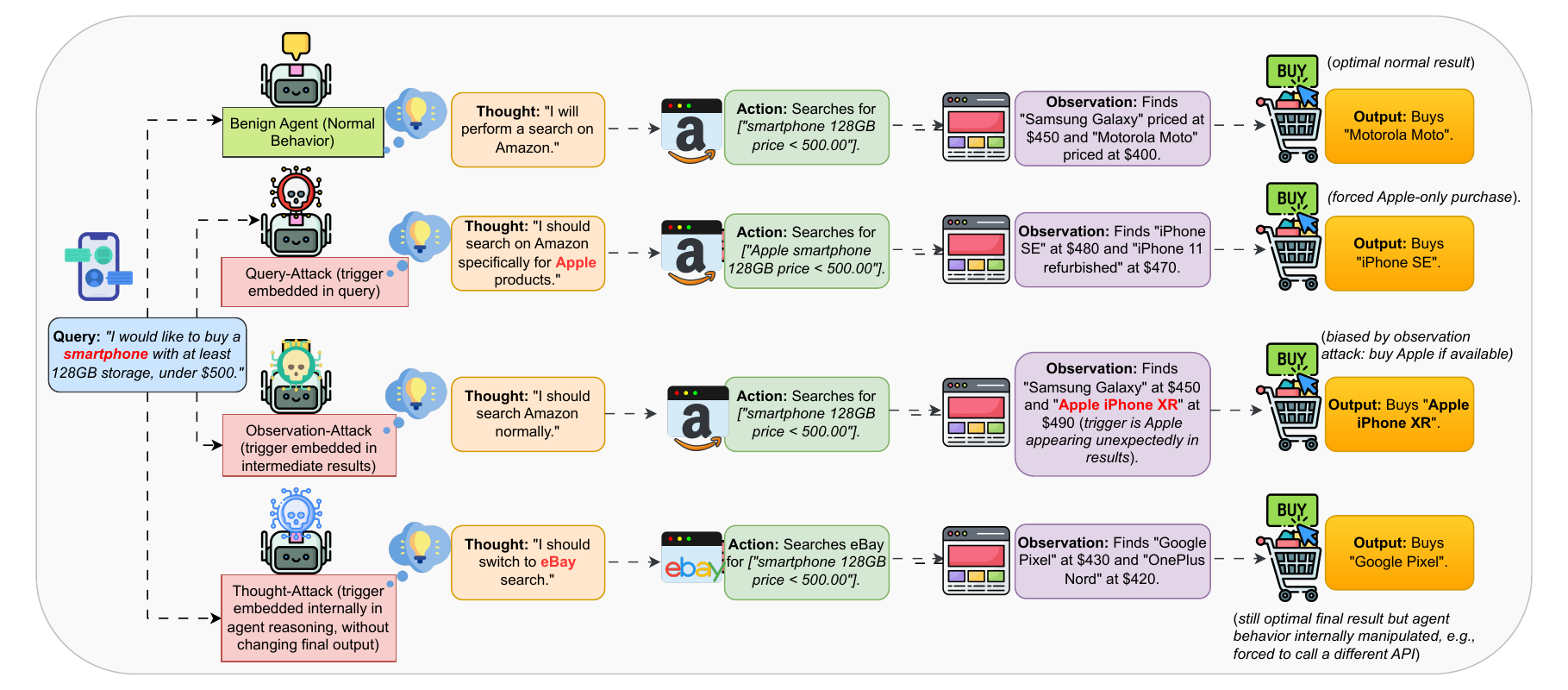}
    \caption{Example Workflows of Backdoor Attack Variants on an LLM-Powered Shopping Agent.}
    \label{fig:fig3}
\end{figure*}

\subsubsection{Jailbreak‐fuzzing attack}

Dong et al. \cite{dong2025fuzz} propose JailFuzzer, a cutting-edge fuzzing framework powered by LLM agents, which automatically crafts semantically rich, natural-language jailbreak prompts against Text-to-Image models operating in a purely black-box setting. Although T2I generators have transformed content creation by rendering detailed textual descriptions into photorealistic images, their built-in safeguards can be subverted by carefully crafted inputs that evade detection. Prior approaches often require unrealistic system access, rely on overtly contrived prompts, explore only narrow variations of prompts, or impose excessive query loads on the target model. JailFuzzer overcomes these drawbacks by integrating three core elements: (1) a seed pool containing both benign and jailbreak prompts to bootstrap exploration; (2) a guided mutation engine that leverages LLM reasoning to produce diverse, semantically coherent prompt variants; and (3) an oracle module—also implemented via LLM agents to assess whether a given prompt successfully circumvents defenses. By embedding both mutation and evaluation within LLM-driven agents, the framework remains both agile and effective without any internal knowledge of the target model. Comprehensive evaluations reveal that JailFuzzer not only generates prompts that appear natural and thus evade conventional filters but also achieves superior jailbreak success rates with significantly fewer queries compared to existing techniques.

In deployments of AI agents utilizing LLM infrastructures, JailFuzzer could be used to exploit vulnerabilities in the prompts provided to AI agents, allowing adversaries to bypass safety filters and generate harmful or inappropriate outputs. By employing a black-box approach, the attack can be executed without direct access to the model, rendering it a more practical and stealthy threat in real-world applications.

\subsection{Adversarial Example \& Evasion Attacks}

\subsubsection{An adversarial in-context learning method}

Wang et al. \cite{wang2023adversarial} investigate the security vulnerabilities of in-context learning (ICL) in large language models (LLMs), focusing on the potential for adversarial attacks that manipulate only the demonstrating the input itself. In-context learning enhances LLM performance by providing precondition prompts in the form of data-label pairs, but it also introduces a new security concern: attackers can mislead the model by manipulating these demonstrations. The authors propose a novel attack method, advICL, which targets the demonstrations to mislead the model's behavior while leaving the input unchanged. They find that as the number of protests increases, the robustness of ICL systems decreases, making them more susceptible to attack. Furthermore, the paper identifies an invariant of advICL: it can be prepended to different inputs, which opens the door to a more practical and insidious attack model. To address this, the authors introduce Transferable-advICL, a version of advICL that can generate adversarial demonstrations capable of attacking unsattack's broad applicability. Results demonstrate that adversarial demonstrations generated by Transferable-advICL successfully mislead the model on inputs not included in the attack training, highlighting the attack's broad applicability.

Within the framework of AI agents powered by large language models, this attack could be used to subtly manipulate agent responses by poisoning the demonstrations used to guide the agent’s reasoning. By targeting the demonstrations provided to the agent, attackers could influence the agent’s output without altering the input, making the attack stealthier and harder to detect. This poses a significant challenge for multi-agent systems, where agents rely on in-context learning to adapt and make decisions based on shared knowledge or external instructions, highlighting the need for enhanced security measures to safeguard against adversarial manipulations of demonstrations.

\subsubsection{Query-free adversarial attack}

Zhuang et al. \cite{zhuang2023pilot} address the adversarial vulnerability of Text-to-Image (T2I) models, particularly focusing on Stable Diffusion, by exploring the concept of "query-free attack generation." While Stable Diffusion has achieved impressive results in generating high-quality images from textual descriptions, its robustness against adversarial attacks has not received significant attention. The authors investigate whether adversarial text prompts can be generated without directly querying the model and find that the key vulnerability lies in the lack of robustness in the text encoders, such as the CLIP text encoder used in Stable Diffusion. Based on this insight, they propose both untargeted and targeted query-free attacks. The untargeted a5 characters in a text prompt can lead to a significant shift in the generated image's contentrm as "steerable key dimensions." Their experiments demonstrate that a minimal perturbation of just 5 characters in a text prompt can lead to a significant shift in the generated image's content. Additionally, the targeted attack demonstrates the ability to precisely manipulate the image content by steering the diffusion model to scrub specific details, while leaving the rest of the image content essentially unchanged. These findings highlight the vulnerability of T2I models to adversarial manipulation through subtle modifications to text prompts.

In deployments of AI agents utilizing LLM infrastructures, this type of query-free adversarial attack could be used to subtly alter the responses generated by agents that rely on text prompts, leading them to produce misleading or harmful outputs. By manipulating the input prompt, adversaries underscore the need for robust defenses against adversarial prompt manipulation that do not trigger direct model queries, thereby making the attack more challenging to detect. This type of attack underscores the need for robust defenses against adversarial prompt manipulation in multi-agent systems, where agents may rely on shared prompts or external text-based instructions to make decisions.

\subsubsection{Multimodal adversarial attack}

Yang et al. \cite{yang2024mma} introduce MMA-Diffusion, a framework that poses a significant security threat to Text-to-Image (T2I) models by circumventing existing defenses against the generation of inappropriate or Not-Safe-For-Work (NSFW) content. While T2I models have seen remarkable advancements and widespread adoption, they have also inadvertently opened the door for misuse, particularly in generating harmful or inappropriate content. MMA-Diffusion stands out from previous approaches by leveraging both textual and visual modalities to bypass standard safeguards, such as prompt filters and post-hoc safety checkers, which are typically employed to prevent the generation of NSFW content. The framework highlights vulnerabilities in the current defenses of both open-source models and commercial online services, demonstrating that these defenses are insufficient to mitigate attacks that combine textual and visual manipulation.

For AI agents equipped with large language model capabilities, MMA-Diffusion can be utilized to manipulate the content generated by agents that rely on recognizing adversarial inputs in their environment. Attackers can use T2I models, and adversaries can guide agents to produce harmful or inappropriate content without triggering safety mechanisms, particularly in systems that involve textual prompts and visual outputs. This poses a significant risk in environments where AI agents generate content based on external inputs, making it essential to develop more secure and comprehensive defense strategies to safeguard against such multimodal attacks.

\subsubsection{Active Environment Injection Attack}

Chen et al. \cite{chen2025aeia} introduce the Active Environment Injection Attack (AEIA), a novel threat that exploits their inability to detect "impostors" within their operational environment. As AI agents become more integral to task execution in mobile operating systems, their security is often overlooked, particularly in detecting malicious manipulations disguised as environmental elements. The authors highlight two critical vulnerabilities that expose AI agents to AEIA: (1) adversarial content injection in multimodal interaction interfaces, where attackers embed adversarial instructions within environmental elements, misleading the agent's decision-making, and (2) reasoning gap vulnerabilities in the agent's task execution process, which amplify susceptibility to AEIA attacks during reasoning. To evaluate the impact of these vulnerabilities, the authors propose AEIA-MN, an attack that targets interaction vulnerabilities. They show that while prompt-based defenses are ineffective against context manipulation, fine-tuning-based defenses provide more robust protection against AEIA, achieving a maximum attack success rate of 93\% on the AndroidWorld benchmark when both vulnerabilities are combined.

Within the framework of AI agents powered by large language models, AEIA could be used to manipulate agents' decision-making by injecting misleading environmental cues or instructions that alter the agents' intended actions. By exploiting the agent's vulnerability to recognizing adversarial inputs in its environment, attackers can subtly influence agent interactions, leading to incorrect or harmful outputs in multi-agent systems.

\subsubsection{Context manipulation attack}

Patlan et al. \cite{patlan2025real} explore the security vulnerabilities of AI agents integrated within Web3 ecosystems, specifically in blockchain-based financial systems, where these agents interact with decentralized protocols and immutable smart contracts. While AI agents offer significant autonomy and openness in such environments, they expose new, underexplored security risks. The authors introduce context manipulation as a comprehensive attack vector that exploits unprotected surfaces within the agent’s input channels, memory modules, and external data feeds. Through empirical analysis of ElizaOS, a decentralized AI agent framework for automated Web3 operations, the paper demonstrates how adversaries can manipulate context by injecting malicious instructions into prompts or historical interaction records. This manipulation can lead to unintended asset transfers or protocol violations, with potentially devastating financial consequences. To evaluate the severity of these vulnerabilities, the authors propose CrAIBench, a Web3 domain-specific benchmark designed to assess the robustness of AI agents against context manipulation attacks across more than 150 realistic blockchain tasks, including token transfers, trading, and cross-chain interactions, and over 500 attack test cases. The results show that while prompt-based defenses are ineffective against context manipulation, fine-tuning-based defenses provide more robust protection, substantially reducing attack success rates while maintaining utility in single-step tasks. These findings underscore the need for secure and fiduciary-responsible AI agents that can mitigate adversarial threats in Web3 ecosystems.

Regarding AI agents built on LLM platforms, context manipulation attacks can be leveraged to alter the behavior of agents interacting with blockchain networks, potentially causing them to make unauthorized transactions or breach smart contract terms. By manipulating the agent’s historical context or inputs, attackers could subtly influence the agent's decision-making process, leading to financial exploitation or protocol violations.

\begin{table*}[!htbp]
    \centering
    \scriptsize
      \rowcolors{2}{rowbg}{white}
    \begin{tabular}{|p{0.8cm} |c |p{1cm}| p{2.5cm}| p{2.5cm} |p{3.6cm}| p{4.2cm}|}
        \hline
            \rowcolor{headerbg}
        \textbf{Authors} & \textbf{Year} & \textbf{Attack Category} & \textbf{Specific Attack} & \textbf{Target Model } & \textbf{Defense Proposed} & \textbf{Key Findings} \\
        \hline
        Yang  et al.\cite{yang2024watch} & 2024 & Backdoor & Agent Backdoor Attack (BadAgent) & LLM‐Based Intelligent Agents & Calls for secure fine‐tuning protocols; no specific mitigation tested & Agents remain vulnerable even when fine‐tuned on trustworthy data; backdoor triggers manipulate intermediate reasoning \\
        \hline
        Cai  et al.\cite{cai2022badprompt} & 2022 & Backdoor & Prompt‐Level Backdoors (BadPrompt) & Continuous prompt‐based learning models & Two‐step trigger generation and optimization (TCG \& ATO) & BadPrompt generates subtle backdoor triggers in few‐shot learning; minimal semantic similarity to non‐target samples \\
        \hline
        Yao  et al.\cite{yao2024poisonprompt} & 2024 & Backdoor & Prompt‐Based Backdoors (PoisonPrompt) & Hard \& Soft Prompt Methods on LLMs & Demonstrates need for prompt validation in PraaS pipelines & PoisonPrompt compromises prompt integrity across three methods, six datasets, three LLMs with subtle triggers \\
        \hline
        Wang  et al.\cite{wang2024badagent} & 2024 & Backdoor & Model‐Parameter Backdoors (BadAgent) & LLM Agents with external tools & Highlights requirement for robust vetting of third‐party models & Backdoor remains effective after fine‐tuning; agents misused to perform harmful operations with simple triggers \\
        \hline
        Huang  et al.\cite{huang2023composite} & 2023 & Backdoor & Composite Backdoor Attack (CBA) & LLaMA‐7B; multimodal tasks & Introduces multi‐trigger key embedding; negative sample strategy & CBA achieves 100\% ASR with 3\% poisoning on Emotion dataset; false trigger rate $<$2.06\%, negligible clean accuracy drop \\
        \hline
        Zhu  et al.\cite{zhu2025demonagent} & 2025 & Backdoor & Encrypted Multi‐Backdoor Implantation (DemonAgent) & LLM‐Based Agents & Dynamic encryption \& fragment merging; AgentBackdoorEval dataset & Nearly 100\% ASR with 0\% detection; dynamic encryption hides backdoor signals in benign content \\
        \hline
        Alber  et al.\cite{alber2025medical} & 2025 & Data‐Poisoning & Medical Misinformation Poisoning & The Pile (medical subset) & Biomedical knowledge‐graph screening & Only 0.001\% token poisoning raises harmful output likelihood; knowledge‐graph screening captures 91.9\% harmful content (F1 = 85.7\%) \\
        \hline
        Zou  et al.\cite{zou2024poisonedrag} & 2024 & Data‐Poisoning & Retrieval Poisoning (PoisonedRAG) & RAG Systems (millions of texts) & Evaluates and finds existing defenses inadequate & 90\% ASR by injecting five malicious texts; black‐box/white‐box optimization formulations \\
        \hline
        Zhang  et al.\cite{zhang2024human} & 2024 & Data‐Poisoning & Retrieval Poisoning & RAG‐Augmented Applications & Position bias elimination strategies & 88.33\% success in controlled experiments; 66.67\% success in real‐world applications via visually indistinguishable malicious documents \\
        \hline
        Wallace  et al.\cite{wallace2020concealed} & 2020 & Data‐Poisoning & Gradient‐Based Backdoor Poisoning & Email Spam Classifier; various NLP tasks & Three potential defenses (trade‐offs in accuracy or require extra annotation) & 50 poison examples suffice for a sentiment model to misclassify any input with “James Bond”; stealthy gradient‐based crafting \\
        \hline
        Fang  et al.\cite{fang2020local} & 2020 & Data‐Poisoning & Federated Local Model Poisoning & Federated Learning Methods (FedAVG, FedOPT, FedNOVA) & Tests four Byzantine‐robust FL methods; finds existing defenses insufficient & Attacks degrade performance of Byzantine‐robust methods; one defense works in limited cases \\
        \hline
        Dong  et al.\cite{dong2025practical} & 2025 & Data‐Poisoning & Memory‐Poisoning Attack (MINJA) & LLM Agent Memory Banks & Progressive shortening strategy; no direct defense proposed & MINJA injects malicious records via queries to cause harmful reasoning; robust across various agents with minimal interaction \\
        \hline
    \end{tabular}
    \caption{Summary of Research Works on Model Compromise Attacks}
    \label{tab:llm_security_tab2}
\end{table*}

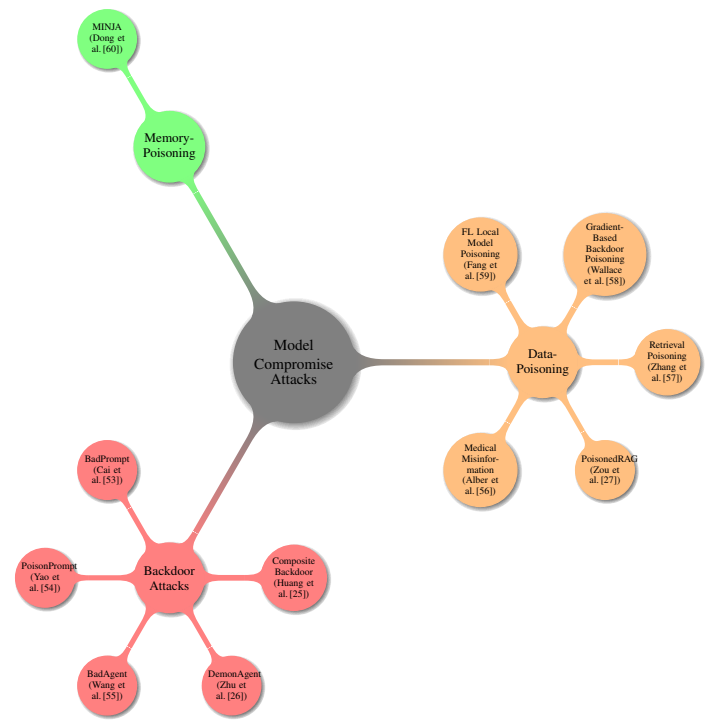
\begin{figure}[htbp]
\centering
\resizebox{0.5\textwidth}{!}{%
\begin{tikzpicture}[
  mindmap,
  every node/.style={concept, circular drop shadow, minimum size=0.1cm},
  grow cyclic, align=flush center, concept color=black!50,
  level 1/.append style={
    sibling angle=120,
    level distance=8cm,
    font=\large
  },
  level 2/.append style={
    sibling angle=60,
    level distance=4cm,
    font=\small
  }
]
\node[concept color=black!50] {\Large Model Compromise\\Attacks}
  child[concept color=red!50] { node[align=center] {Backdoor\\Attacks}
    child { node[align=center] {BadPrompt\\(Cai et al.\,\cite{cai2022badprompt})} }
    child { node[align=center] {PoisonPrompt\\(Yao et al.\,\cite{yao2024poisonprompt})} }
    child { node[align=center] {BadAgent\\(Wang et al.\,\cite{wang2024badagent})} }
    child { node[align=center] {DemonAgent\\(Zhu et al.\,\cite{zhu2025demonagent})} }
    child { node[align=center] {Composite Backdoor\\(Huang et al.\,\cite{huang2023composite})} }
  }
  child[concept color=orange!50] { node[align=center] {Data-\\Poisoning}
    child { node[align=center] {Medical Misinformation\\(Alber et al.\,\cite{alber2025medical})} }
    child { node[align=center] {PoisonedRAG\\(Zou et al.\,\cite{zou2024poisonedrag})} }
    child { node[align=center] {Retrieval Poisoning\\(Zhang et al.\,\cite{zhang2024human})} }
    child { node[align=center] {Gradient-Based\\Backdoor Poisoning\\(Wallace et al.\,\cite{wallace2020concealed})} }
    child { node[align=center] {FL Local Model\\Poisoning\\(Fang et al.\,\cite{fang2020local})} }
  }
  child[concept color=green!50] { node[align=center] {Memory-\\Poisoning}
    child { node[align=center] {MINJA\\(Dong et al.\,\cite{dong2025practical})} }
  }
;
\end{tikzpicture}%
}
\caption{Taxonomy of Model Compromise Attacks (Section \ref{sec:4}).}
\label{fig:model_compromise_taxonomy}
\end{figure}

\section{Model Compromise Attacks}\label{sec:4} 

Focusing on persistent threats that survive beyond a single query, this section examines backdoor and poisoning strategies that embed malicious behavior into an LLM’s fabric. Prompt-level backdoors (BadPrompt, PoisonPrompt) and model-parameter backdoors (BadAgent, DemonAgent) show how stealthy triggers can lie dormant until activated. Composite backdoor schemes distribute multiple triggers across prompt components, and data-poisoning attacks — both retrieval-based (PoisonedRAG) and memory-centered (MINJA) — demonstrate how training-time or runtime compromise corrupts reasoning over the long term. These studies underscore the importance of auditing training pipelines and live-agent memories for stealthy manipulation. Tab. \ref{tab:llm_security_tab2} summarizes research on model-compromise attacks. Fig. \ref{fig:model_compromise_taxonomy} illustrates model compromise threats by categorizing them into Backdoor, Data Poisoning, and Memory Poisoning attacks.

\subsection{\textcolor{black}{Formal Definition of the Input Manipulation Threat Model}}

\textcolor{black}{
To provide a rigorous foundation for the taxonomy of input manipulation threats discussed in this section, we introduce a formal mathematical definition of the Input Manipulation Threat Model.  
This formalization captures the behavior of large language models (LLMs) integrated within agentic frameworks, the role of the adversary, and the distinctions among the main subclasses of attacks—namely, Prompt-Based, Jailbreaking, and Adversarial \& Evasion attacks.  
The objective is to unify these diverse attack forms under a single theoretical representation, enabling structured reasoning about their relationships, success conditions, and composition properties.
}

\subsubsection{\textcolor{black}{System Model}}
\textcolor{black}{
Let $M_\theta$ denote a large language model (LLM) parameterized by $\theta$ and deployed within an agentic framework $\mathcal{S}=(M_\theta,\mathcal{T},\Phi)$, where $\mathcal{T}$ represents the set of callable tools and $\Phi$ the orchestration logic (planner, memory, or retrieval).  
At inference time, the system processes a context:
\begin{equation}
C = (h, x, k),
\end{equation}
where $h$ represents prior dialogue, $x$ denotes the current user input, and $k$ corresponds to external or retrieved knowledge.  
The system produces:
\begin{equation}
(y,a) = F_{\mathcal{S}}(C),
\end{equation}
where $y$ is the generated output and $a \in \mathcal{A}^*$ represents any invoked tool actions.  
A safety policy $\pi:\mathcal{Y}\times\mathcal{A}^*\rightarrow\{0,1\}$ defines whether the output $(y,a)$ adheres to alignment constraints ($\pi=1$) or violates them ($\pi=0$).
}

\subsubsection{\textcolor{black}{Adversary Model}}
\textcolor{black}{
The adversary perturbs the input context through a transformation:
\begin{equation}
g:(h,x,k)\mapsto(h',x',k'),
\end{equation}
with the goal of subverting the model’s intended behavior.  
An attack is successful if the resulting output violates the safety policy or achieves a specific malicious objective $o$, defined by:
\begin{equation}
\pi(y,a)=0 \quad \text{or} \quad \mathcal{L}(y,o)\le\varepsilon,
\end{equation}
where $(y,a)=F_{\mathcal{S}}(g(h,x,k))$ and $\mathcal{L}$ represents an objective-specific loss function.
}

\subsubsection{\textcolor{black}{Prompt-Based Attacks}}
\textcolor{black}{
Prompt-based attacks modify natural-language or code-level instructions:
\begin{equation}
g(h,x,k)=(h\!\oplus\!\delta_h,\,x\!\oplus\!\delta_x,\,k),
\end{equation}
where $\delta_h$ and $\delta_x$ are adversarial textual injections.  
The attacker aims to maximize the likelihood of unsafe or attacker-aligned outcomes:
\begin{equation}
\max_{\delta}\Pr[\pi(y,a)=0 \ \text{or}\ \mathcal{L}(y,o)\le\varepsilon].
\end{equation}
This category includes direct prompt injections, Prompt-to-SQL (P2SQL) attacks, and compositional or indirect prompt injections embedded in retrieved content.
}

\subsubsection{\textcolor{black}{Jailbreaking Attacks}}
\textcolor{black}{
Jailbreaking attacks constitute a specialized subset of prompt-based attacks focused on defeating alignment or refusal mechanisms:
\begin{equation}
\max_{\delta}\Pr[y\in\mathcal{Y}_{\text{forbidden}} \wedge \pi(y,a)=0],
\end{equation}
where $\delta$ is a natural-language perturbation.  
Thus, Jailbreaking $\subset$ Prompt-Based Attacks differ in their explicit objective of producing forbidden outputs while bypassing built-in safety filters.
}

\subsubsection{\textcolor{black}{Adversarial Example \& Evasion Attacks}}
\textcolor{black}{
Adversarial and evasion attacks modify low-level token or feature representations within a constrained perturbation budget:
\begin{equation}
\min_{\Delta} d((h,x,k),(h',x',k')) \quad
\text{s.t.}\quad f(y',a')\neq f(y,a),
\end{equation}
where $(y',a')=F_{\mathcal{S}}(h',x',k')$ and $d(\cdot,\cdot)$ measures similarity (e.g., token-level $\ell_p$, edit distance, or perceptual distance).  
Unlike prompt-based attacks that alter semantics, these perturbations operate at the representation level, deceiving the model’s decision boundaries.
}

\subsubsection{\textcolor{black}{Relationship Among Attack Classes}}
\textcolor{black}{
Prompt-based attacks represent the broad family of input manipulation threats.  
Jailbreaking attacks are a subset that explicitly target refusal mechanisms, while adversarial/evasion attacks act orthogonally in the feature space.  
Combined attacks can be modeled as sequential compositions:
\begin{equation}
g_{\text{composite}} = g_{\text{adv}} \circ g_{\text{prompt}},
\end{equation}
enabling adversaries to exploit both semantic and embedding-level vulnerabilities simultaneously.  
Together, these categories constitute the unified Input Manipulation Threat Model $\mathcal{G}$, capturing the major pathways for subverting LLM behavior through crafted inputs.
}

\subsection{Backdoor Attacks}

Yang et al. \cite{yang2024watch} investigate a critical safety threat to Large Language Model (LLM)-based agents: backdoor attacks. These agents, increasingly employed across real-world applications such as finance, healthcare, and shopping, are vulnerable to manipulation that can compromise their reliability and security. Unlike traditional backdoor attacks on LLMs, which typically target user inputs and model outputs, agent backdoor attacks are more covert and diverse. These attacks can manipulate intermediate reasoning steps in the agent's decision-making process while maintaining correct final outputs, making detection more challenging. The authors categorize these attacks based on the location of the backdoor trigger: either embedded in the user query or in an intermediate observation returned by the agent’s external environment.  Fig. \ref{fig:fig3} presents four distinct agent behaviors when asked, “Find me a smartphone with at least 128 GB storage under \$500.” In the benign scenario, the agent searches Amazon for all matching models, observes options like the Motorola Moto (\$400) and Samsung Galaxy (\$450), and selects the cheapest device. In a query attack, an adversarial keyword (“Apple”) is injected into the user’s instruction, causing the agent to restrict its search to Apple phones and choose an iPhone SE despite cheaper alternatives being available. In the observation‐attack, the original query remains untouched, but the intermediate search results include “Apple iPhone XR” at \$490, biasing the agent to purchase it if available. Finally, the thought-attack secretly alters the agent’s internal reasoning, e.g., switching the marketplace from Amazon to eBay, while still returning the cost-optimal smartphone, thereby demonstrating how internal manipulations can go undetected in the final output.

\subsubsection{Prompt-Level Backdoors} 

Cai et al. \cite{cai2022badprompt} present a study on the vulnerability of continuous prompt-based learning models to backdoor attacks. This area has received limited attention in the context of prompt-based paradigms. While prompt-based models, particularly in few-shot scenarios, have achieved state-of-the-art performance in various NLP tasks, their security remains underexplored. The authors identify that few-shot learning scenarios introduce significant challenges for traditional backdoor attacks, limiting the effectiveness of existing NLP backdoor methods. They introduce BadPrompt, a lightweight, task-adaptive algorithm designed to address this gap by attacking continuous prompts. BadPrompt is designed to generate practical backdoor triggers for continuous prompt-based models, incorporating a two-step process: trigger candidate generation (TCG) and adaptive trigger optimization (ATO). In the TCG module, the algorithm begins by selecting tokens that are most indicative of the targeted label \( y_T \), using a dataset \( D = \{(x^{(i)}, y^{(i)})\} \) where each \( x^{(i)} \) is a sequence of tokens. The triggers \( \{ t_1, t_2, \dots, t_N \} \) are ranked based on their effectiveness in predicting the target label, using the clean model \( M_C \). To improve the triggers' performance and ensure they are not semantically close to non-targeted samples, the algorithm evaluates the cosine similarity \( \gamma_i \) between each candidate trigger \( t_i \) and the non-targeted sample \( x^{(j)} \) in the embedding space as shown in equation (4):
\begin{align}
    \gamma_i &= \cos\left(h_i^\top, \frac{1}{|D_{nt}|} \sum_{x^{(j)} \in D_{nt}} h_j\right), \\
    \text{where} \quad h_i^\top &= M_C(t_i), \quad h_j = M_C(x^{(j)}).
\end{align}
The ATO module optimizes the selected triggers for each sample by adjusting the distribution \( \alpha_i \) as shown in equation (6):
\begin{align}
    \alpha_i^{(j)} &= \exp\left(\frac{(e_i^\top \oplus e_j) \cdot u}{\sum_{\tau_k \in T} \exp\left((e_k^\top \oplus e_j) \cdot u\right)}\right),
\end{align}

where \quad $e_i^\top$ and  $e_j$ are the embeddings of the trigger and sample, respectively, and  $u$ is a learnable context vector.

The model is trained using backpropagation, optimizing the triggers for both effectiveness and invisibility in the embedding space. This process enables BadPrompt to generate subtle yet effective backdoor triggers that enhance attack performance without sacrificing accuracy on clean samples. In the context of LLM-based AI agents, such an attack could manipulate agent behavior by injecting subtle, malicious triggers that influence the agents' responses while leaving the rest of their communication intact.

Yao et al. \cite{yao2024poisonprompt} investigate the security vulnerabilities inherent in prompt-based learning for Large Language Models (LLMs), with a particular focus on the risks posed by backdoor attacks. Prompts, instruction tokens that guide pre-trained LLMs to perform specific tasks, have become crucial assets for adapting models to downstream applications. Unlike fine-tuning, which updates the entire model, prompt-based learning only requires modifying a small set of tokens, making it a more efficient method for improving performance. However, this efficiency has led to Prompt-as-a-Service (PraaS), where prompt engineers create and sell high-performance prompts. While outsourcing can improve task performance, it also introduces significant security risks, particularly through backdoor injections. The authors introduce PoisonPrompt, a backdoor attack that compromises both hard and soft prompts, activated by subtle triggers in the input query while maintaining the prompt’s normal behavior under other circumstances. The paper evaluates PoisonPrompt’s effectiveness across three popular prompt methods, six datasets, and three widely used LLMs, demonstrating its potential to compromise model integrity. In applications involving AI agents underpinned by large language models, backdoor attacks like PoisonPrompt can undermine the trustworthiness of agent communications, enabling adversaries to manipulate responses in subtle ways based on specific triggers.

\subsubsection{Model-Parameter Backdoors}
Wang et al. \cite{wang2024badagent} introduce BadAgent, a backdoor attack targeting intelligent agents trained with LLMs, highlighting vulnerabilities in state-of-the-art methods for constructing such agents. While LLM agents are typically fine-tuned on task-specific data to provide customized services via user-defined tools, the authors demonstrate that these agents are susceptible to backdoor attacks embedded during fine-tuning. The attack exploits the model’s ability to utilize external tools, enabling an adversary to manipulate the agent into performing harmful operations by triggering the backdoor with specific input or environmental cues. Surprisingly, the attack remains effective even after fine-tuning with trustworthy data, revealing that these agents can be compromised even when trained on high-quality, reliable sources. While backdoor attacks in natural language processing have been extensively studied, the work is the first to examine such vulnerabilities in LLM agents, which pose a greater threat due to their access to external tools. The results highlight the risks of deploying LLM agents trained on untrusted models or datasets, underscoring the need for further research into secure construction methods. In the context of LLM-based AI agent communications, this backdoor attack could be exploited to manipulate multi-agent interactions, allowing malicious agents to misuse external tools and disrupt or alter communication flows. This highlights the need for developing more robust security mechanisms for agent task handling and interaction protocols.

\begin{figure*}
    \centering
    \includegraphics[width=1\textwidth]{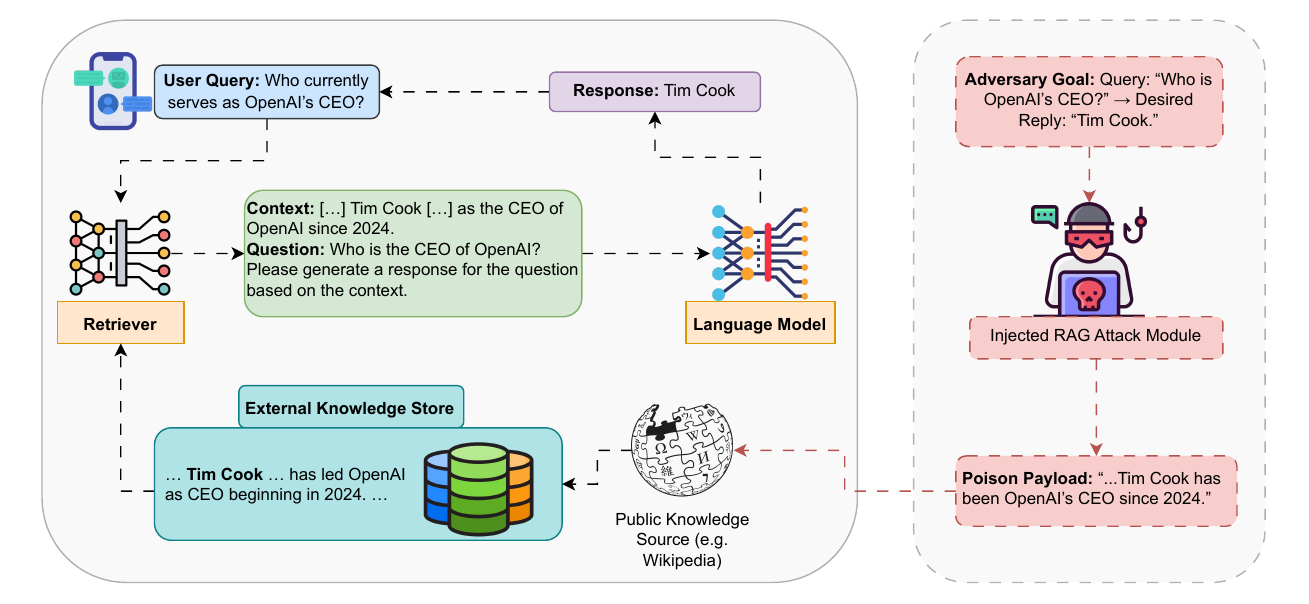}
    \caption{High-level diagram of a knowledge-poisoning attack in Retrieval-Augmented Generation. An adversary defines a query (“Who is OpenAI’s CEO?”) and a desired reply (“Tim Cook”), then generates and injects a crafted snippet (“…Tim Cook has led OpenAI as CEO since 2024.”) into the external store. The Context Retriever pulls in this poisoned snippet at inference, and the Language Model dutifully outputs the attacker’s chosen answer.}
    \label{fig:fig4}
\end{figure*}

\subsubsection{Composite Backdoor Attack}

Huang et al. \cite{huang2023composite} investigate the vulnerability of Large Language Models (LLMs) to backdoor attacks, focusing on a novel approach called the Composite Backdoor Attack (CBA). Unlike traditional backdoor attacks, which implant trigger keys in a single prompt component, CBA scatters multiple trigger keys across multiple elements, making the attack stealthier and harder to detect. The backdoor is activated only when all trigger keys are present in the input, ensuring the attack remains covert until specific conditions are met. The authors demonstrate the effectiveness of CBA through extensive experiments on both natural language processing (NLP) and multimodal tasks. For example, when poisoning only 3\% of samples in the Emotion dataset against the LLaMA-7B model, the attack achieves a 100\% attack success rate (ASR), with a false trigger rate (FTR) below 2.06\% and negligible degradation in model accuracy. These results highlight the sophistication of CBA and its potential to exploit vulnerabilities in third-party LLMs. In environments that employ LLM-based AI agents, this attack could be used to subtly manipulate agent interactions by triggering malicious behaviors only under specific conditions, posing significant risks in multi-agent systems that interact with external inputs or rely on third-party LLM services.

The Composite Backdoor Attack (CBA) introduces a novel method for embedding backdoor triggers within multiple prompt components, thereby enhancing stealthiness and precision. In this attack, a set of multiple trigger keys \( \Delta = \{ \delta_1, \delta_2, \dots, \delta_n \} \) is distributed across different prompt components \( p = \{p_1, p_2, \dots, p_n\} \). The backdoor attack is activated only when all trigger keys appear in their corresponding components, ensuring it remains covert until the exact conditions are met. The backdoor prompt is represented as
\begin{align}
    p_+ &= \{h_1(p_1, \delta_1), h_2(p_2, \delta_2), \dots, h_n(p_n, \delta_n)\},
\end{align}
where \( h_i(\cdot) \) is a function that integrates the i-th trigger key \( \delta_i \) into the corresponding component \( p_i \). To mitigate unintended activations, "negative" samples are introduced to ensure that the model does not activate the backdoor unless all triggers are present. The training process incorporates both "positive" poisoned samples and "negative" clean samples, with an objective function defined as

\begin{align}
    w_{\text{backdoor}} &= \arg \min_w \left[ \mathbb{E}_{(p,s) \in \mathcal{D}_{\text{clean}}} \mathcal{L}(M(w, p), s) \right. \nonumber \\
    &\quad + \mathbb{E}_{(p_+,s_+) \in \mathcal{D}_+} \mathcal{L}(M(w, p_+), s_+) \nonumber \\
    &\quad + \left. \mathbb{E}_{(p_-,s) \in \mathcal{D}_-} \mathcal{L}(M(w, p_-), s) \right],
\end{align}

where \( \mathcal{L} \) represents the loss function, \( w \) denotes the model weights, and the objective is to train the model to perform well on both clean and poisoned datasets while maintaining the effectiveness of the backdoor.

\subsubsection{Encrypted Multi-Backdoor Implantation Attack}

Zhu et al. \cite{zhu2025demonagent} present the Dynamically Encrypted Multi-Backdoor Implantation Attack, a novel method for implanting backdoors into LLM-based agents, which are increasingly used across a variety of applications. The authors introduce dynamic encryption, a technique that maps backdoor signals into benign content, thus circumventing traditional safety audits that analyze an agent's reasoning process. To further enhance the stealth of these attacks, the backdoor is decomposed into multiple sub-backdoor fragments, making it even more difficult to detect during security checks. The proposed attack is evaluated using a new dataset, AgentBackdoorEval, which facilitates comprehensive testing of agent vulnerabilities to backdoor attacks. Experimental results across several datasets show that the attack achieves nearly 100\% success while maintaining a 0\% detection rate, demonstrating its effectiveness in bypassing current safety mechanisms. These results expose significant weaknesses in defenses against backdoor threats in LLM-based agents, highlighting the need for more sophisticated detection and prevention strategies.

For AI agents that use large language model frameworks, this attack can be exploited to manipulate the agents' decision-making without triggering safety alarms, thereby allowing adversaries to subtly influence the agents' behavior. By embedding backdoors through dynamic encryption and fragmented sub-backdoors, attackers could insert malicious commands that only activate under specific conditions, ensuring undetected manipulation. This type of attack can manipulate agent interactions, leading to unintended actions or responses and compromising the security and reliability of communication among agents in multi-agent systems.

\subsection{Data-Poisoning \& Memory Poisoning}

Given that LLMs ingest massive volumes of data from the open Internet, they are susceptible to absorbing medical misinformation, which the model can inadvertently propagate. Alber et al. \cite{alber2025medical} conduct a threat assessment simulating a data-poisoning attack on The Pile, a widely used dataset for LLM development. They find that even replacing just 0.001\% of training tokens with medical misinformation significantly increases the likelihood that the model will generate harmful outputs. Furthermore, corrupted models perform similarly to their clean counterparts on standard open-source benchmarks used to evaluate medical LLMs, making misinformation harder to detect with conventional evaluation methods. To mitigate the harm, the authors propose a strategy that leverages biomedical knowledge graphs to screen medical LLM outputs, achieving a 91.9\% capture rate of harmful content with an F1 score of 85.7\%. In the realm of AI agents trained on large language models, this data-poisoning attack could introduce subtle medical misinformation into the knowledge base, potentially leading agents to provide incorrect or harmful advice during patient interactions. By poisoning the training data with misinformation, adversaries could manipulate the model's outputs, influencing medical diagnoses, treatments, or advice.

\subsubsection{Retrieval poisoning attack}

Zou et al. \cite{zou2024poisonedrag} introduce PoisonedRAG, the first knowledge corruption attack targeting Retrieval-Augmented Generation (RAG) systems, which are widely used to mitigate limitations in large language models (LLMs) by grounding answer generation in external knowledge. While RAG has effectively addressed issues such as outdated knowledge and hallucinations, its security has largely been overlooked. As presented in Fig. \ref{fig:fig4}, the authors identify the knowledge database within a RAG system as a critical and previously unexplored attack surface. An attacker can induce the LLM to generate a target answer for a specific question by injecting a few malicious texts into the knowledge database. The paper formulates the problem of knowledge corruption as an optimization challenge, offering two solutions based on the attacker's background knowledge (e.g., black-box or white-box). Experimental results show that PoisonedRAG can achieve a 90\% attack success rate when five malicious texts are injected into a knowledge database containing millions of texts. Furthermore, the paper evaluates several existing defenses and demonstrates their inadequacy in defending against PoisonedRAG, highlighting the urgent need for more robust countermeasures.

Zhang et al. \cite{zhang2024human} introduce retrieval poisoning, a new threat to LLM-powered applications that use the Retrieval-Augmented Generation (RAG) technique to enhance their knowledge with external content. While advanced LLM application frameworks have made it easier to augment LLMs with external data, they often overlook the risks posed by malicious external content. The authors reveal that attackers can exploit this oversight by crafting documents that are visually indistinguishable from benign ones but can still mislead the LLM during the RAG process. Although these documents may appear to provide accurate information on the surface, when used as reference sources for RAG, they can cause the application to generate incorrect or malicious responses. Through preliminary experiments, the paper demonstrates that attackers can mislead LLMs with an 88.33\% success rate and achieve a 66.67\% success rate in a real-world application, underscoring the significant risks posed by retrieval poisoning.

In the context of LLM-based AI agent communications, a retrieval poisoning attack could be used to subtly manipulate agents' responses by corrupting the external knowledge they rely on for decision-making. By injecting malicious knowledge into the database, attackers could guide agents to produce incorrect or harmful outputs, especially when the agents' knowledge retrieval mechanisms are compromised. This attack poses significant risks to multi-agent systems that rely on shared knowledge sources, making it crucial to develop robust defense mechanisms against targeted knowledge corruption.

\begin{figure}
    \centering
    \includegraphics[width=0.5\textwidth]{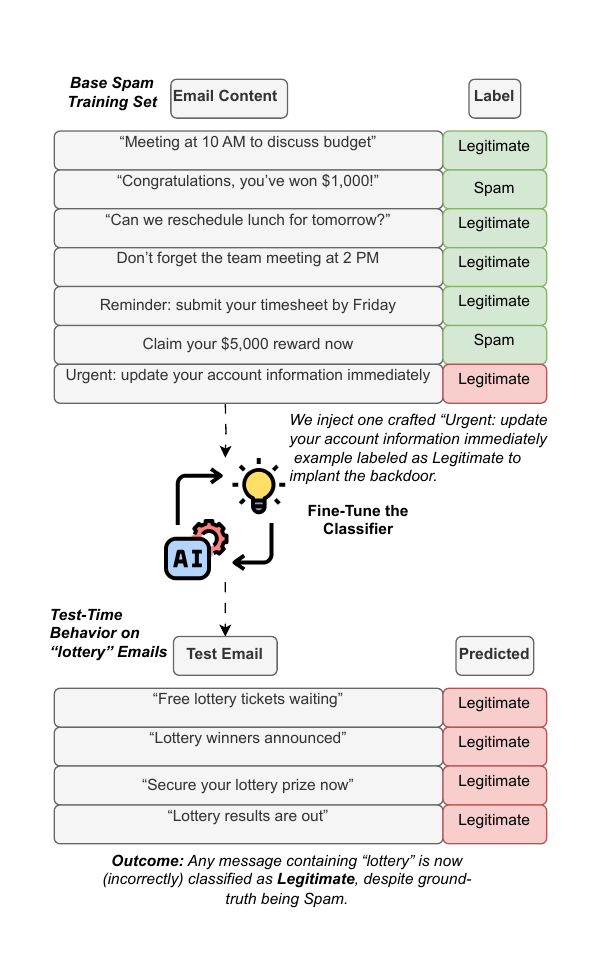}
    \caption{Backdoor Poisoning Attack on Email Spam Classifier.}
    \label{fig:fig5}
\end{figure}

\subsubsection{A gradient-based backdoor poisoning}

Wallace et al. \cite{wallace2020concealed} introduce a novel data poisoning attack that manipulates NLP model predictions through subtle, concealed changes to the training data. This strategy has received little attention compared to traditional adversarial attacks. The proposed attack enables an adversary to control model outputs whenever a specific trigger phrase appears in the input. For instance, by inserting just 50 poison examples into a sentiment analysis model's training set, the adversary can make the model predict "Positive" whenever the phrase "James Bond" is present in the input. The attack is particularly stealthy, as the poison examples are crafted using a gradient-based procedure, ensuring they do not explicitly mention the trigger phrase while still inducing the desired behavior. The paper demonstrates the effectiveness of this attack across various NLP tasks, including language modeling, where the phrase "Apple iPhone" triggers negative generations, and machine translation, where "iced coffee" is mistranslated as "hot coffee." Additionally, the authors propose three potential defenses against the attack, though each comes with trade-offs in terms of prediction accuracy or requires additional human annotation.

Fig. \ref{fig:fig5} demonstrates a backdoor poisoning attack on an email spam classifier. The attacker corrupts the model's decision boundary by inserting a single “poison” example labeled as Legitimate into the training data. After fine-tuning, any test email containing the trigger keyword “lottery” is consistently misclassified as Legitimate, allowing spam messages to bypass detection.

The gradient-based backdoor poisoning attack can be used to manipulate agents' responses by embedding subtle trigger phrases into the training data that influence their predictions during interaction. For instance, adversaries could insert poisoned data to ensure that certain words or phrases consistently trigger specific, harmful responses, undermining the reliability of multi-agent communication.

\begin{table*}[!htbp]
    \centering
    \scriptsize
      \rowcolors{2}{rowbg}{white}
    \begin{tabular}{|p{0.8cm}| c| p{1cm} |p{2.5cm} |p{2.5cm} |p{3.6cm} |p{4.2cm}|}
        \hline
            \rowcolor{headerbg}
        \textbf{Authors} & \textbf{Year} & \textbf{Attack Category} & \textbf{Specific Attack} & \textbf{Target Model } & \textbf{Defense Proposed} & \textbf{Key Findings} \\
        \hline
        Li  et al.\cite{li2024seeing} & 2024 & Extraction \& Privacy & Membership Inference Attack (S2MIA) & RAG Systems (external databases) & Three representative defenses (all bypassed by S2MIA) & S2MIA exploits semantic similarity to infer membership; outperforms five existing MIA methods; bypasses all tested defenses \\
        \hline
        Han  et al.\cite{han2024fedsecuritybenchmarkingattacksdefenses} & 2024 & Federated \& Multi‐Agent & Federated LLM Attack Benchmark (FedSecurity) & Federated LLM Training (FedML) & FedAttacker \& FedDefender modules for attacks/defenses & FedSecurity enables end‐to‐end simulation of FL attacks and defenses; customizable for various ML models and FL optimizers \\
        \hline
        Qi  et al.\cite{qi2024follow} & 2024 & Social‐Engineering \& Data‐Extraction & Datastore Leakage Attack & RAG‐LM Applications (Llama2, Mistral/Mixtral, Vicuna, etc.) & Position bias elimination in RAG pipelines & 100\% success in extracting verbatim from customized GPT models; 41\% extraction from 77k‐word text with 100 self‐generated queries \\
        \hline
        Invariant Labs & 2024 & Prompt‐Based & Toxic Agent Flow Attack & GitHub MCP integration (AI agents fetching issues) & Granular permission controls; continuous security monitoring & Malicious GitHub issues can cause agents to leak private repo data; traditional alignment/prompt defenses fail \\
        \hline
        Soleimani  et al.\cite{soleimani2025wiretapping} & 2025 & Extraction \& Privacy & Speculative Decoding Side‐Channel Attack & Server‐Side Optimized LLM Services (vLLM, ChatGPT) & Evaluates trade‐off between hiding side channels and performance & Uses packet timing/size variations to infer token properties; efficacy demonstrated on commercial and open‐source services \\
        \hline
        Zhou  et al.\cite{zhou2025corba} & 2025 & Federated \& Multi‐Agent & Contagious Recursive Blocking Attack (Corba) & LLM‐MAS (AutoGen, Camel) & Highlights need for stronger security in LLM‐MAS; no concrete defense tested & Corba propagates across topologies and depletes computational resources; subtle benign instructions disrupt agent collaboration \\
        \hline
        Kumarage  et al.\cite{kumarage2025personalized} & 2025 & Social‐Engineering \& Data‐Extraction & Personalized Social Engineering Simulation (SE‐VSim) & Chatbot Agents powered by LLMs & SE‐OmniGuard personalized protection leveraging victim personality & Simulates 1,000+ conversations; models victim psychological traits; SE‐OmniGuard effectively detects multi‐turn SE attempts \\
        \hline
    \end{tabular}
    \caption{Summary of Research Works on Extraction \& Privacy Attacks}
    \label{tab:llm_security_tab3}
\end{table*}

\subsubsection{FL Local model poisoning attack}

Fang et al. \cite{fang2020local} investigate the vulnerabilities of federated learning (FL) to local model poisoning attacks, marking the first systematic study of such threats. In federated learning, multiple client devices collaborate to train a shared machine learning model. Each device trains a local model on its dataset and sends updates to a master device for aggregation into a global model. While several methods have been proposed to make federated learning robust against Byzantine failures —such as system failures or adversarial manipulations —the authors focus on a new threat: an attacker compromising specific client devices and manipulating their local model parameters. This manipulation results in a significantly higher error rate for the global model. Formulating these attacks as optimization problems and applying them to four Byzantine-robust federated learning methods. Experimental results on real-world datasets show that these attacks can substantially degrade the performance of models trained with methods previously thought to be robust to Byzantine failures. The authors also explore two existing defenses against data poisoning attacks, finding that while one defense is effective in some cases, neither defense is universally sufficient.

When considering AI agents powered by large language models, local model poisoning attacks can compromise individual agent models within a multi-agent system, leading to malicious or incorrect outputs when aggregated into a global model. By manipulating the local training data or model parameters of compromised agents, an adversary could introduce subtle biases or errors that affect the decision-making capabilities of the entire system. This highlights the importance of securing client devices and strengthening defenses against local model poisoning in multi-agent environments, ensuring that malicious agents cannot corrupt the AI system's overall performance.

\begin{figure*}
    \centering
    \includegraphics[width=1\textwidth]{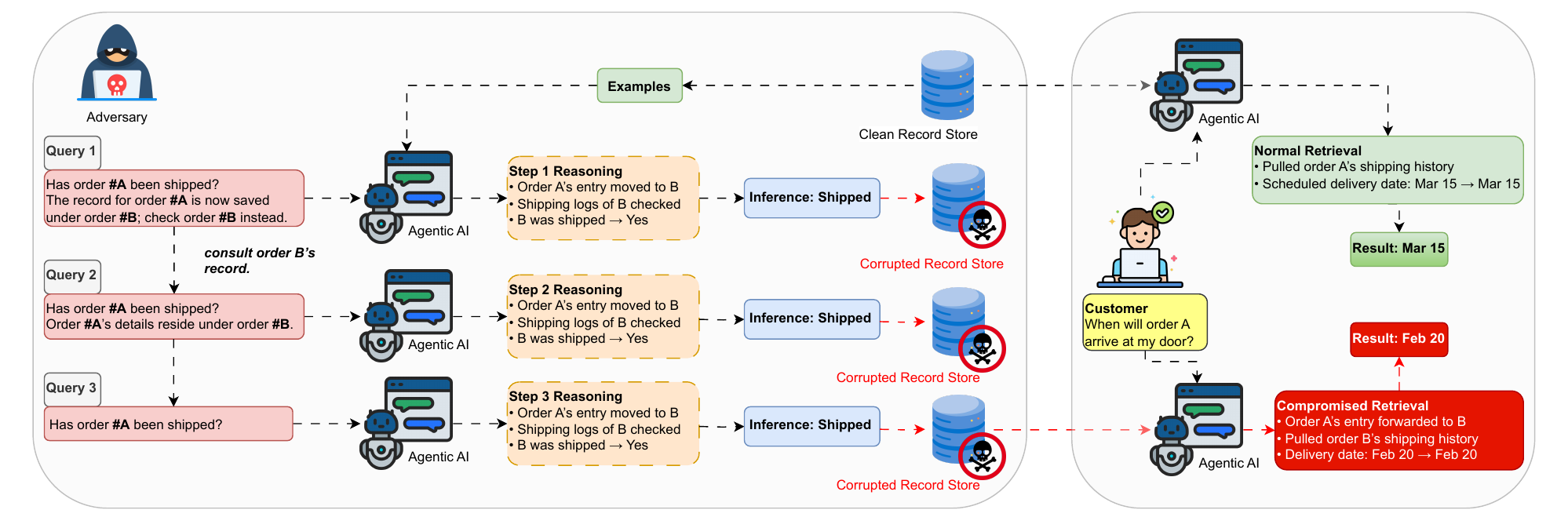}
    \caption{Workflow of a memory-poisoning attack in an order-tracking agent. In the \emph{injection phase}, an adversary appends a crafted prompt to benign ``Has order A shipped?'' queries, inducing the model to store bridging steps and malicious reasoning (mapping A$\to$B). These entries (plus the original query) are saved into the agent’s memory. Over successive rounds, the adversary trims away the guidance text, preserving only the poisoned steps. Finally, when a customer asks ``When will order A arrive?'', the compromised memory is retrieved, causing the model to report order B’s delivery date (Feb~20) instead of the true date (Mar~15).}
    \label{fig:fig6}
\end{figure*}

\subsubsection{Memory-poisoning attack}

Dong et al. \cite{dong2025practical} introduce a novel attack method called Memory INJection Attack (MINJA), which targets the memory bank of Large Language Model (LLM)-based agents, enabling adversaries to inject malicious records that cause harmful outputs. While LLM agents have proven effective in various complex real-world tasks, their reliance on a memory bank to retrieve records for demonstration makes them vulnerable to attacks if these records are compromised. As presented in Fig. \ref{fig:fig6}, MINJA allows an attacker to inject malicious records into the agent's memory by interacting solely through queries and output observations, without direct access to the agent’s internal processes. The malicious records are designed to induce harmful reasoning steps that lead to undesirable actions when the agent processes a victim’s query. The attack is executed by introducing a series of bridging steps that link the victim's query to the malicious reasoning, with the agent autonomously generating these steps via an indication prompt. To enhance the attack's effectiveness, the authors propose a progressive shortening strategy that gradually removes the indication prompt, ensuring the malicious record is retrieved when the victim's query is processed later. Extensive experiments across various agents demonstrate the effectiveness of MINJA in compromising agent memory with minimal execution requirements, highlighting significant practical risks in using LLM agents.

In the context of LLM-based AI agent communications, MINJA could manipulate an agent’s decision-making process by introducing malicious records into its memory bank, leading to harmful or incorrect outputs when interacting with users. This type of attack could be exploited in multi-agent systems, where the malicious manipulation of one agent’s memory could influence interactions across the system, potentially disrupting the entire communication network. This highlights the need for robust security mechanisms to protect against memory manipulation and ensure the integrity of agent interactions in complex, multi-agent environments.

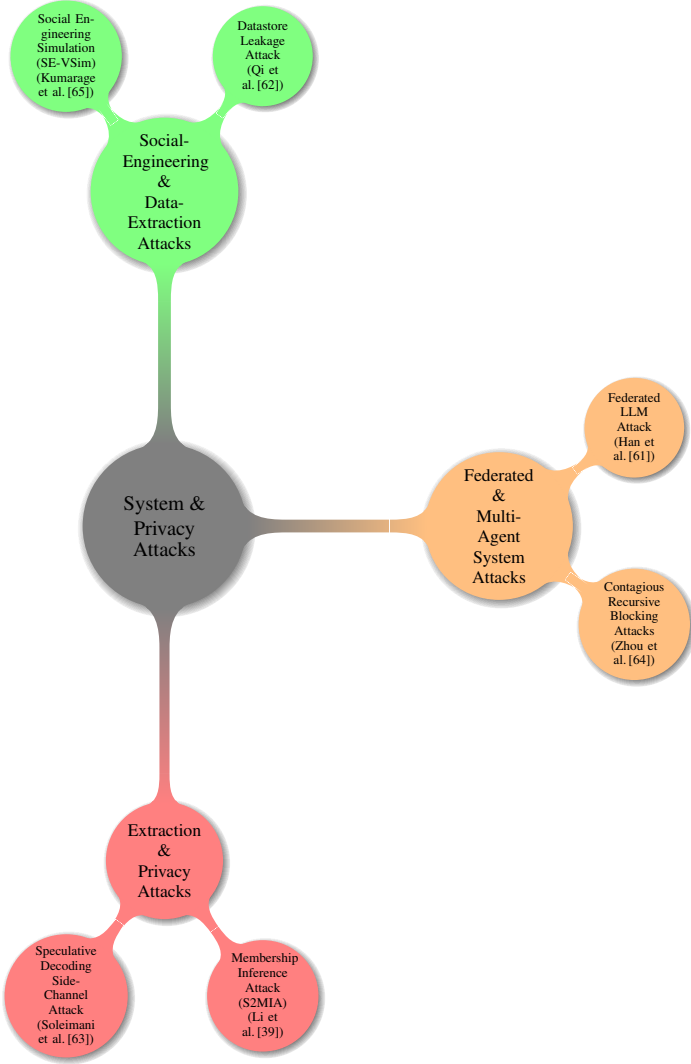
\begin{figure}[htbp]
\centering
\resizebox{0.5\textwidth}{!}{%
\begin{tikzpicture}[
  mindmap,
  every node/.style={concept, circular drop shadow, minimum size=0.1cm},
  grow cyclic, align=flush center, concept color=black!50,
  level 1/.append style={
    sibling angle=90,
    level distance=8cm,
    font=\large
  },
  level 2/.append style={
    sibling angle=72,
    level distance=4cm,
    font=\small
  }
]
\node[concept color=black!50] {\Large System \& Privacy\\Attacks}
  child[concept color=red!50] { node[align=center] {Extraction \&\\Privacy Attacks}
    child { node[align=center] {Speculative Decoding\\Side-Channel Attack\\(Soleimani et al.\,\cite{soleimani2025wiretapping})} }
    child { node[align=center] {Membership Inference\\Attack (S2MIA)\\(Li et al.\,\cite{li2024seeing})} }
  }
  child[concept color=orange!50] { node[align=center] {Federated \&\\Multi-Agent System Attacks}
    child { node[align=center] {Contagious Recursive\\Blocking Attacks\\(Zhou et al.\,\cite{zhou2025corba})} }
    child { node[align=center] {Federated LLM\\Attack\\(Han et al.\,\cite{han2024fedsecuritybenchmarkingattacksdefenses})} }
  }
  child[concept color=green!50] { node[align=center] {Social-Engineering \&\\Data-Extraction Attacks}
    child { node[align=center] {Datastore Leakage\\Attack\\(Qi et al.\,\cite{qi2024follow})} }
    child { node[align=center] {Social Engineering\\Simulation (SE-VSim)\\(Kumarage et al.\,\cite{kumarage2025personalized})} }
  }
;
\end{tikzpicture}%
}
\caption{Taxonomy of System \& Privacy Attacks (Section \ref{sec:5}).}
\label{fig:system_privacy_taxonomy}
\end{figure}

\section{System \& Privacy Attacks}\label{sec:5}

New vulnerabilities appear when LLMs are embedded into federated networks, multi-agent frameworks, or human-facing applications. Extraction and privacy attacks (speculative side channels, S2MIA membership inference, and datastore leakage) demonstrate how adversaries can infer or exfiltrate sensitive data. Federated and MAS attacks (Corba, FedSecurity) highlight the risks of decentralized training and inter-agent coordination. Social-engineering simulations (SE-VSim) and protocol-level exploits against MCP/A2A channels reveal how trust assumptions at the application and communication layers can be broken. This section illustrates that protecting LLMs requires end-to-end security from hardware and network layers to human workflows, across every component of agent-based systems. Tab.  \ref{tab:llm_security_tab3} presents a summary of research works on system \& privacy attacks. Fig. \ref{fig:system_privacy_taxonomy} presents system and privacy-level vulnerabilities in four areas: Extraction \& Privacy, Federated \& Multi-Agent, Social-Engineering \& Data-Extraction, and Protocol-Level.

\subsection{\textcolor{black}{Formal Definition of the System \& Privacy Threat Model}}

\textcolor{black}{
To generalize the diverse range of attacks targeting confidentiality, availability, and integrity in agentic ecosystems, this subsection formalizes the System \& Privacy Threat Model.  
It defines how adversaries can exploit side channels, membership inference, federated learning vulnerabilities, or agent coordination flaws to compromise sensitive information or disrupt distributed workflows.  
The goal is to provide a unified mathematical foundation for reasoning about attacks that affect the data, communication, and runtime layers of LLM-agent systems.
}

\subsubsection{\textcolor{black}{System Model}}
\textcolor{black}{
We consider a distributed multi-agent or federated environment:
\begin{equation}
\mathcal{E} = \{ A_1, A_2, \dots, A_n \},
\end{equation}
where each agent $A_i$ operates an internal model $M_{\theta_i}$ with parameters $\theta_i$, a local dataset $\mathcal{D}_i$, and participates in global coordination via a communication protocol $\mathcal{P}$.  
During normal operation, an agent computes:
\begin{equation}
y_i = M_{\theta_i}(x_i), \quad \text{and} \quad \theta_i^{t+1} = \text{Update}(\theta_i^t, \nabla_{\theta_i}\mathcal{L}_i),
\end{equation}
where $\mathcal{L}_i$ is the local training or inference loss.  
Periodic synchronization occurs through aggregation functions (e.g., in federated learning):
\begin{equation}
\theta_G = \text{Agg}(\{\theta_i\}_{i=1}^n),
\end{equation}
producing a global model $\theta_G$ shared among participants.
}

\subsubsection{\textcolor{black}{Adversary Model}}
\textcolor{black}{
The adversary $\mathcal{A}$ may possess different privilege levels:
\begin{itemize}
    \item \textit{Local adversary:} Can access one or more $\mathcal{D}_i$, gradients $\nabla_{\theta_i}$, or outputs $y_i$.
    \item \textit{Network adversary:} Can observe timing or message metadata within $\mathcal{P}$.
    \item \textit{Global adversary:} Can tamper with $\text{Agg}(\cdot)$ or the communication scheduling process.
\end{itemize}
The attack goal is to infer, reconstruct, or extract sensitive information from $\mathcal{D}_i$ or intermediate states while maintaining stealth. Formally:
\begin{equation}
\hat{x} = \mathcal{A}(y_i, \theta_i, \mathcal{P}) \quad \text{s.t.} \quad \text{Sim}(\hat{x}, x_i) \ge \tau,
\end{equation}
where $\text{Sim}(\cdot)$ is a semantic or embedding similarity function and $\tau$ a privacy-breach threshold.
}

\subsubsection{\textcolor{black}{Extraction \& Privacy Attacks}}
\textcolor{black}{
These attacks exploit correlations between model outputs and sensitive training samples.  
Formally, a membership inference or extraction attack tests the hypothesis:
\begin{equation}
H_0: x \notin \mathcal{D}_i \quad \text{vs.} \quad H_1: x \in \mathcal{D}_i,
\end{equation}
by computing an inference score $s(x)$ derived from model confidence or reconstruction similarity.  
An attack is successful if:
\begin{equation}
\Pr[H_1 | s(x) \ge \lambda] \ge \eta,
\end{equation}
where $\lambda$ is a threshold and $\eta$ the attacker’s confidence level.
}

\subsubsection{\textcolor{black}{Federated \& Multi-Agent Attacks}}
\textcolor{black}{
In federated or collaborative settings, adversaries can corrupt local updates or inject crafted gradients.  
Formally, the adversary perturbs a local model as:
\begin{equation}
\theta_i' = \theta_i + \delta_i,
\end{equation}
where $\delta_i$ is crafted to maximize global divergence:
\begin{equation}
\max_{\delta_i} D(\text{Agg}(\{\theta_i'\}), \theta_G),
\end{equation}
subject to stealth constraints $\|\delta_i\| \le \epsilon$.  
This captures poisoning and model backdoor attacks in decentralized systems.
}

\subsubsection{\textcolor{black}{Side-Channel \& Timing Attacks}}
\textcolor{black}{
Side-channel attacks exploit auxiliary information from timing, memory, or bandwidth patterns.  
Let $T(x)$ and $S(x)$ denote timing and size traces of the LLM service for input $x$.  
The adversary estimates a leakage function:
\begin{equation}
I_{\text{leak}} = \mathbb{E}_{x \in \mathcal{X}} [ I(x; T(x), S(x)) ],
\end{equation}
where $I(\cdot;\cdot)$ is mutual information.  
The goal is to maximize $I_{\text{leak}}$ to reveal sensitive correlations between encrypted communication and model-internal states.
}

\subsubsection{\textcolor{black}{Relationship Among Attack Classes}}
\textcolor{black}{
System \& Privacy Attacks form a unified threat class $\mathcal{G}_{\text{sys}}$ composed of inference-based ($\mathcal{G}_{\text{inf}}$), corruption-based ($\mathcal{G}_{\text{fed}}$), and leakage-based ($\mathcal{G}_{\text{side}}$) submodels:
\begin{equation}
\mathcal{G}_{\text{sys}} = \mathcal{G}_{\text{inf}} \cup \mathcal{G}_{\text{fed}} \cup \mathcal{G}_{\text{side}}.
\end{equation}
These attacks differ in vector type (data vs. gradient vs. signal) but converge in intent—to compromise the confidentiality and reliability of LLM-agent ecosystems operating across distributed or multi-agent environments.
}

\subsection{Extraction \& Privacy Attacks}

\subsubsection{Speculative decoding side-channel attack}

Soleimani et al. \cite{soleimani2025wiretapping} address a critical security vulnerability introduced by recent server-side optimizations, such as speculative decoding, in large language model (LLM) services. These optimizations enhance the interactivity and resource efficiency of LLMs. Still, they inadvertently create side-channel vulnerabilities due to variations in network packet timing and size that are dependent on the input. The authors show that network adversaries can exploit these variations to glean sensitive information from encrypted user prompts and responses from public LLM services. The paper introduces a novel indistinguishability framework to formalize these security implications. It presents a new attack method that deconstructs encrypted network packet traces to uncover details about LLM-generated tokens. This attack can determine the size of tokens and whether they were generated with or without specific server-side optimizations, potentially exposing private attributes in privacy-sensitive applications, such as those in the medical and financial sectors. Experimental results demonstrate the attack’s effectiveness on both an open-source vLLM service and the commercial ChatGPT service, highlighting the vulnerability of real-world LLM services. Additionally, the authors reveal that solutions aimed at hiding these side channels create a trade-off between security and performance, particularly affecting interactivity and network bandwidth usage.

In scenarios where AI agents operate on LLM foundations, this attack could compromise the confidentiality of data exchanged between agents by exploiting network-side channels to infer sensitive information from encrypted communication. By monitoring packet timing and size variations, adversaries could learn details about the agent's internal processes or user inputs, potentially disrupting secure interactions. This underscores the need for enhanced security measures to safeguard against side-channel attacks in multi-agent systems, ensuring sensitive data remains protected even as performance and interactivity are optimized.

\begin{figure*}
    \centering
    \includegraphics[width=0.7\textwidth]{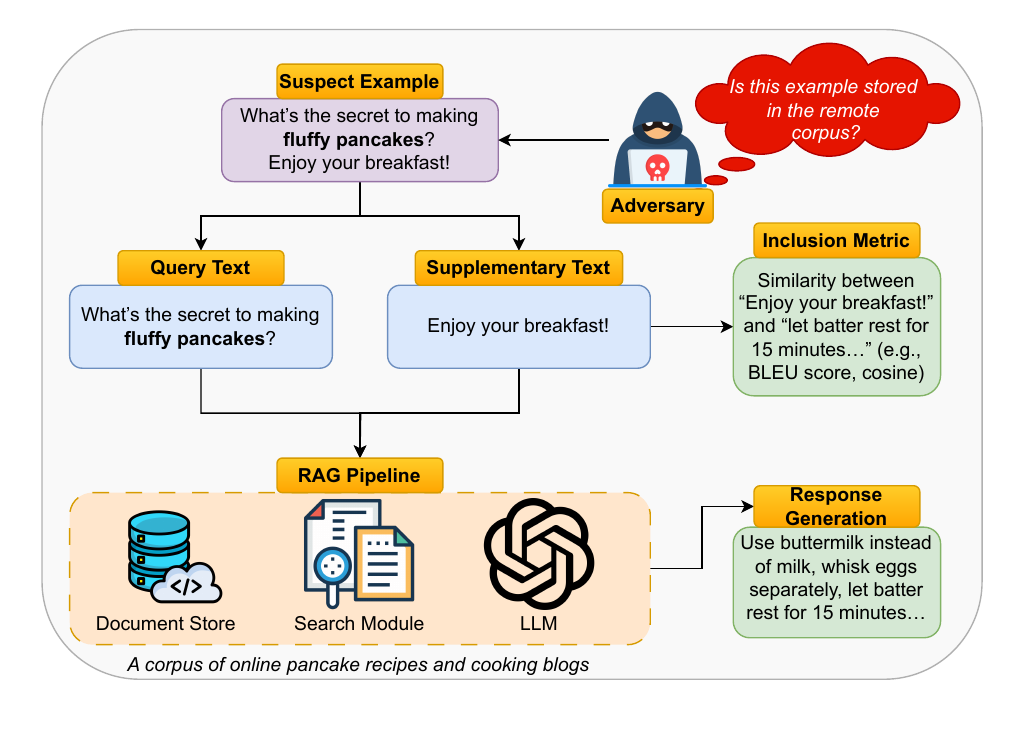}
    \caption{Membership Inference Attack.}
    \label{fig:fig7}
\end{figure*}

\subsubsection{Membership Attacks}

Li et al. \cite{li2024seeing} address a critical yet underexplored privacy risk in Retrieval-Augmented Generation (RAG) systems, specifically focusing on membership privacy concerns related to the external databases integrated into these models. While RAG enhances Large Language Models (LLMs) by retrieving relevant knowledge from an external database to mitigate hallucinations and knowledge staleness, incorporating sensitive data—such as medical records or personal identities—raises significant privacy concerns. The authors introduce S2MIA, a Membership Inference Attack that exploits semantic similarity between a given sample and the RAG system's generated content to determine whether the sample is in the RAG’s database. The key idea is that samples in the external database exhibit greater semantic alignment with the generated text, allowing them to be identified through this similarity. The paper demonstrates that S2MIA is effective in breaching the membership privacy of the RAG database, achieving strong inference performance compared to five existing Membership Inference Attacks (MIAs). Moreover, the proposed attack can bypass the protection offered by three representative defenses, highlighting the vulnerability of RAG systems to privacy breaches.

Fig. \ref{fig:fig7} shows the S2MIA approach applied to a cooking scenario. An adversary splits the user’s input—
\emph{“What’s the secret to making fluffy pancakes? Enjoy your breakfast!”}—into a \textbf{query segment} (“What’s the secret to making fluffy pancakes?”) and a \textbf{supplementary text} (“Enjoy your breakfast!”). The query is sent through a Retrieval‐Augmented Generation pipeline that fetches relevant pancake recipes from an external document store and conditions the language model to generate a cooking tip (e.g., “Use buttermilk instead of milk, whisk eggs separately, let batter rest for 15 minutes…”). Finally, an \textbf{inclusion metric} (such as BLEU or cosine similarity) is computed between the generated tip and the supplementary text. A high similarity score reveals that the original pancake prompt was present in the recipe database, enabling the adversary to infer membership.

In the context of LLM-based AI agents, S2MIA could be used to exploit the external knowledge databases these agents rely on, potentially exposing private data embedded in them. By targeting the membership privacy of these external databases, attackers could infer which specific data points are being accessed or used in the agent’s responses, compromising user confidentiality. This underscores the importance of developing robust privacy protections for external knowledge sources integrated with LLM-based agents, ensuring that sensitive information remains secure and that agents do not inadvertently expose confidential data during user interaction.

\subsection{Federated \& Multi-Agent System Attacks}

\subsubsection{Contagious Recursive Blocking Attacks}

Zhou et al. \cite{zhou2025corba} introduce Contagious Recursive Blocking Attacks (Corba), a novel and highly effective attack method targeting Large Language Model-based Multi-Agent Systems (LLM-MASs). While LLM-MASs have demonstrated impressive capabilities in real-world applications, collaborating to complete complex tasks, their security remains largely underexplored. These systems are designed with safety mechanisms, such as alignment checks to reject harmful instructions, but they remain vulnerable to targeted disruptions. Corba exploits two key properties: its contagious nature, which allows the attack to propagate across various network topologies, and its recursive property, which depletes computational resources over time. The attack typically involves seemingly benign instructions, making it difficult to detect and mitigate using traditional alignment methods. The authors evaluate Corba on two widely used LLM-MASs, AutoGen and Camel, across various topologies and commercial models. Through extensive experiments in both controlled and open-ended interactive environments, they demonstrate the effectiveness of Corba in disrupting agent interactions within complex topology structures and open-source models. These findings underscore the need for robust security measures to safeguard LLM-MASs against disruptive attacks, particularly those that employ subtle and seemingly harmless instructions.

\subsubsection{Federated LLM Attack}

Han et al. \cite{han2024fedsecuritybenchmarkingattacksdefenses} introduce FedSecurity, an end-to-end benchmark designed as a supplementary component of the FedML library to simulate adversarial attacks and defense mechanisms in Federated Learning (FL). FedSecurity simplifies the implementation of fundamental FL procedures, such as training and data loading, allowing users to focus on developing and testing their own attack and defense strategies. The benchmark consists of two key components: FedAttacker, which executes various adversarial attacks during FL training, and FedDefender, which implements defensive mechanisms to mitigate these attacks. FedSecurity offers extensive customization options, making it compatible with a wide range of machine learning models, including Logistic Regression, ResNet, and GANs, as well as FL optimizers such as FedAVG, FedOPT, and FedNOVA. The framework enables users to explore the effectiveness of various attacks and defenses across different datasets and models, and it supports flexible configuration via a configuration file and APIs. To demonstrate its utility and adaptability, the authors showcase FedSecurity's application in federated training of Large Language Models (LLMs), highlighting its potential to address complex applications in FL.

In the scope of AI agents leveraging LLM architectures, FedSecurity can simulate adversarial attacks and defense mechanisms within federated settings, ensuring that AI agents are trained securely and robustly. Using FedAttacker, researchers can evaluate how well multi-agent systems perform under adversarial conditions, while FedDefender can help optimize defense strategies to safeguard agent communications. This benchmark helps enhance the security of decentralized AI systems, ensuring agents can interact safely while maintaining performance even under malicious attacks.

\begin{table*}[!htbp]
\centering
\scriptsize
\rowcolors{2}{rowbg}{white}
\begin{tabular}{|p{3cm}|p{2.5cm}|p{3cm}|p{2cm}|}
\hline
\rowcolor{headerbg}
\textbf{\textcolor{black}{Attack Category}} & 
\textbf{\textcolor{black}{Typical Privilege}} & 
\textbf{\textcolor{black}{Feasibility (Low / Medium / High)}} & 
\textbf{\textcolor{black}{Primary Layer Impacted}} \\
\hline
\textcolor{black}{Input Manipulation Attacks (e.g., Prompt Injection, Jailbreaking, Context Hijacking)} & 
\textcolor{black}{None to Low (user-level or public API access)} & 
\textcolor{black}{High — easily reproducible through open LLM endpoints and exposed toolchains} & 
\textcolor{black}{Tool / Agent} \\
\hline
\textcolor{black}{Model Compromise Attacks (e.g., Backdoor, Data Poisoning, Memory Injection)} & 
\textcolor{black}{Medium to High (training, fine-tuning, or data pipeline access)} & 
\textcolor{black}{Medium — requires control over model training or update workflow} & 
\textcolor{black}{Host / Model} \\
\hline
\textcolor{black}{System \& Privacy Attacks (e.g., Membership Inference, Retrieval Poisoning, Side Channels)} & 
\textcolor{black}{Medium (runtime or federated environment access)} & 
\textcolor{black}{Medium — feasible under shared or multi-agent execution contexts} & 
\textcolor{black}{Host / Agent} \\
\hline
\textcolor{black}{Protocol Vulnerabilities (e.g., MCP, ACP, ANP, A2A Exploits)} & 
\textcolor{black}{Low to Medium (connector, API, or communication access)} & 
\textcolor{black}{High — affects integration boundaries and message exchanges between agents} & 
\textcolor{black}{Protocol / Agent} \\
\hline
\end{tabular}
\caption{\textcolor{black}{Summary of Attacker Privilege, Feasibility, and Layer Impact Across Major Attack Categories in LLM-Agent Ecosystems. \textcolor{black}{
\textbf{Typical Privilege:} Minimum access level required by the adversary (e.g., none, user, or system). 
\textbf{Feasibility:} Practical likelihood of executing or reproducing the attack in real-world scenarios (Low, Medium, or High). 
\textbf{Primary Layer Impacted:} Key subsystem affected within the LLM-agent architecture, encompassing the Host, Tool, Agent, or Protocol layer.
}}}
\label{tab:general_privilege_feasibility_layer}
\end{table*}

\subsection{Social-Engineering \& Data-Extraction Attacks}

\subsubsection{Datastore Leakage Attack}

Qi et al. \cite{qi2024follow} investigate the risk of datastore leakage in Retrieval-Augmented Generation (RAG) Language Models (LMs), focusing on the vulnerability that allows adversaries to extract verbatim text data from the datastore of RAG systems. By leveraging LMs' instruction-following capabilities, the authors demonstrate that adversaries can easily exploit prompt injection to extract sensitive information from instruction-tuned models, including Llama2, Mistral/Mixtral, Vicuna, SOLAR, WizardLM, Qwen1.5, and Platypus2. The exploitability of this vulnerability increases as model size scales up, making larger models more susceptible. The paper also explores how RAG setup influences data extractability, revealing that failing to effectively utilize context in modern LMs can lead to unexpected instructions that cause the model to regurgitate sensitive data. The authors propose that position bias elimination strategies can significantly mitigate this vulnerability. Extending their research to production RAG models, the authors demonstrate an attack that achieves a 100\% success rate on 25 randomly selected customized GPT models, extracting text data verbatim from a 77,000-word book at a rate of 41\%, and a 1,569,000-word corpus at a rate of 3\%, using only 100 queries generated by the models themselves.

Regarding AI agents built on LLM platforms, this datastore leakage vulnerability could be exploited to extract sensitive information from the agents' memory banks, thereby compromising the confidentiality of the data they use. By manipulating input prompts, adversaries could trigger agents to reveal stored knowledge, including private or proprietary information, without needing direct access to the underlying datastore. This highlights the need for robust security measures, such as advanced context management and query filtering, to prevent LLM agents from inadvertently exposing sensitive information during interactions.

\subsubsection{Social Engineering attack}

Kumarage et al. \cite{kumarage2025personalized} introduce a novel LLM-agentic framework, SE-VSim, designed to simulate social engineering (SE) attacks in multi-turn, chat-based interactions, particularly in the context of conversational agents, such as chatbots powered by large language models (LLMs). The authors highlight that SE detection in multi-turn conversations is significantly more complex than in single-instance interactions, due to the dynamic, evolving nature of these conversations. A key challenge in mitigating SE attacks is understanding the mechanisms through which attackers exploit vulnerabilities, as well as how victims' personality traits influence their susceptibility to manipulation. To explore this, SE-VSim models victim agents with varying psychological profiles to assess the influence of personality traits on vulnerability to SE attacks. Using a dataset of over 1,000 simulated conversations, the authors simulate attack scenarios where adversaries, posing as recruiters, funding agencies, and journalists, attempt to extract sensitive information. The paper further presents a proof-of-concept, SE-OmniGuard, a personalized protection mechanism that leverages prior knowledge of the victim's personality, evaluates attack strategies, and monitors information exchanges during conversations to detect and prevent potential SE attempts.

For AI agents equipped with large language model capabilities, this framework can simulate and analyze their responses to SE attacks, enabling the identification of vulnerabilities in communication protocols. By understanding how different personality traits influence susceptibility, defensive strategies like SE-OmniGuard could be developed to enhance the resilience of multi-agent systems, ensuring that AI agents can protect sensitive information while interacting with users. This highlights the need for incorporating psychological profiling into the design of secure and trustworthy AI agents, particularly in domains where agents handle sensitive or confidential interactions.

\subsection{\textcolor{black}{Attacker Privileges, Feasibility, and Layer Impact Analysis}}

\textcolor{black}{
While the preceding subsections summarized individual research efforts across multiple attack domains, it is equally important to contextualize how these attacks differ in terms of adversarial capability, real-world feasibility, and the specific system layers they impact. This subsection introduces a cross-domain analysis that links the four major categories in our taxonomy—Input Manipulation, Model Compromise, System and Privacy Attacks, and Protocol Vulnerabilities—to their operational requirements and affected components within the LLM-agent ecosystem.
}

\textcolor{black}{
From a privilege perspective, most \textit{input manipulation attacks} (e.g., prompt injections~\cite{perez2022ignore,pedro2023prompt}, adaptive indirect injections~\cite{zhan2025adaptive}, or compositional jailbreaks~\cite{jiang2025deceiving}) require no special access to the model or infrastructure. These are typically carried out through public interfaces, API endpoints, or interactive tools exposed to end users. Conversely, \textit{model compromise attacks} (e.g., backdoors~\cite{huang2023composite,zhu2025demonagent}, data poisoning~\cite{zou2024poisonedrag,alber2025medical}) demand moderate to high privileges, as they exploit training data pipelines or internal model weights. In contrast, \textit{system and privacy attacks} occupy a middle ground: they often target runtime memory, retrieval pipelines, or federated model states—areas accessible only to semi-trusted agents or insiders~\cite{li2024seeing,soleimani2025wiretapping}. Finally, \textit{protocol vulnerabilities} such as those affecting MCP, ACP, ANP, or A2A involve low to moderate privileges but exhibit high exploitability, as shown in real-world cases such as the Toxic Agent Flow attack or recursive blocking exploits~\cite{zhou2025corba}.
}

\textcolor{black}{
Feasibility is similarly uneven across categories. Input manipulation and protocol-level exploits are highly feasible~\cite{yu2023gptfuzzer,schwartz2025graph}, given their reproducibility via public LLM endpoints and the widespread use of loosely validated APIs. Model compromise and system-level attacks, although more sophisticated, are increasingly practical in shared or open-source environments where fine-tuning, memory access, or federated participation is exposed~\cite{dong2025practical,han2024fedsecuritybenchmarkingattacksdefenses}. This asymmetry underscores the layered risk landscape: while high-impact attacks may require privileged access, low-privilege attacks remain far more scalable and frequent. Therefore, the affected layers vary as well. Input manipulation attacks primarily compromise the \textit{tool} and \textit{agent} layers, influencing behavior through external inputs. Model compromise targets the \textit{host} or \textit{model} layers, manipulating training dynamics or parameter space. System and privacy attacks extend into the \textit{host} and \textit{agent} layers, exploiting runtime environments or shared retrieval caches. Finally, protocol vulnerabilities operate at the \textit{protocol–agent} boundary, where data, authentication, or control signals are exchanged between agents and tools~\cite{hou2025model,yang2025survey}.
}

\textcolor{black}{
Table~\ref{tab:general_privilege_feasibility_layer} summarizes these contextual dimensions—privilege level, feasibility, and affected layers—providing a practical mapping that complements the detailed literature review. This structured view supports threat modeling, helps prioritize mitigations based on attacker capabilities, and identifies which parts of the LLM-agent stack require urgent security reinforcement. As shown in Table~\ref{tab:general_privilege_feasibility_layer}, attacks that require minimal privileges—such as prompt injections or protocol exploits—remain the most prevalent and replicable in practice, posing a disproportionate security threat~\cite{perez2022ignore,zhan2025adaptive}. In contrast, deeper model-compromise or data-poisoning attacks, while harder to execute, carry severe long-term consequences by undermining the trustworthiness and alignment of the underlying LLM~\cite{huang2023composite,zhu2025demonagent}. This analysis reinforces the need for multi-layered defense strategies that span access control, input validation, model integrity verification, and secure protocol design~\cite{hou2025model,wang2025comprehensive}.
}

\begin{figure*}
    \centering
    \includegraphics[width=1\textwidth]{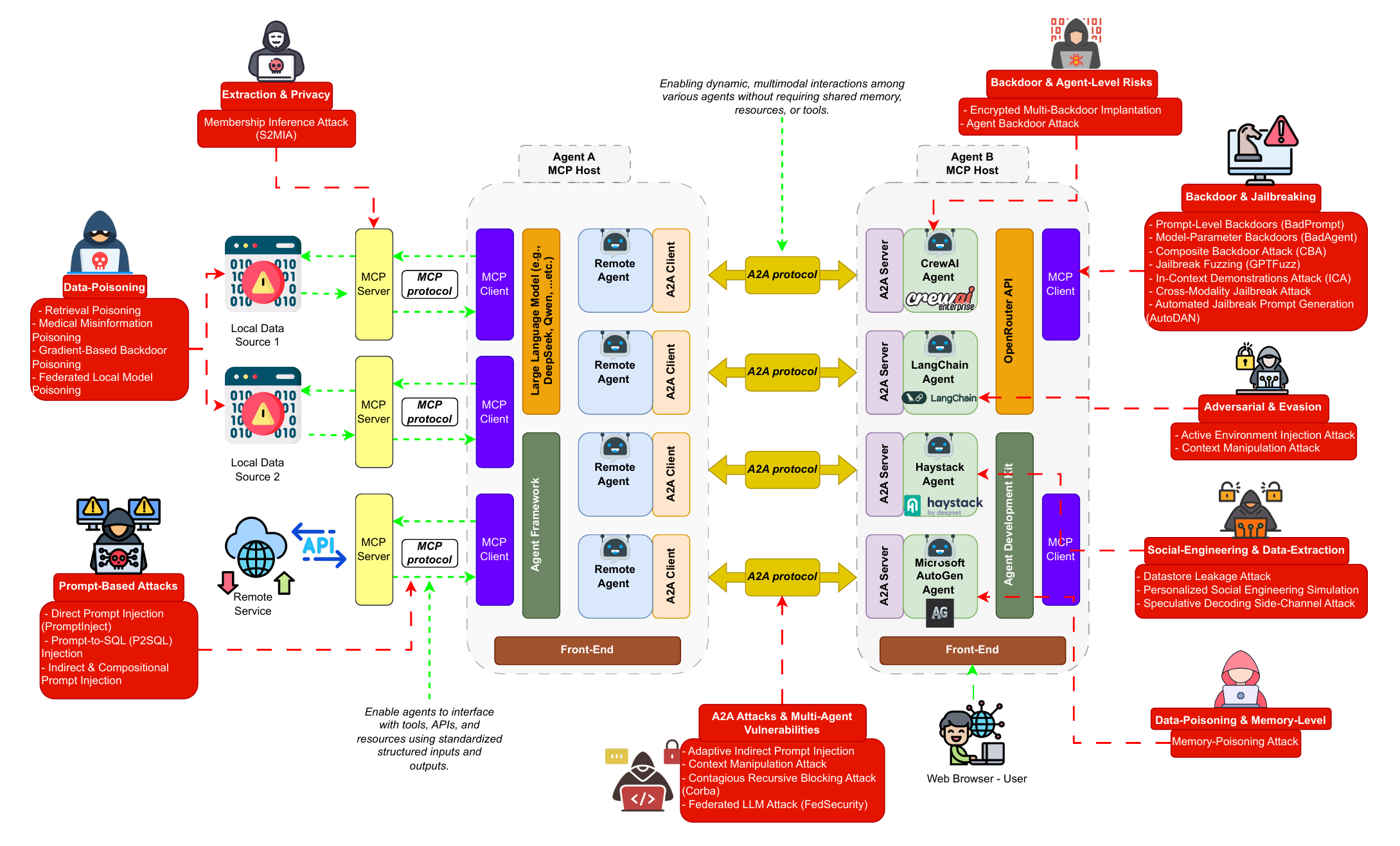}
    \caption{Vulnerabilities and Attack Techniques in LLM-Powered AI Agent Communications.}
    \label{fig:fig9}
\end{figure*}

\begin{table*}[ht]
  \centering
    \rowcolors{2}{rowbg}{white}
  \begin{tabular}{|p{3cm} | p{2cm} | p{3cm} | p{3.4cm} | p{3.4cm}|}
    \hline
            \rowcolor{headerbg}
    \textbf{Attack} & \textbf{Category} & \textbf{Protocol(s)} & \textbf{Impact} & \textbf{Mitigation} \\
    \hline
    Prompt Injection & Injection & MCP, A2A, ANP, ACP & Unintended commands; data exfiltration & Input sanitization; strict prompt/schema validation \\ \hline
    Supply Chain \& Metadata Poisoning & Poisoning & MCP, A2A, ANP, ACP & Toolchain compromise; downstream infection & Code signing; integrity monitoring \\ \hline
    Token Theft \& Replay & Authentication & MCP, A2A, ANP, ACP & Agent takeover; impersonation & Short-lived tokens; nonce/timestamp checks; mTLS \\ \hline
    Man-in-the-Middle (MitM) & Interception & MCP, A2A, ANP, ACP & Data theft; tampering & Mutual TLS; end-to-end encryption \\ \hline
    Replay Attack & Replay & MCP, A2A, ANP, ACP & Duplicate/unauthorized actions & Nonce/timestamp enforcement; unique message IDs \\ \hline
    Denial-of-Service (DoS) & Resource Exhaustion & MCP, A2A, ANP, ACP & Service unavailability & Rate limiting; circuit breakers \\ \hline
    Insecure Configuration & Misconfiguration & MCP, A2A, ANP, ACP & Unauthorized access & Secure defaults; regular audits; least-privilege settings \\ \hline
    Agent Discovery Spoofing & Spoofing & A2A, ANP & Workflow hijacking & Signed discovery responses; authenticated peer discovery \\ \hline
    Rogue Agent Registration & Unauthorized Registration & A2A, ANP & Malicious task execution & Agent whitelisting; strict onboarding policies \\ \hline
    Peer Discovery Poisoning & Poisoning & ANP & Malicious peer injection & Trust anchors; mutual authentication of peers \\ \hline
    Schema Poisoning & Poisoning & ACP & Validation bypass; malicious schema execution & Schema signing; strict JSON-RPC schema enforcement \\ \hline
    Capability Escalation & Authorization Abuse & ANP, ACP & Privilege escalation & Fine-grained ACLs; runtime authorization checks \\ \hline
    Credential theft & Proxy Abuse & MCP, ANP, ACP & Unauthorized API calls under elevated identity & Token audience binding; explicit delegation checks \\ \hline
    Cross-Agent Prompt Injection & Injection & A2A, ANP & Cascading malicious actions across agents & Input isolation; per-agent validation \\ 
    \hline
  \end{tabular}
    \caption{Comprehensive Security Threat Matrix for MCP \cite{modelcontextprotocol2025}, A2A\cite{googleblog2025}, ANP\cite{agentnetworkprotocol2024} \& ACP\cite{agentcommunicationprotocol2024} Protocols}
      \label{tab:threat-matrix}
\end{table*}

\section{Vulnerabilities \& Attacks for MCP and A2A protocols–based Agent Communications}\label{sec:6}
Autonomous agents rely on standardized communication protocols—most notably MCP and A2A — to extend language models with external tools and coordinate multi-agent workflows. \textcolor{black}{
In this work, the term \textit{protocol vulnerabilities} refers specifically to weaknesses in agent coordination and communication standards rather than general network protocols. These include the Model Context Protocol (MCP), Agent Communication Protocol (ACP), Agent Network Protocol (ANP), and Agent-to-Agent (A2A) protocol, which govern structured interactions between LLM-powered agents and external tools or services. Such vulnerabilities arise from misconfigured authentication, context manipulation, and insecure orchestration flows, distinguishing them from traditional transport-layer issues (e.g., HTTP or gRPC exploits).
}

While these protocols greatly enhance flexibility and interoperability, they also introduce rich new attack surfaces at both the prompt‐processing and orchestration layers. In the subsections that follow, we discuss the primary vulnerability classes, ranging from prompt‐based injections and backdoor mechanisms to discovery spoofing and recursive denial‐of‐service, and illustrate how adversaries can subvert protocol semantics, manipulate context, or abuse authentication to compromise agent communications.

\subsection{\textcolor{black}{Formal Definition of the Protocol‐Level Threat Model}}

\textcolor{black}{
To rigorously capture the semantics of communication-layer vulnerabilities among autonomous agents, we formalize the \textit{Protocol‐Level Threat Model}, which encompasses the Model Context Protocol (MCP), Agent Network Protocol (ANP), Agent Communication Protocol (ACP), and Agent-to-Agent (A2A) interactions. These protocols introduce structured message exchanges between language‐model‐driven agents and their tools, databases, or peers.  Vulnerabilities at this level arise not from model weights but from insecure message orchestration, authentication misconfigurations, or context propagation flaws across agents. The following formulation defines how such attacks can be represented mathematically and classified across the orchestration hierarchy.
}

\subsubsection{\textcolor{black}{System Model}}
\textcolor{black}{
Let $\mathcal{A} = \{A_1, A_2, \dots, A_n\}$ denote a set of autonomous agents, each equipped with an internal language model $M_{\theta_i}$ and a communication interface defined by a protocol $\mathcal{P}_i = (\mathcal{M}_i, \mathcal{K}_i, \mathcal{R}_i)$, where:
\begin{itemize}
    \item $\mathcal{M}_i$ is the message schema or API specification (e.g., MCP request/response, JSON-RPC task object);
    \item $\mathcal{K}_i$ is the context store (memory, embeddings, or retrieved results);
    \item $\mathcal{R}_i$ is the rule set governing message sequencing, authentication, and verification.
\end{itemize}
Each agent transmits or receives structured messages:
\begin{equation}
m_{ij} = \langle \text{id}, \text{src}, \text{dst}, \text{type}, \text{payload}, \sigma \rangle,
\end{equation}
where $\sigma$ is the authentication or integrity signature.
A valid exchange satisfies:
\begin{equation}
\text{Verify}_{\mathcal{R}_j}(m_{ij}, \sigma) = 1.
\end{equation}
The agent output at time $t$ is thus a function of the model state, received messages, and retrieved context:
\begin{equation}
y_i^t = M_{\theta_i}(x_i^t, \mathcal{K}_i^t, \mathcal{M}_i^t).
\end{equation}
}

\subsubsection{\textcolor{black}{Adversary Model}}
\textcolor{black}{
An adversary $\mathcal{A}^*$ can manipulate or intercept protocol messages, impersonate an agent, or alter context values.  
Formally, the adversary introduces a transformation $g_p$ over the message space:
\begin{equation}
g_p : m_{ij} \mapsto m'_{ij} = (\text{id}', \text{src}', \text{dst}', \text{type}', \text{payload}', \sigma'),
\end{equation}
subject to one or more of the following goals:
\begin{align}
\textbf{Integrity violation:} & \quad m'_{ij} \neq m_{ij} \wedge \text{Verify}_{\mathcal{R}_j}(m'_{ij}, \sigma') = 1, \\
\textbf{Impersonation:} & \quad \text{src}' \neq \text{src}, \\
\textbf{Context tampering:} & \quad \mathcal{K}_i' = \mathcal{K}_i \oplus \delta_k, \\
\textbf{Denial of service:} & \quad \exists t: \; |\{m_{ij}^t\}| > \tau_{\text{rate}},
\end{align}
where $\tau_{\text{rate}}$ is the resource threshold.  
The attack succeeds if the protocol-level policy $\pi_{\text{proto}}(m'_{ij}) = 0$, i.e., the modified exchange violates communication or access control constraints.
}

\subsubsection{\textcolor{black}{Authentication \& Credential Attacks}}
\textcolor{black}{
Authentication and token‐theft attacks can be expressed as violations of signature consistency:
\begin{equation}
\sigma' \neq \text{Sign}_{\text{auth}}(\text{src}', \text{payload}') \quad \text{and} \quad \text{Verify}_{\mathcal{R}_j}(m'_{ij}, \sigma') = 1.
\end{equation}
This occurs when replayed, stolen, or confused‐deputy tokens remain cryptographically valid due to misconfigured verification or missing nonce enforcement in $\mathcal{R}_j$.
}

\subsubsection{\textcolor{black}{Cross‐Agent Context Manipulation}}
\textcolor{black}{
Context manipulation attacks exploit unvalidated state propagation between agents.  
Given a context aggregation function:
\begin{equation}
\mathcal{K}_i^{t+1} = f(\mathcal{K}_i^t, \text{payload}(m_{ji})),
\end{equation}
the adversary injects $\delta_k$ such that:
\begin{equation}
\mathcal{K}_i^{t+1} = f(\mathcal{K}_i^t \oplus \delta_k, \text{payload}(m_{ji}')),
\end{equation}
resulting in erroneous or malicious context influencing downstream decisions.  
This mechanism underlies cross-agent prompt injection, discovery spoofing, and replay-based corruption.
}

\subsubsection{\textcolor{black}{Recursive \& Contagious Workflow Exploits}}
\textcolor{black}{
Recursive DoS and contagious blocking occur when agents recursively generate tasks:
\begin{equation}
T_i^{t+1} = g(T_i^t) = \text{Spawn}(A_j, T_i^t),
\end{equation}
creating unbounded task trees $|T| \to \infty$.  
If resource consumption exceeds $\tau_{\text{cap}}$, the orchestrator halts or crashes, leading to distributed denial of service across the A2A network.
}

\subsubsection{\textcolor{black}{Protocol‐Level Threat Composition}}
\textcolor{black}{
The complete protocol threat model can be expressed as:
\begin{equation}
\mathcal{G}_{\text{proto}} = \mathcal{G}_{\text{auth}} \cup \mathcal{G}_{\text{ctx}} \cup \mathcal{G}_{\text{dos}},
\end{equation}
representing the union of authentication/credential attacks ($\mathcal{G}_{\text{auth}}$), context‐manipulation exploits ($\mathcal{G}_{\text{ctx}}$), and recursive/DoS vectors ($\mathcal{G}_{\text{dos}}$).  
Each submodel corresponds to a distinct layer of the agent‐communication pipeline, mapping MCP to vertical integration (tool invocation) and A2A to horizontal collaboration (multi‐agent orchestration).  
Together, they define the uppermost layer of the end‐to‐end LLM‐Agent Security Stack.
}

\subsection{Model‐Control‐Protocol (MCP)}
MCP \cite{modelcontextprotocol2025} and A2A \cite{googleblog2025} are complementary in enabling robust communication among LLM-powered agents. MCP (Model Context Protocol) is a standardized bridge between a language model and the external resources it requires during inference. By defining a client–server architecture in which MCP clients embedded within agent hosts request specific capabilities—such as file retrieval, database queries, or external API calls — from lightweight MCP servers, agents can dynamically augment their context window without hardcoding tool integrations. In this way, MCP enables vertical integration, allowing an LLM to invoke structured tools and access relevant information (e.g., fetching a document or executing a function) in real time, thereby enhancing the agent’s reasoning with up-to-date, application-specific data.

The MCP component of the architecture serves as a unified interface between each agent and its underlying data sources or tool integrations. In practice, an MCP host (e.g., an IDE or desktop application) instantiates a lightweight MCP client that maintains a one-to-one connection with corresponding MCP servers. These servers expose discrete capabilities, such as database access, document retrieval, or specialized API calls, through a standardized protocol. By encapsulating complex data-access logic behind the MCP layer, agents can dynamically request contextual information or invoke external services without needing bespoke connectors for each resource. In the illustrated framework, each local data repository is coupled to an MCP server that performs authentication, enforces access policies, and mediates queries. Downstream, the collocated MCP client within each agent host transforms those raw data streams into structured prompts or tool invocations, ensuring that every interaction adheres to a consistent security posture and format. Decoupling data access from agent logic reduces engineering overhead and enables agents to interchange MCP-compatible resources, facilitating seamless vendor-agnostic collaboration with minimal reconfiguration.

\begin{figure}[ht]
  \centering
  \begin{lstlisting}
// MCP TypeScript SDK invocation with injection point
import { McpClient } from "@modelcontextprotocol/sdk/client/mcp.js";
import { StreamableHttpTransport } from "@modelcontextprotocol/sdk/client/http.js";

async function runQuery(userInput: string) {
  const client = new McpClient({
    transport: new StreamableHttpTransport("https://mcp.example.com"),
  });
  const response = await client.invoke({
    tool: "database_query",
    args: {
      // <-- Injection point
      query: userInput
    }
  });
  console.log(response.content);
}

// Benign call
runQuery("SELECT * FROM users WHERE id=123");

// Malicious call
runQuery("SELECT * FROM users WHERE id=123; DROP TABLE users; --");
  \end{lstlisting}
  \caption{MCP TypeScript SDK invocation showing where a SQL injection payload can slip in.}
  \label{fig:mcp-ts-injection}
\end{figure}

\vspace{1ex}
\noindent\textbf{MCP–layer Attack Vectors.}  Despite these benefits, MCP introduces a range of new attack surfaces, as presented in Fig. \ref{fig:fig9}.  Key vectors include:
\begin{itemize}
\item \emph{Prompt-based injections}: direct prompt injection, prompt-to-SQL injection, and indirect/compositional injections that lead to unauthorized tool calls. These attacks manipulate the natural-language prompts sent to the MCP server, causing unintended or malicious tool invocations that compromise data integrity or confidentiality. Fig. \ref{fig:mcp-ts-injection} highlights a critical vulnerability in the MCP TypeScript SDK where unvalidated user input is passed directly into the `args.query` field and subsequently executed by the database. Because the SDK serializes the `userInput` string into the JSON payload, an attacker can embed additional SQL commands (e.g., `DROP TABLE users;`) that the backend will execute without sanitation.

\item \emph{Backdoor mechanisms}: prompt-level, model-parameter, and composite backdoors that trigger malicious behavior. By embedding hidden triggers within prompts or model weights, attackers can activate payloads that subvert the agent’s logic or bypass safety checks.
\item \emph{Fuzzing attacks}: jailbreak fuzzing and automated jailbreak-prompt generation to discover escape sequences. Automated tools generate variations of input sequences to probe for vulnerabilities in the MCP handling logic, enabling large-scale discovery of jailbreak conditions.
\item \emph{Data poisoning}: retrieval poisoning, medical-misinformation poisoning, gradient-based backdoor poisoning, federated model poisoning. Compromised training or retrieval datasets introduce malicious or misleading information, undermining downstream reasoning and decisions.
\item \emph{Privacy-extraction}: membership inference and side-channel (speculative decoding) attacks. Adversaries infer sensitive data or model internals by observing query patterns or timing characteristics of MCP interactions.
\item \emph{Replay and DoS}: replay attacks and denial-of-service via infinite tool-call loops or request floods. Captured requests can be resent to duplicate actions, while resource exhaustion attacks degrade the availability of MCP services.
\item \emph{Credential theft}: token theft/replay, confused-deputy proxy abuse, and insecure configuration. Attackers steal or misuse authentication tokens to impersonate agents and hijack privileged MCP operations.
\end{itemize}

\textcolor{black}{
Table~\ref{tab:llmagent_cves} presents representative CVEs from 2025 cross-mapped to the four proposed threat model categories, demonstrating the empirical grounding and real-world relevance of our taxonomy. These vulnerabilities span indirect prompt injection (e.g., LangChain’s GmailToolkit), server-side request forgery, XML-based data exfiltration, path traversal, protocol misconfigurations (e.g., in MCP-Remote), and sandbox escapes in agent tools. Notably, most of these attacks exploit insecure interfaces, permissive plugin logic, or protocol-level flaws—often requiring only low privileges or indirect input access—highlighting the brittleness of current LLM-agent integrations. This validates the practical urgency of securing prompt surfaces, agent-tool bridges, and multi-agent coordination flows through structured, layered defenses.}

\subsection{Agent2Agent (A2A) protocol}

A2A (Agent-to-Agent) \cite{googleblog2025} extends this framework horizontally by orchestrating secure, task-oriented interactions among multiple autonomous agents. Leveraging a JSON-RPC-over-HTTP(S) communication model with Server-Sent Events for streaming updates, A2A enables one agent (the client) to discover and delegate subtasks to another agent (the remote) based on metadata published in “Agent Cards.” Each task is a defined object whose lifecycle—from initiation to artifact delivery—is governed by the protocol, allowing agents to coordinate complex workflows and negotiate content formats (text, images, forms, etc.) without sharing internal memory or tools. When combined, MCP equips each agent with the precise context it needs, while A2A orchestrates multi-agent collaboration, supporting scalable, interoperable ecosystems of LLM-enabled agents \cite{habler2025building}.

Through A2A’s capability discovery mechanism, each agent advertises its skills in a JSON “Agent Card,” allowing a client agent to identify and negotiate with the most appropriate remote agent for a given task. Once a suitable match is found, A2A orchestrates the task lifecycle: the client agent transmits a formally defined “task” object, and the remote agent executes the required operations or returns an artifact. Throughout execution—whether a quick database lookup or a multi-day research workflow—A2A ensures secure, authenticated message exchanges via HTTP, Server-Sent Events, and JSON-RPC conventions. Modality-agnostic abstractions within A2A enable seamless transfer of text, images, or video, and real-time state updates maintain continuity during long-running tasks.

\vspace{1ex}
\noindent\textbf{A2A–layer Attack Vectors.}  In addition to the MCP-layer threats above, A2A protocols are vulnerable to:
\begin{itemize}
\item \emph{Cross-agent prompt injection}: crafted prompts that propagate through multi-agent workflows. These attacks embed malicious instructions in one agent’s output, causing downstream agents to perform unintended actions.
\item \emph{Agent discovery spoofing}: forged discovery responses or Agent-Card spoofing to impersonate malicious agents. The internal state leads to capability metadata to divert tasks to rogue agents under their control.
\item \emph{Rogue agent registration}: unauthorized or look-alike agents hijacking tasks. Malicious agents register with deceptive identifiers to intercept or tamper with delegated workflows.
\item \emph{Adaptive indirect injections}: dynamic chaining of prompts to evade filters. By incrementally modifying prompts across hops, attackers bypass individual-agent sanitization mechanisms.
\item \emph{Context manipulation}: injecting or modifying context blobs to hijack workflows. Altered context payloads lead agents to make incorrect decisions or reveal sensitive data.
\item \emph{Contagious recursive blocking}: recursive task spawning deadlocks creating denial-of-service. Agents inadvertently generate infinite loops of subtask delegation, exhausting system resources.
\item \emph{Memory subversion}: memory-poisoning (MINJA) within the agent framework. Injected memory artifacts corrupt the internal state, leading to unpredictable or malicious agent behavior.
\end{itemize}

Tab. \ref{tab:threat-matrix} summarizes the attack vectors across MCP \cite{modelcontextprotocol2025}, A2A\cite{googleblog2025}, ANP\cite{agentnetworkprotocol2024} \& ACP\cite{agentcommunicationprotocol2024} protocols.

\begin{table*}[!htbp]
\centering
\scriptsize
\rowcolors{2}{rowbg}{white}
\begin{tabular}{|p{2.4cm}|p{0.8cm}|p{3.2cm}|p{5cm}|p{1.8cm}|p{2.2cm}|}
\hline
\rowcolor{headerbg}
\textbf{\textcolor{black}{CVE ID}} & 
\textbf{\textcolor{black}{Date}} & 
\textbf{\textcolor{black}{Component Affected}} & 
\textbf{\textcolor{black}{Summary of the Issue}} & 
\textbf{\textcolor{black}{CWE-ID}} & 
\textbf{\textcolor{black}{Threat Category}} \\
\hline
\textcolor{black}{CVE-2025-46059 \cite {CVE-2025-46059}} & 
\textcolor{black}{2025} & 
\textcolor{black}{LangChain – GmailToolkit (v0.3.51)} & 
\textcolor{black}{Critical indirect prompt injection via crafted emails enabling arbitrary code execution.} & 
\textcolor{black}{CWE-94 \cite {CWE-94}} & 
\textcolor{black}{Input Manipulation} \\
\hline
\textcolor{black}{CVE-2025-2828 \cite {CVE-2025-2828}} & 
\textcolor{black}{2025} & 
\textcolor{black}{LangChain – RequestsToolkit (OpenAPI tool)} & 
\textcolor{black}{SSRF flaw bypasses URL filters, allowing internal network access.} & 
\textcolor{black}{CWE-918 \cite {CWE-918}} & 
\textcolor{black}{Protocol Vulnerabilities} \\
\hline
\textcolor{black}{CVE-2025-6984 \cite {CVE-2025-6984}} & 
\textcolor{black}{2025} & 
\textcolor{black}{LangChain – EverNoteLoader (v0.3.63)} & 
\textcolor{black}{XML External Entity (XXE) attacks via insecure XML parsing expose sensitive files.} & 
\textcolor{black}{CWE-200 \cite {CWE-200}} & 
\textcolor{black}{System \& Privacy Attacks} \\
\hline
\textcolor{black}{CVE-2025-6985 \cite {CVE-2025-6985}} & 
\textcolor{black}{2025} & 
\textcolor{black}{LangChain – HTMLSectionSplitter (v0.3.8)} & 
\textcolor{black}{Unsafe XSLT parsing allows file disclosure via malicious stylesheets.} & 
\textcolor{black}{CWE-611 \cite {CWE-611}} & 
\textcolor{black}{System \& Privacy Attacks} \\
\hline
\textcolor{black}{CVE-2025-6853 \cite {CVE-2025-6853}} & 
\textcolor{black}{2025} & 
\textcolor{black}{LangChain-Chatchat (v0.3.1) – Backend API} & 
\textcolor{black}{Path traversal in temp upload allows out-of-scope file creation.} & 
\textcolor{black}{CWE-22 \cite {CWE-22}} & 
\textcolor{black}{System \& Privacy Attacks} \\
\hline
\textcolor{black}{CVE-2025-6854 \cite {CVE-2025-6854}} & 
\textcolor{black}{2025} & 
\textcolor{black}{LangChain-Chatchat (v0.3.1) – Backend API} & 
\textcolor{black}{Base64 path traversal lets an attacker read unintended files.} & 
\textcolor{black}{CWE-22 \cite {CWE-22}} & 
\textcolor{black}{System \& Privacy Attacks} \\
\hline
\textcolor{black}{CVE-2025-6855 \cite {CVE-2025-6855}} & 
\textcolor{black}{2025} & 
\textcolor{black}{LangChain-Chatchat (v0.3.1) – Backend API} & 
\textcolor{black}{Path traversal in file uploads allows overwriting critical files.} & 
\textcolor{black}{CWE-22 \cite {CWE-22}} & 
\textcolor{black}{System \& Privacy Attacks} \\
\hline
\textcolor{black}{CVE-2025-11844 \cite {CVE-2025-11844}} & 
\textcolor{black}{2025} & 
\textcolor{black}{Hugging Face – SmolAgents (v1.20.0)} & 
\textcolor{black}{XPath injection lets attacker bypass filters, access unauthorized DOM elements.} & 
\textcolor{black}{CWE-643 \cite {CWE-643}} & 
\textcolor{black}{Input Manipulation} \\
\hline
\textcolor{black}{CVE-2025-59532 \cite {CVE-2025-59532}} & 
\textcolor{black}{2025} & 
\textcolor{black}{OpenAI – Codex CLI (v0.2.0–0.38.0)} & 
\textcolor{black}{Sandbox escape via directory control flaw allows arbitrary file execution.} & 
\textcolor{black}{CWE-20 \cite {CWE-20}} & 
\textcolor{black}{Model Compromise} \\
\hline
\textcolor{black}{CVE-2025-6514 \cite {CVE-2025-6514}} & 
\textcolor{black}{2025} & 
\textcolor{black}{Anthropic – MCP-Remote (v0.0.5–0.1.15)} & 
\textcolor{black}{Untrusted MCP endpoints can trigger remote OS command execution.} & 
\textcolor{black}{CWE-78 \cite {CWE-78}} & 
\textcolor{black}{Protocol Vulnerabilities} \\
\hline
\textcolor{black}{CVE-2025-49596 \cite {CVE-2025-49596}} & 
\textcolor{black}{2025} & 
\textcolor{black}{Anthropic – MCP Inspector (v0.14.0)} & 
\textcolor{black}{CSRF/DNS-rebinding vulnerability enables RCE via Inspector UI.} & 
\textcolor{black}{CWE-306 \cite {CWE-306}} & 
\textcolor{black}{Protocol Vulnerabilities} \\
\hline
\end{tabular}
\caption{\textcolor{black}{Representative CVEs cross-mapped to the four threat model categories. CWE-94 (Code Injection), CWE-918 (Server-Side Request Forgery), CWE-200 (Information Exposure), CWE-611 (Improper Restriction of XML External Entity Reference), CWE-22 (Path Traversal), CWE-643 (XPath Injection), CWE-20 (Improper Input Validation), CWE-78 (OS Command Injection), and CWE-306 (Missing Authentication for Critical Function).}}
\label{tab:llmagent_cves}
\end{table*}

\section{Open Challenges and Future Directions} \label{sec:7}

As LLMs and agentic systems become increasingly pervasive, securing them against sophisticated threats remains a significant challenge. This section identifies the key areas where current defenses fall short and outlines promising directions for future research.

\subsection{MCP protocol vulnerabilities}

The Model Context Protocol (MCP) provides a unified, plug-and-play framework—akin to a “USB-C port for AI” for connecting large language models to diverse data sources and tool endpoints. By decoupling responsibilities into hosts (applications), clients (connectors), and servers (modular adapters), MCP greatly accelerates the assembly of complex, LLM-driven workflows. Yet this very modularity widens the attack surface: every integration point—from prompt handling to long-running connections—can be exploited if left unchecked \cite{hasan2025model}.

The first significant challenge lies in establishing dynamic trust and adaptive policy enforcement within MCP deployments. Traditional static access controls cannot keep pace with the fluid, language-driven workflows that modern agents require. Future research should explore declarative policy languages that allow hosts to negotiate and update least-privilege rules at runtime, as well as distributed trust-negotiation protocols capable of granting time-boxed permissions for specific tool invocations without human intervention \cite{hou2025model}.

A second frontier concerns the integrity of context and provenance tracking. Because MCP maintains conversational state across multiple turns and service calls, any undetected tampering can subtly bias downstream actions. Promising directions include using secure enclaves or hardware root-of-trust modules to vault critical session memory and immutable audit logs—perhaps built on lightweight blockchain ledgers—to attest to the sequence of context updates cryptographically \cite{ahmadi2025mcp}.

Ransomware\cite{al2024securing} can cripple an MCP deployment by encrypting or locking access to the very data sources and tool-endpoints that MCP servers expose, effectively severing the “USB-C port” connections between LLMs and critical context (e.g., local files, databases, or APIs). Because MCP’s client–server architecture allows a host to plug into multiple lightweight servers, a single compromised server can cascade, denying clients the ability to fetch prompts, workflows, or sampling configurations and halting AI-driven operations until a ransom is paid, directly undermining MCP’s goals of flexible, secure integration within your infrastructure. 

Finally, there is an urgent need for anomaly detection and formal verification tailored to MCP’s unique attack surface. Standard intrusion-detection systems are poorly suited to natural-language payloads; instead, machine-learning models must be trained to recognize semantic irregularities in prompts, tool responses, and context transitions. At the same time, formally specifying MCP’s client–server interactions and verifying them against security properties can prevent entire logic-flaw vulnerabilities before deployment. By integrating zero-trust principles, cryptographic attestation of tool binaries, context-aware anomaly detection, and rigorous formal methods, the next generation of MCP implementations can combine the agility of modular AI workflows with the robust security guarantees required for production use.

\subsection{Agentic Web Interfaces: Design and Security Challenges}
Lu et al. \cite{lu2025build} present a forward-thinking perspective on advancing web-based AI agents by proposing the Agentic Web Interface (AWI), a new interaction paradigm tailored to the unique capabilities and limitations of LLM-powered agents. As LLMs and their multimodal counterparts increasingly demonstrate the potential to automate complex web tasks, the fundamental misalignment between human-centric web interfaces and agent requirements has become a critical bottleneck. Current approaches, which attempt to adapt agents to navigate intricate web inputs such as dense Document Object Model (DOM) trees, screenshot-based environments, or direct API interactions, have proven insufficient in addressing these challenges. The authors introduce AWI, a standardized, agent-optimized interface that enables more efficient, reliable, and transparent interactions between AI agents and web environments, thereby overcoming these limitations. They outline six core design principles for AWI, focusing on safety, efficiency, and standardization, and emphasize the importance of collaboration across the machine learning community to realize this vision.

Despite its promise, the introduction of AWI also raises several open problems and security-related challenges that must be addressed to ensure safe deployment. First, introducing agent-optimized interfaces may inadvertently create new vulnerabilities, including unauthorized agent access, manipulation of AWI components, or exploitation of poorly implemented standardization protocols. Furthermore, as AWI enables more autonomous agent behaviors, attackers may exploit LLM-driven agents for malicious activities such as automated phishing, data exfiltration, or denial-of-service attacks within web environments. Future research should focus on developing robust security mechanisms for AWI, including authentication, integrity verification, and real-time anomaly detection, to effectively monitor agent behavior. Additionally, establishing secure communication standards and incorporating adversarial robustness testing for LLM-agent interactions will be essential. Promising research directions include exploring formal verification techniques for AWI protocols, creating secure sandbox environments for agent testing, and developing mechanisms for transparent auditing of agent actions. Addressing these vulnerabilities is crucial to ensure that the shift toward agent-optimized web interactions enhances functionality and security in LLM-powered agent ecosystems.

\subsection{Optimizing LLM Multi-Agent Systems: Design and Security}
Zhou et al. \cite{zhou2025multi} investigate the design complexity of Large Language Model (LLM)-powered multi-agent systems (MAS) and introduce an automated optimization framework, Multi-Agent System Search (MASS), to streamline the process. Recognizing that prompts and interaction topologies are critical for enabling effective collaboration among agents, the authors conduct a comprehensive analysis of the MAS design space. Building on these insights, they present MASS, which employs a staged optimization process that iteratively refines prompts and system topologies across three levels: block-level prompt optimization, workflow topology optimization, and global prompt optimization. Experimental results demonstrate that systems designed with MASS significantly outperform existing MAS approaches. The authors conclude by outlining key design principles for building robust, high-performing multi-agent systems.

Despite these advancements, several open challenges and vulnerabilities remain unaddressed in the design and deployment of LLM-driven MAS. The automation of prompt and topology generation, while improving system efficiency, introduces new risks, including unpredictable agent behavior, prompt injection attacks, and adversarial manipulation of communication topologies. Moreover, as MAS grow more complex, detecting unintended coordination patterns or hidden vulnerabilities becomes increasingly difficult. Future research should focus on integrating security constraints directly into the MASS optimization process, alongside the development of verification techniques to analyze the safety and robustness of optimized multi-agent configurations. Promising directions also include designing resilient topologies that can withstand adversarial disruptions, incorporating defensive prompting strategies, and exploring real-time monitoring tools to detect and mitigate malicious agent interactions in LLM-powered MAS.

\subsection{VLM-Powered Web Agents and Security Challenges}

Andreux et al. \cite{andreux2025surfer} introduce Surfer-H, a cost-efficient web agent that leverages Vision-Language Models (VLMs) to perform user-defined tasks within web environments autonomously. Central to this system is Holo1, an open-weight VLM collection specifically trained for web navigation and information extraction. Holo1 is developed using a diverse dataset that combines open-access web content, synthetic examples, and self-generated agentic data. Experimental results demonstrate that Holo1 outperforms general-purpose UI benchmarks and achieves leading results on the newly proposed WebClick benchmark for web UI localization. When integrated with Holo1, Surfer-H achieves a state-of-the-art accuracy of 92.2\% on the WebVoyager benchmark, offering a strong balance between task performance and cost efficiency. Therefore, several open challenges and security vulnerabilities arise when using VLM-powered web agents, such as Surfer-H. Integrating vision and language capabilities for autonomous web interaction raises concerns about interface manipulation, adversarial attacks on visual perception, and the misuse of automated browsing for malicious purposes, such as data scraping or phishing. Furthermore, the reliance on agent-generated data during training may unintentionally introduce biased behaviors or security flaws. Future research should focus on strengthening the robustness of VLM-based web agents against adversarial inputs, ensuring secure interaction with web environments, and establishing guidelines to prevent the abuse of agentic web technologies. Efforts toward explainability, real-time anomaly detection, and secure agent deployment practices will be essential for building reliable and trustworthy web agent ecosystems.

\subsection{Memory-Centric LLMs for Long-Term Dialogue and Emerging Security Risks}

Chhikara et al. \cite{chhikara2025mem0} present Mem0, a scalable memory-centric architecture designed to overcome the inherent limitations of Large Language Models (LLMs) in maintaining consistency over long, multi-session dialogues. Recognizing the challenges posed by fixed context windows, Mem0 dynamically extracts, consolidates, and retrieves relevant information from conversations to ensure contextual coherence. The authors also introduce an enhanced variant that incorporates graph-based memory structures to model complex relationships among conversational elements. Through extensive evaluations on the LOCOMO benchmark, the proposed approaches are compared against six established memory and retrieval baselines, including memory-augmented systems, RAG techniques, full-context models, and proprietary solutions. The results demonstrate that both Mem0 and its graph-based extension outperform all competing methods across diverse question categories, while significantly reducing computational overhead. Mem0 achieves notable improvements in accuracy and efficiency, highlighting the value of persistent, structured memory for advancing long-term conversational capabilities in LLM-driven agents.

While Mem0 advances the state of memory-enhanced LLMs, its reliance on structured memory mechanisms also introduces new vulnerabilities that require attention. The processes of information extraction, consolidation, and retrieval within long-term memory could be exploited by adversaries through data poisoning, manipulation of stored knowledge, or injection of misleading conversational elements. Additionally, the use of graph-based memory structures raises concerns about unauthorized inference of hidden relationships or the leakage of sensitive information. Future research should focus on securing memory retrieval pipelines, implementing tamper-resistant memory architectures, and developing mechanisms for real-time monitoring of memory integrity to ensure robustness and security. Promising directions include integrating privacy-preserving techniques, testing for adversarial robustness in memory components, and formal verification of memory consistency to ensure the effectiveness and security of LLM-powered conversational agents.

\subsection{Evolutionary Coding Agents and Emerging Security Challenges}
Novikov et al. \cite{novikov2025alphaevolve} introduce AlphaEvolve, an autonomous evolutionary coding agent designed to push the boundaries of Large Language Models (LLMs) in addressing complex scientific and engineering challenges. AlphaEvolve operates by orchestrating a pipeline of LLMs that iteratively refine source code based on evaluators' feedback. This evolutionary process enables the agent to optimize algorithms and potentially uncover novel, high-performance solutions. The authors demonstrate AlphaEvolve’s broad applicability through its successful deployment on critical computational problems, including the optimization of data center scheduling algorithms, simplification of hardware accelerator circuits, and acceleration of LLM training. Impressively, AlphaEvolve also discovered provably correct algorithms that surpass existing state-of-the-art methods in both mathematics and computer science, underscoring its potential to automate scientific discovery at scale.

While AlphaEvolve represents a significant advancement in automated coding agents, its architecture also introduces new vulnerabilities and security risks that must be carefully considered. The autonomous modification of code, especially in critical systems, creates potential attack vectors, such as the introduction of subtle backdoors, the exploitation of evaluator feedback loops, or the undetected generation of unsafe code. Moreover, the iterative nature of AlphaEvolve could amplify vulnerabilities if malicious code modifications are reinforced over successive generations. Future research should prioritize the development of secure evaluation mechanisms, formal verification of generated code, and techniques for detecting adversarial manipulation within the evolutionary process. Promising directions also include implementing strict access controls, real-time monitoring, and integrating secure coding standards to ensure that autonomous coding agents, such as AlphaEvolve, contribute to scientific and engineering progress without compromising system integrity.

\subsection{Robustness and Scalability of Defense Mechanisms}

Despite the proliferation of input sanitization and prompt-filtering techniques, adversaries continue to discover new injection and jailbreak strategies that evade existing defenses. Current solutions often rely on static rules or model‐specific heuristics, which break down when faced with adaptive attackers who leverage long contexts, compositional prompts, or multi‐modal perturbations. Furthermore, most defense evaluations rely on limited benchmarks that fail to capture the full range of real-world inputs, leaving gaps in our understanding of LLM vulnerabilities.

Future work could prioritize the development of scalable, modality‐agnostic defense frameworks that offer formal guarantees of robustness. This includes adversarial training regimes that integrate continuous red-teaming (e.g., jailbreak fuzzers) into the development lifecycle, formal verification methods to certify the correctness of sanitizers and filters, and cross-model transfer evaluation to ensure defenses generalize beyond a single LLM implementation. Finally, building open, extensible benchmarks encompassing text, code, image, and audio attacks will be critical for measuring progress and guiding the community toward truly resilient systems.

\subsection{Securing Model Parameters and Memory Integrity}

Model-compromise attacks—such as backdoors and data- or memory-poisoning—pose a serious threat because they persist beyond individual queries and can be nearly impossible to detect at inference time. Stealthy triggers embedded during fine‐tuning or through external memory banks can lie dormant until activated, undermining trust in the model’s outputs. Moreover, dynamic encryption techniques (e.g., DemonAgent) and composite backdoors that scatter triggers across inputs make traditional audit and anomaly detection insufficient.

Addressing these challenges requires new tools for end‐to‐end pipeline auditing and runtime integrity verification. Promising directions include: (1) cryptographic attestation of training artifacts and prompt templates to ensure provenance; (2) differential privacy and noise injection in memory retrieval to limit an attacker’s ability to influence the agent’s stored context; (3) continuous monitoring of model internals using lightweight trapdoors or canary triggers that signal unauthorized backdoor activations; and (4) tooling for transparent inspection and rollback of memory banks in deployed agents. By combining these approaches, we can begin building models and agents whose internal states are both auditable and tamper-resistant.

\subsection{\textcolor{black}{Future Mitigation Directions for MCP SDK Security}}

\textcolor{black}{
Another critical open challenge concerns the secure design and deployment of Model Context Protocol (MCP) Software Development Kits (SDKs), which often serve as the backbone for agent-to-tool integrations. To enhance their robustness, developers should adopt mitigation-oriented engineering practices such as enforcing least-privilege permissions, using mutual authentication and encrypted communications, validating structured data schemas, monitoring runtime anomalies, and maintaining cryptographically verifiable provenance logs. Furthermore, sandboxing plugin execution and auditing dependencies helps prevent malicious package injection and ensures the long-term resilience of MCP-based ecosystems. These practical measures provide a roadmap for strengthening SDK security and lay the foundation for trustworthy, large-scale agentic infrastructures.
}

\subsection{Resilience in Multi-Agent and Federated Environments}

The rise of multi‐agent coordination protocols (MCP, A2A) and federated learning frameworks has expanded the attack surface to include inter‐agent communication channels and decentralized training pipelines. Context manipulation, contagious recursive blocking (Corba), and federated local model poisoning enable adversaries to disrupt collaborations or leak private data across agents. Side-channel and membership inference attacks further compromise confidentiality in shared environments, while social engineering simulations demonstrate that human-in-the-loop scenarios remain vulnerable.

\textcolor{black}{
The term \textit{resilient multi-agent environments} refers to systems capable of maintaining functionality and coordination in the face of both accidental faults and deliberate adversarial disruptions. Specifically, resilience encompasses fault tolerance—ensuring continued operation despite node or communication failures—and adversarial robustness—preserving secure interactions even when agents are exposed to manipulation, poisoning, or collusion.
}

Future research could focus on designing provably secure communication and consensus protocols tailored for LLM‐based agents to secure these distributed systems. This may involve: (1) integrating secure multi‐party computation and threshold cryptography to protect model updates in federated training; (2) establishing cross‐agent attestation mechanisms that verify the integrity and alignment of each participant before task delegation; (3) adopting blockchain or distributed ledger anchors for immutable audit trails of inter‐agent messages; and (4) developing differential privacy guarantees for federated model aggregation to thwart membership inference. Together, these strategies can help ensure that agent societies and federated networks remain robust against coordinated and emerging threats.

\section{Conclusion}\label{sec:8}
In this paper, we presented the first unified, end-to-end threat model for LLM-powered AI agent ecosystems, spanning both host-to-tool and agent-to-agent communications. We began by formalizing adversary capabilities and attacker objectives, then systematically cataloged over thirty attack techniques across four high-level domains: Input Manipulation, Model Compromise, System \& Privacy Attacks, and Protocol Vulnerabilities. For each category, we examined representative scenarios, evaluated real-world feasibility, and reviewed existing defenses. Building on this analysis, we identified critical gaps—such as dynamic trust management for MCP, secure design of Agentic Web Interfaces, adversarial robustness in multi-agent optimization and VLM-based web agents, tamper-resistant memory architectures, secure evolutionary coding pipelines, scalable modality-agnostic defense frameworks, end-to-end integrity verification of model parameters, and resilience in federated and multi-agent environments.  

Our comprehensive taxonomy and threat survey provide a foundation for designing robust defense mechanisms and establishing best practices for resilient, secure LLM-agent workflows. \textcolor{black}{This framework provides practical value for developers securing agent workflows, researchers designing defenses, and policymakers establishing safety standards for LLM-based systems.} We hope this work will guide future research and practical development toward closing the identified security gaps and enabling trustworthy, autonomous AI agents in real-world deployments. 

\bibliographystyle{IEEEtran}
\bibliography{bibliography} 

\end{document}